\documentclass[a4paper]{JHEP3}
\usepackage[centertags]{amsmath}
\usepackage{amssymb}
\usepackage{fixltx2e}
\MakeRobust{\eqref} 
\usepackage{graphicx}
\usepackage{subfigure}
\preprint{TIFR/TH/13-08}

\newcommand{\Tr}{\text{Tr}} 
\newcommand{\ra}{\rangle}
\newcommand{\la}{\langle}

\def \beal#1 {\begin{align}#1\end{align}}

\def\Tr{\mathrm{Tr}}

\def\[{\left[}
\def\]{\right]}
\def\({\left(}
\def\){\right)}

\def\={\stackrel{\bullet}{=}}

\def\Tr{\mathrm{Tr}}

\def\[{\left[}
\def\]{\right]}
\def\({\left(}
\def\){\right)}

\def\cF{{\cal F}}

\def \be {\begin{equation}}
\def \ee {\end{equation}}
\def \bea {\begin{eqnarray}}
\def \eea {\end{eqnarray}}

\usepackage{enumerate}

\title{
Duality and Higher Temperature 
Phases of Large $N$ Chern-Simons Matter 
Theories on $S^2 \times S^1$}

\author{
Tomohisa Takimi, 
\\
$^{a)}$Department of Theoretical Physics, Tata Institute of Fundamental Research,
Homi Bhabha Road, Mumbai 400005, India\\
{\small \tt E-mail: 
takimi(at)theory.tifr.res.in}
}
\abstract{It has been recently demonstrated that the thermal partition function of any large $N$ Chern-Simons gauge theories on $S^2 \times S^1$, coupled to fundamental matter, reduces
to a capped unitary matrix model. The matrix models corresponding to several 
specific matter Chern-Simons theories at temperature $T$ were determined in \cite{Jain:2013py}.
The large $N$ saddle point equations for these theories were determined in the 
same paper, and were solved in the low temperature phase. In this paper we 
find exact solutions for these saddle point equations in three other phases of these theories and thereby explicitly determine the free energy of the corresponding theories at all values of $T^2/N$. As anticipated on general grounds in 
\cite{Jain:2013py}, our results are in perfect agreement with conjectured
level rank type bosonization dualities between pairs of such theories.}

\begin{document}
\section{Introduction}
Recently, three dimensional field theories have become 
fascinating in the context of AdS/CFT (or dS/CFT) correspondences
which do not necessarily rely on supersymmetry~\cite{Klebanov:2002ja}.
It has been recently conjectured that level $k$ 
Chern-Simons theories with matter in the 
fundamental/bifundamental representations admit a dual 
description governed by 
parity violating
Vasiliev's higher spin 
equations~\cite{Vasiliev:1990en,Vasiliev:2003ev}
(see \cite{Giombi:2012ms} for more detail and references) 
at finite values of the 
't~Hooft coupling $\lambda=\frac{N}{k}$ \cite{Giombi:2011kc,Chang:2012kt}. 

Sometimes, analysis of 
boundary field theories can be applied to 
study bulk gravity theories through the 
AdS/CFT correspondence when the direct analysis of the 
gravity theory is difficult. 
For example, it is known that the AdS/CFT correspondence
maps deconfinement transitions of large $N$ gauge 
theories on spheres to gravitational phase transitions involving 
black hole nucleation~\cite{Witten:1998zw}.
This observation has motivated the 
intensive study of the phase structures as well as the 
deconfinement phase transitions of large $N$
p+1 dimensional Yang Mills theories (coupled to adjoint and fundamental
matter) on $S^p$. 
Now, we can expect similar application 
of the Chern-Simons matter theories 
to know the phase structures of  parity violating Vasiliev's higher spin gravity
theories.

From this motivation, in \cite{Jain:2013py},
the finite temperature phase structure of renormalized level $k$ U($N$) Chern-Simons (CS) theories coupled to a finite number of fundamental fields on $S^2 \times S^1$ 
in the t' Hooft limit $N \to \infty$, $k \to \infty$ with 
$\lambda=\frac{N}{k}$ fixed were studied.
The phase structure was studied by using a simple toy large $N$ 
Gross-Witten-Wadia (GWW) matrix integral, and it turned out that there is a
rich phase structure caused by the summation over the $U(1)$ flux sectors in 
the Chern-Simons theory. 
In particular, 
the eigenvalues must be discretized and
the eigenvalue density function must be saturated from 
upper bound $\frac{1}{2\pi\lambda}$.
This saturation was first suggested by \cite{Aharony:2012ns}.
This discretization generates the new phases so-called the ``upper gap phase" 
and the ``two gap phase", which are absent in Yang-Mills theories.
\footnote{
The same upper bound for the density function appears in two dimensional Yang-Mills theory ($p=1$), in which case the situation becomes similar due to the fact that there are no propagating degrees of freedom for the gauge field. 
A phase transition relevant to this upper bound was studied in 2d (q-deformed) Yang-Mills theory on $S^2$ \cite{Douglas:1993iia,Arsiwalla:2005jb,Caporaso:2005ta,Jafferis:2005jd}, which we will see from Chern-Simons theory. 
}
The level-rank duality
has also been confirmed in the GWW toy model in \cite{Jain:2013py}
(see \cite{Kapustin:2010xq,Kapustin:2010mh} for a relatively 
recent discussion of level-rank duality and references to earlier work),
but the phase structures of the actual Chern-Simons matter theories have
not been worked out. Moreover, we have not had the 
perfect analytic proof of the duality, even in the GWW model.

In this paper we will work out the study of the phase structure 
of several Chern-Simons matter theories:
(1) the CS theory minimally coupled to the fundamental fermion 
\cite{Giombi:2011kc},
(2) the 
CS theory coupled to massless critical 
bosons~\cite{Aharony:2011jz,Aharony:2012nh},
and (3) the ${\cal N} = 2$ supersymmetric CS theory with a 
single fundamental chiral multiplet~\cite{Ivanov:1991fn,Gaiotto:2007qi}
(SUSY CS matter theory).
We will also provide 
the perfect 
analytic proof of the level-rank duality (Giveon-Kutasov type duality) 
on the $S^2 \times S^1$ 
between the regular fermion theory
and the critical boson theory. Moreover we will 
prove the level-rank self-duality
in both the SUSY CS matter theory and the GWW matrix integration as well.
We will also supply several useful formulae
which can be used for any analysis related to the duality 
in general CS matter theories on $S^2 \times S^1$.

\subsection{Outline of this paper}
The rest of this paper is organized as follows:
In section 2 we will review the Landau-Ginzburg 
description of the partition function of Chern-Simons theories
coupled to fundamental matters as a preliminary, and supply techniques
to evaluate eigenvalue densities as well as partition functions
by using complex line integration in the complex plane.
In section 3, we will elaborate the phase structure 
of the regular fermion theory and the critical boson theory,
confirming that the level-rank duality between them
is in agreement.
In section 4, we will study the phase structure of the ${\cal N} = 2$
supersymmetric CS matter theory and examine the Giveon-Kutasov type 
duality.
We will confirm the self-duality of the theory.
In section 5, we will give the analytic proof of 
the level-rank self-duality in the two gap 
phase of the GWW toy model. 
Section 6 is devoted to the summary and discussion.

The content of the appendix is as follows:
We put several definitions of the cut regions and cut functions in
appendix~\ref{Ap:CUT}.
We give the proof of important formulae used in the proof of the 
duality and others in appendix \ref{Ap:Form}.
In appendix~\ref{Ap:LZ-sec}, we provide the detailed
analysis of the large $\zeta$ behavior of the eigenvalue density functions.
In appendix \ref{Ap:LZ-sec},
we also prove that the eigenvalue distributions at $\zeta \to \infty$
converge to the universal distribution \eqref{evd 0}.

\section{Preliminaries}
\subsection{High temperature effective action of the Chern-Simons-fundamental 
matter theory on $S^2 \times S^1$}
We will consider the level $k$ $U(N)$ 
fundamental matter Chern-Simons theories in the 't Hooft large $N$ limit, 
with\footnote{
We use the dimensional reduction regulation scheme throughout this paper. 
In this case $|k|=|k_{YM}|+N$, where $k_{YM}$ is the level of the Chern-Simons theory regulated by including an infinitesimal Yang Mills term in the action,
and $k$ is the level of the theory regulated in the dimensional reduction scheme
\cite{Pisarski:1985yj,Chen:1992ee}. 
For this reason this class of Chern-Simons theories may be well-defined only in the range $|\lambda| \leq 1$.
Hereafter we assume that 
the 't~Hooft coupling is always positive for simplicity. 
Our results are easily generalized to the case where 
$\lambda$ is negative by taking the absolute value of it. 
}
\begin{equation}
V_2 T^2 = N \zeta, \qquad \lambda = \frac{N}{k} \label{eq:zeta-N-T}
\end{equation}
where $\zeta$ and $\lambda$ are held fixed in the large $N$ limit.
These theories have recently been studied intensively at finite 
$\lambda$~\cite{Giombi:2011kc,Aharony:2011jz,Maldacena:2011jn,Maldacena:2012sf,Banerjee:2012gh,Chang:2012kt,Aharony:2012nh,Jain:2012qi,Yokoyama:2012fa,Banerjee:2012aj,GurAri:2012is,Aharony:2012ns}, and thermal partition functions 
of these theories are studied in 
\cite{Shenker:2011zf,Giombi:2011kc,Chang:2012kt}.

We will consider the partition function of these theories
\begin{equation}
\label{purecsa0}
Z_{\text{CS}}=\int DA D \mu~
e^{i \frac{k}{4\pi} \Tr  \int \left( AdA + \frac{2}{3} A^3 \right)
 -S_{matter} }
\end{equation}
where $D\mu$ denotes the integration measure of matter fields.
To calculate it, we will integrate the matter fields first, and obtain the 
effective action which depends on the gauge fields.
According to the study in \cite{Jain:2013py},
the partition function is given as
\begin{equation}
\label{purecsa}
Z_{\text{CS}}=\int DA 
e^{i \frac{k}{4\pi} \Tr  \int \left( AdA + \frac{2}{3} A^3 \right)
 -S_{eff}(U) }
\end{equation}
where $S_{eff}(U)$ is a function of the 
two dimensional holonomy fields $U(x)$ around the 
thermal circle $S^1$.
Here $x$ of $U(x)$ indicates a point on the $S^2$, and its eigenvalue is 
$e^{i \alpha_{m}(x)}$ where $m$ is the index of $N \times N$ 
matrix.
The effective action $S_{eff}(U)$ would be
obtained by summing up all vacuum graphs 
which involve at least one matter field.
The graphs which do not involve any matter fields are not summed yet,
but they are summed at the stage of the path integration in 
\eqref{purecsa}.

To calculate \eqref{purecsa}, 
it is useful to observe that $S_{eff}(U)$ is ultralocal 
in large $N$ as well as in high temperature.
The $S_{eff}(U)$ would be 
expanded as a series of local operators,
\begin{equation}\label{ltdf}
S_{eff}(U) = \int d^2x \left( T^2 \sqrt{g} v(U)   + v_1(U) \Tr D_i U D^i U        
+ \ldots \right) .
\end{equation}
Note that 
the scaling \eqref{eq:zeta-N-T} converts higher temperature expansion
\eqref{ltdf} into expansion in inverse power of $N$.
Here the first term is leading ${\cal O}(N^2)$ and the second term is 
${\cal O}(N^1)$ and terms $\ldots$ are further suppressed at large $N$.
\footnote{Every term in the effective 
action \eqref{ltdf}  is of order $N$ at fixed $T$, as the action is generated by integrating out fundamental 
fields.} 
Then at large $N$, the effective action is simplified to be 
the leading term 
\begin{equation}
S_{eff}(U) = \int d^2x T^2 \sqrt{g} v(U).
\label{Eq:EA}
\end{equation}
This \eqref{Eq:EA} is originally suggested in \cite{Aharony:2012ns},
which is a consequence of the entire effect of matter loops on gauge 
dynamics at temperature $\sqrt{N}$ and at leading order in $N$.
By \eqref{Eq:EA}, the partition function would be given by
\begin{equation}
\begin{split}
Z_{\text{CS}}&=\int DA e^{i \frac{k}{4\pi} \Tr  \int \left( AdA + \frac{2}{3} A^3 \right)
 -T^2 \int d^2x \sqrt g  ~ v(U)  }
\\
&=
\la 
e^{-T^2 \int d^2x \sqrt g  ~ v(U(x))}  
\ra_{N,k} 
\\
&=
\la 
e^{-V_2T^2 v(U)  }
\ra_{N,k} = \la 
e^{-N\zeta v(U)  }
\ra_{N,k}
\end{split}
\label{purecs-2}
\end{equation}
where 
\begin{equation}
\la \Psi \ra_{N,k}
\end{equation}
is the expectation value of $\Psi$ in the 
pure $U(N)$ Chern-Simons theory at level $k$. 
Since the Chern-Simons theory is topological, 
expectation values become independent of $x$, then 
the $x$ dependence is removed in the step
from the second line to the third line of 
\eqref{purecs-2}. 
Here $V_2$ is the volume of $S^2$.

\subsection{Path integration to obtain the partition function, 
deconfinement phase transition and 
further phase transition caused by flux}
We can perform the calculation of \eqref{purecs-2}
in the same manner as \cite{Blau:1993tv}.
You can see the details of how to compute the path integration
in \cite{Jain:2013py}.
Keeping the effective potential term $N \zeta v(U)$ intact,
we achieve the following value 
\begin{equation}
\begin{split}
Z_{\text{CS}}
&=
\int \prod_{j =1}^{N} d\alpha_{j}
 \left(
\prod_{m \ne l}
2 \sin \left(
\frac{\alpha_{m}(n_m) - \alpha_{l}(n_l)}{2}
\right)
\right) 
e^{- N\zeta v(U)}
\left(\sum_{M_j = - \infty}^{\infty}
e^{ik M_j \alpha_j} \right)
\\
&=
\int \prod_{j =1}^{N} d\alpha_{j}
 \left(
\prod_{m \ne l}
2 \sin \left(
\frac{\alpha_{m}(n_m) - \alpha_{l}(n_l)}{2}
\right)
\right) 
e^{- N\zeta v(U)}
\sum_{n \in \mathbb{Z}}
\delta(k \alpha_j - 2 \pi n)
\end{split}
\end{equation}
where $M_j$ are constant units of flux in the $U(1)_j$ $(j = 1, \ldots N)$
factors.
As we can see in the last line,
the summation over flux $M_j$ in $e^{i k M_j \alpha_j}$ 
constrains $\alpha_j$ to take 
discrete values
\begin{equation}
\alpha_j = \frac{2 \pi n_j}{k}.
\label{discrete constraints}
\end{equation}
Then the partition function would be 
\begin{equation}
\label{csmintg}
Z_{\text{CS} }=\prod_{m=1}^N \sum_{n_m=-\infty}^\infty 
\left[ \left(
\prod_{l \neq m}2 \sin \left(\frac{\alpha_l(\vec n)-\alpha_m(\vec n)}{2}
\right) 
\right)
e^{-N\zeta v(U)}\right]
\end{equation} 
where the summation over $n_m$ is restricted so that no two $n_m$ are 
allowed to be equal. 

\subsection{Eigenvalue density}
We will estimate \eqref{csmintg} in the 't Hooft limit, 
$k \to \infty, N \to \infty, \lambda = \frac{N}{k}: \text{fixed}$.
The summation over the eigenvalue \eqref{csmintg} in the limit
is dominated by the saddle point configuration of the eigenvalues at
large $N$.
The saddle points minimize the following potential
\begin{equation}\begin{split}\label{vef}
V(U)-\sum_{m \neq l} \ln 2 \sin \frac{\alpha_m-\alpha_l}{2} 
\end{split}
\end{equation}
where $V(U) = N \zeta v(U)$.
The saddle points are obtained by solutions of the following equation
\begin{equation}\label{vefe}
V'(\alpha_m)=\sum_{m\neq l}\cot \frac{\alpha_m-\alpha_l}{2} .
\end{equation}
In terms of eigenvalue density function $\rho(\alpha)$, 
the equation would be represented by
\begin{equation}
\label{vefec}
V'(\alpha_0)= N {\cal P}  \int d \alpha \cot \frac{\alpha_0-\alpha}{2} 
\rho(\alpha) \end{equation}
where $\rho(\alpha) = \frac{1}{N} 
\sum_{m=1}^{N} \delta(\alpha - \alpha_m)$.

We can see that this saddle point equation problem is 
very similar to the Gross-Witten-Wadia (GWW) problem~\cite{Gross:1980he,Wadia:2012fr,Wadia:1980cp} in the Yang-Mills theory on $S^p \times S^1$.
In a usual GWW problem in Yang-Mills theories, a thermal 
partition function obtained after integrating out all massive modes 
is given by the integration over the single unitary matrix as
\begin{equation}
\label{YM_GWW}
Z_{YM} =
\prod_{m=1}^{N}
\int^{\infty}_{-\infty} d \alpha_m
\left[
\prod_{l \ne m} 2 \sin \left(
\frac{\alpha_l - \alpha_m}{2}\right)
e^{-V_{YM}(U)}
\right],
\end{equation}
and there is a competition between 
potential $V_{YM}(U)$
that tends to clump eigenvalues, and the measure
factor (Vandermonde determinant) which tends to repel them.
\footnote{
The effective potential $V_{YM} (U)$ was computed in free gauge theories
\cite{Sundborg:1999ue, Aharony:2003sx}; it has also been evaluated at
higher orders in perturbation theory in special examples
\cite{Aharony:2005bq, Aharony:2006rf,Papadodimas:2006jd, Mussel:2009uw}.
At least in
perturbation theory \cite{Aharony:2003sx} and perhaps beyond
\cite{AlvarezGaume:2005fv,AlvarezGaume:2006jg}, 
the potential $V_{YM}(U)$
is an analytic function of $U$.
}
At low temperature with small $\zeta$, 
the repulsive force by the measure factor is stronger.
Then the eigenvalue density would have support everywhere on 
$-\pi\le |\alpha| \le \pi$.
But on the other hand, in the high enough temperature, the attracting force
caused by the potential $V_{YM}(U)$ becomes stronger than 
the repulsive force, and the eigenvalue will be clumped. 
Then in the high temperature the eigenvalue distribution 
has support only on a finite arc, 
there would be a domain of eigenvalues so-called
"lower gap" such that 
$\{ \alpha | \alpha \sim \alpha + 2\pi, \rho(\alpha) = 0\}$,
and phase transition would occur.
At extremely high temperature $T \to \infty$, the eigenvalue density
function would be clumped to be a delta function $\rho(\alpha) \sim 
\delta (\alpha)$.

Also in the current Chern-Simons case \eqref{csmintg}, 
such a competition between 
$V(U)$ and measure term exists.
But in this case, unlike the 
usual Yang-Mills case \eqref{YM_GWW}, 
there is a constraint on the 
eigenvalue \eqref{discrete constraints}.
\eqref{discrete constraints} 
saturates the eigenvalue density from above as
\begin{equation}
\rho(\alpha) \le \frac{k}{2 \pi} \times \frac{1}{N} = \frac{1}{2 \pi \lambda}
\label{Ine-CS}
\end{equation}
while the eigenvalue density in the Yang-Mills theory
is not bounded from above (is bounded only from below). 
Saddle points of 
\eqref{YM_GWW} do not always obey the inequality
\eqref{Ine-CS}, then the Yang-Mills solutions violating 
\eqref{Ine-CS} will not be the solution of the 
Chern-Simons saddle point equations 
\eqref{vefec} and \eqref{csmintg}.
Instead, the current Chern-Simons theory admits new classes of 
solutions saturating the upper bound of the inequality over several arcs
along the unit circle.
Hence in this Chern-Simons case, there can exist not only the "lower gaps"
but also the "upper gaps", which are arcs over which the upper bound of 
\eqref{Ine-CS} are saturated.

Hence the phase structures of the current Chern-Simons theories 
\eqref{csmintg} would be different from the one of usual Yang-Mills theory 
\eqref{YM_GWW}.
In a usual Yang-Mills theory case, 
there are only two phases, 
"no gap phase" and "lower gap phase".
In the no gap phase, the eigenvalue has support everywhere on the 
unit circle in the complex plane, on the other hand in the lower gap
phase, the eigenvalue has support only on a finite arcs.
On the other hand,  
in the current 
Chern-Simons case, due to the existence of 
upper bound \eqref{Ine-CS},
not only the no gap and lower gap phases,
but also "upper gap" and "two gap" phases exist.
In the upper gap phase, there is one upper gap where the upper bound of 
\eqref{Ine-CS} is saturated, 
and the eigenvalue density has support everywhere on the unit circle.
In the two gap phase, there is one upper gap as well as 
one lower gap where the eigenvalue density vanishes.

In the Chern-Simons case, 
the eigenvalue distribution is in the no gap phase
in the low temperature, and 
if we increase the temperature, eigenvalue density starts to clump to 
$\alpha = 0$, and then we can expect lower gaps or upper gaps will show up.
In the small 't Hooft coupling region $\lambda < \lambda_c$, 
since the upper bound in 
\eqref{Ine-CS} is large enough, lower gaps will show up
before the maximum of the eigenvalue density function
reaches the upper bound. Then it transits to lower gap phase first.
As we increase the temperature further, the maximum of the 
eigenvalue density $\rho$ 
eventually reaches the upper bound $\frac{1}{2\pi\lambda}$ 
and it transits to the 
two gap phase.
On the other hand, at large $\lambda$ with 
$\lambda > \lambda_c$, the upper bound in 
\eqref{Ine-CS} is small, then the maximum of the eigenvalue reaches
the upper bound before the lower gap shows up.
Then in the large $\lambda$ region, it transits to upper gap phase first.
If we increase the temperature further, since the eigenvalue density 
keeps clumping, lower gaps will show up and it transits to two gap phase.
In the large $\zeta$ limit, the eigenvalue density will have a 
universal configuration
\begin{equation}
\begin{split}
\rho(\alpha) &= \frac{1}{2 \pi \lambda}~~~(|\alpha| < \pi \lambda)
\\
&= 0 ~~~~~~~(|\alpha| > \pi \lambda)
\end{split}
\label{evd 0}
\end{equation}
which is the nearest thing to a $\delta$ function
permitted 
by the effective 
Fermi statistics of the eigenvalues 
$0 \le \rho(\alpha) \le \frac{1}{2\pi\lambda}$.
This is in perfect agreement with the results of 
\cite{Aharony:2012ns} in which they used the Hamiltonian methods
\cite{Douglas:1994ex}.

\begin{figure}[tbp]
  \begin{center}
  \subfigure[]{\includegraphics[scale=.5]{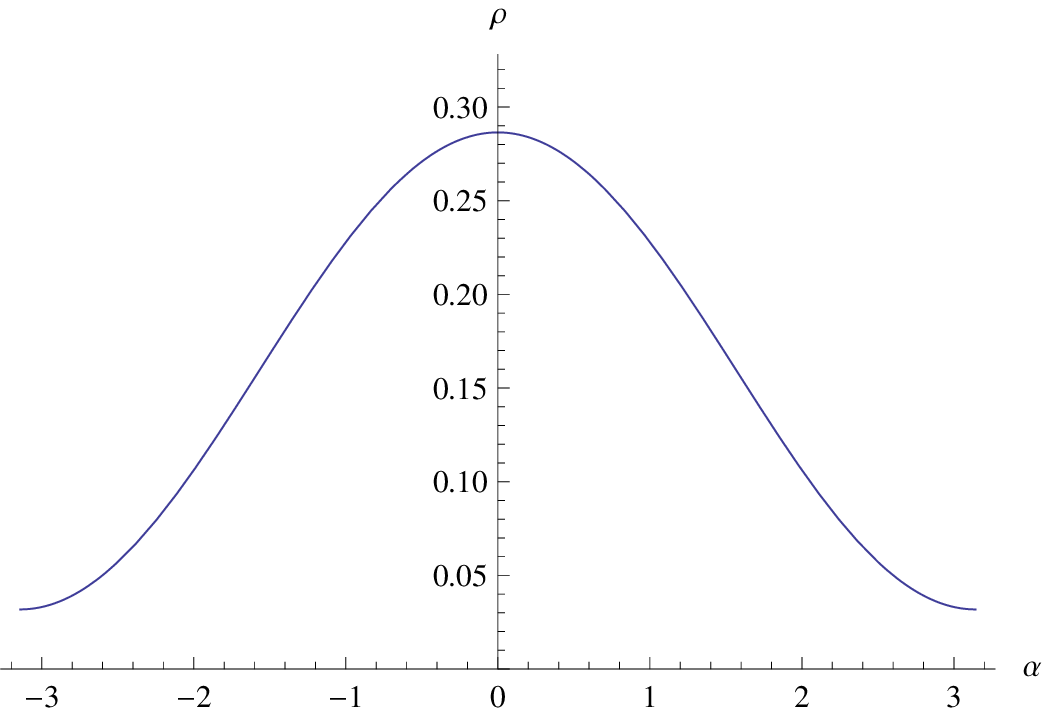}
\label{nocut}
  }
  \qquad\qquad
  \subfigure[]{\includegraphics[scale=.5]{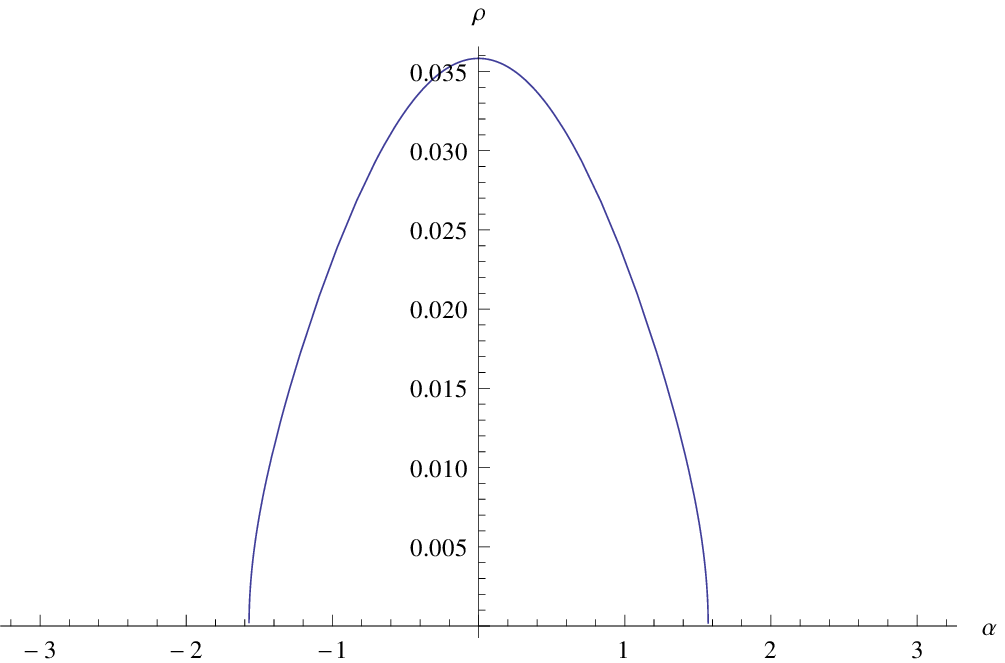}
\label{gwwfig}
  }
\\

 \subfigure[]{\includegraphics[scale=.5]{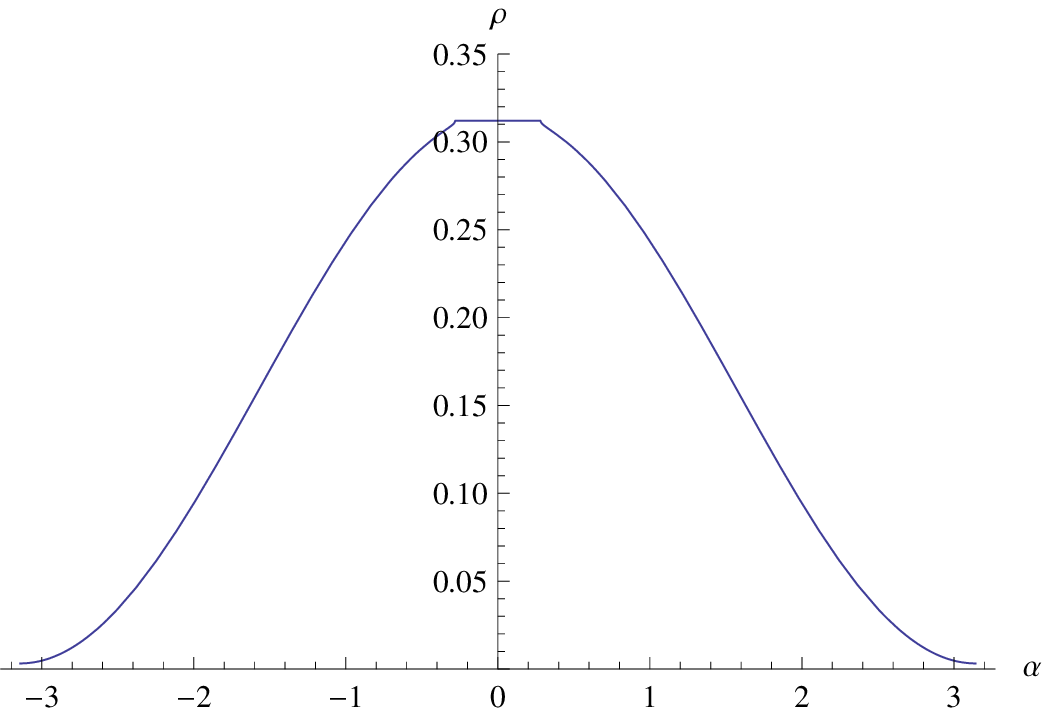}
\label{lbig2}
}
  \qquad\qquad
  \subfigure[]{\includegraphics[scale=.5]{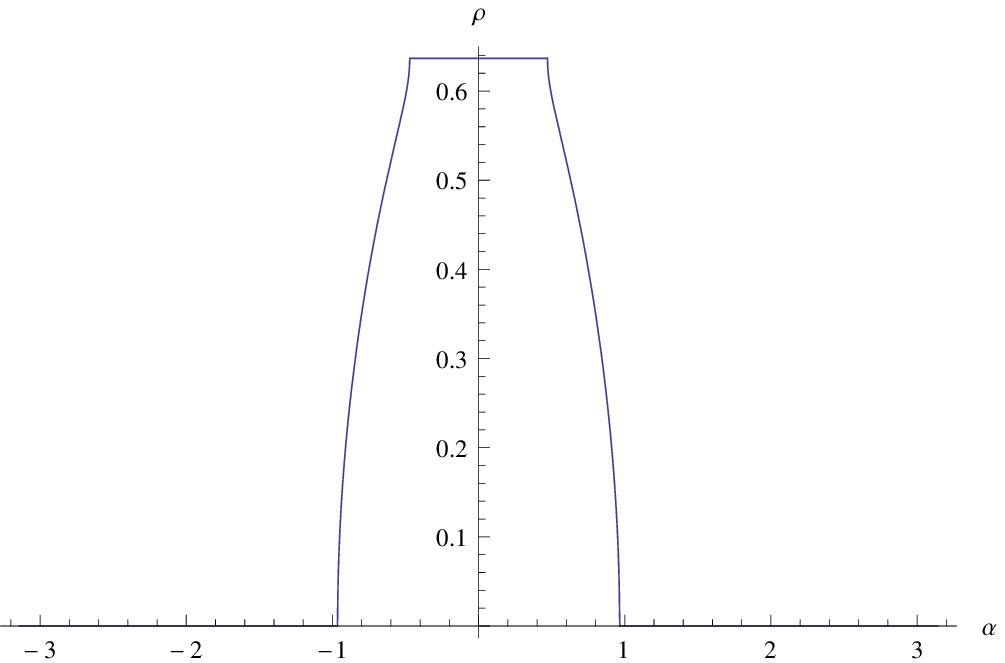}
\label{twocut}
  }
  \end{center}
  \vspace{-0.5cm}
  \caption{The eigenvalue distribution $\rho(\alpha)$
in the no gap phase~Fig.\ref{nocut}, 
in the lower gap phase~Fig.\ref{gwwfig}, 
in the upper gap phase~Fig.\ref{lbig2} at $\lambda = 0.51$, 
and in the two gap phase~Fig.\ref{twocut}.
These are the graphs calculated in the GWW toy model.
We can see that the eigenvalue density in the upper gap phase
~Fig.\ref{lbig2} is saturated from above as $\rho(\alpha) = \frac{1}{2 \times 0.51 \pi}$.  We can also see that the two gap 
density function~Fig.\ref{twocut} has both a lower gap and an upper gap.}
 \end{figure}

For your reference, 
you can see 
the graph of the eigenvalue density function in each phase listed
in Figs.\ref{nocut} $\sim$ \ref{twocut}.
You can also see the phase diagrams of Chern-Simons matter theories
in Figs.\ref{fig:phase-RF} \ref{fig:phase-CB} and \ref{fig:phase-SUSY}.

\subsection{How to calculate the eigenvalue density in lower gap, upper gap and two gap phases}
In this subsection we will review how to obtain the eigenvalue density in the
lower gap, the upper gap and the two gap phases.
To calculate the eigenvalue density, we will decompose the 
eigenvalue density to $\rho(\alpha) = \rho_{0}(\alpha) + \psi(\alpha)$.
By substituting it into the saddle point equation we can rewrite the saddle 
point equation as 
\begin{equation}\label{vvec} \begin{split}
&N {\cal P} \int d \alpha ~\psi(\alpha) 
 \cot \left( \frac{\alpha_0-\alpha}{2} \right)
=U(\alpha_0)\\
&\int d \alpha~ \psi(\alpha)=A[\rho_o]\\
&U(\alpha)=V'(\alpha)-N {\cal P} \int d \theta ~\rho_0(\theta)
\cot \left( \frac{\alpha-\theta}{2} \right) 
\\
&A[\rho_0]=1-  \int d \alpha ~\rho_0(\alpha).
\end{split}
\end{equation}
Here $\psi(\alpha)$ is nonzero in the compliment of both upper gaps and lower
gaps.
Starting from this, we turn to the complex variables
$$z=e^{i \alpha}, ~~~z_0=e^{i \alpha_0}, ~~~A_i=e^{ia_i}, ~~~B_i=e^{ib_i}.$$
Note that 
$$d \alpha=\frac{dz}{iz}, 
~~~\cot \frac{\alpha_0-\alpha}{2}=i \frac{z_0+z}{z_0-z} .$$
Here \eqref{vvec}
will be rewritten in terms of these variables as
\begin{equation}\label{fovv} \begin{split}
&N {\cal P} \int \frac{d z}{z}   \frac{z_0+z}{z_0-z}
\psi(z) =U(z_0),\\
&  \int \frac{dz}{iz} \psi(z)=A
\end{split}
\end{equation}
where the integrals in \eqref{fovv} run counterclockwise over the unit circle 
in the complex plane. 
Suppose that the solution to this equation, $\psi(z)$, has 
support on $n$ connected arcs on the unit circle in the complex plane.
We denote the beginning and endpoints of these arcs by $A_i=e^{i a_i}$ and 
$B_i=e^{i b_i}$ 
($i =1 \ldots n$).
These are the endpoints of lower gaps or upper gaps as well.
Our convention is that the points 
$A_1$, $B_1$, $A_2$, $B_2$ $\ldots$ $A_n$, $B_n$ sequentially 
follow each other counterclockwise on the unit circle. We refer to the 
$n$ arcs $(A_i,B_i)$ as `cuts'. The arcs $(B_i, A_{i+1})$ (as also 
$(B_n, A_1)$ ) are referred to as gaps. By definition, $\psi$ has support
only along the cuts on the unit circle.

We use the (as yet unknown) function $\psi(z)$ to define an analytic function 
$\Phi(u)$ on the complex plane
\begin{equation}\label{gwws1}
\Phi(u)=\sum_{i} \int_{A_i}^{B_i} \frac{dz}{iz} \frac{u+z}{u-z}\psi(z) 
\end{equation}
where the integral is taken counterclockwise along the $n$ cuts, 
$(A_i, B_i)$ of the unit circle in the complex plane.
It is easy to see that the analytic function $\Phi(u)$ is discontinuous 
along these $n$ cuts. Let $\Phi(z)^+$ denote the limit of 
$\Phi(u)$ as it approaches a cut from $|u|>1$, and let $\Phi^-(z)$ 
denote the limit of $\Phi(u)$ as it approaches a cut from $|u|<1$.
Then
\footnote{We use analogous notation for other analytic functions below. 
Let $z$ denote a complex number 
of unit norm. The symbol $F^+(z)$ will denote the limit of 
$F(u)$ as $u \to z$ from above (i.e. from $|u| >1$), while $F^-(z)$ is 
the limit of the same function  as $u \to z$ from below (i.e. from $|u|<1$). 
Note that along a cut $F^-(z)=-F^+(z)$.}
\begin{equation}\label{disc}
\Phi^+(z)-\Phi^-(z)= 4 \pi \psi(z)
\end{equation}
and 
\begin{equation}\label{princ}
\Phi^+(z)+\Phi^-(z)= 2  
\sum_i 
{\cal P}
\int_{A_i}^{B_i} \frac{d\omega}{i\omega} 
\frac{z+\omega}{z-\omega}\rho(\omega)= \frac{2 U(z)}{i N}
\end{equation}
(the first equality is the definition of the principal value, while 
the second equality follows using \eqref{fovv}).
Moreover it follows immediately from \eqref{fovv} and \eqref{gwws1} that  
\begin{equation}\label{asymp}
\lim_{u \to \infty} \Phi(u)= A.
\end{equation}

We are now posed with the 
problem of determining $\Phi(u)$ given its principal value along a cut. 
To determine the $\Phi(u)$, we introduce following "cut function"
$h(u)$ as
\begin{equation}\label{hz}
h(u)=\sqrt{(A_1-u)(B_1-u)(A_2-u)(B_2-u) \ldots (A_n-u)(B_n-u)}.
\end{equation} 
We define $h(u)$ to have cuts precisely on the $n$ arcs on the unit 
circle that extend from $A_i$ to $B_i$. 
This definition fixes the function $h(u)$ up to an overall sign. This sign 
will cancel out in our solution for $\Phi$ below, and so is uninteresting. 
For future use
we note that when $u=e^{i \alpha}$ ,
\begin{equation}\label{hsous}
h^2(u)=\prod_{m=1}^n 4 e^{i \frac{a_m+b_m}{2}} e^{i \alpha}
 \left(\sin^2 \left( \frac{a_m-b_m}{4} \right)
- \sin^2\left( \frac{\alpha}{2}-\frac{a_m+b_m}{4}
\right) \right). {}\; 
\end{equation} 
We use the function $h(z)$ to define a new function, $H(z)$, via the equation
$$\Phi(z)=h(z) H(z).$$
Using the fact that $h^+(z)=-h^-(z)$ along the cut, \eqref{princ} turns
into 
\begin{equation}\label{princd}
H^+(z)-H^-(z)= \frac{2 U(z)}{i N h^+(z)}.
\end{equation}
Here we assume that $H(u)$ is holomorphic except in the cut region.
From \eqref{asymp} and \eqref{hz},
$H(v) ={\cal O}(\frac{1}{v^n})$ at large $v$ so that 
\begin{equation}\label{ci}
\int_{C_\infty} dv \frac{H(v)}{2 \pi i (v-u)}=0
\end{equation}
where the contour $C_{\infty}$ 
runs counterclockwise over a very large circle at 
infinity. 
Since $H(v)$ is holomorphic except on the cut, by the Cauchy's theorem,
\begin{equation}
0 = \oint_{C_\infty} dv \frac{H(v)}{2 \pi i (v-u)} - \oint_{C_{cuts}} 
dv \frac{H(v)}{2 \pi i (v-u)} - H(u) 
= - \oint_{C_{cuts}} 
dv \frac{H(v)}{2 \pi i (v-u)}- H(u).
\end{equation}
The contour 
$C_{cuts}$ are the loops 
enclosing each of the $n$ cuts $(A_i, B_i)$, but not 
the point $u$. 
Then from \eqref{princd}, 
\begin{equation}
H(u) = -\oint_{C_{cuts}} dv \frac{H(v)}{2 \pi i (v-u)}
= \frac{1}{\pi} \int_{L_{arcs}} dz \frac{U(z)}{ N h^+(z)(z-u)}
=
\frac{1}{2\pi} \oint_{C_{cuts}} dv 
\frac{U(v)}{ N h(v)(v-u)}.
\end{equation}
In the equation above, $v$ is a variable on the complex plane while $z$ 
is a variable on the unit circle of the complex plane. 
The integration region $L_{arcs}$ runs counterclockwise 
along $n$ cuts on the unit circle (This is {\it not} loop !), 
integration region for each cut is from
$A_i$ to $B_i$.
The third equality 
uses $h^+(z)=-h^-(z)$ together with the assumption that 
$U(z)$ has no singularities on the $n$ cuts.

Now by applying the Cauchy's theorem for the integrand
$\frac{U(v)}{ N h(v)(v-u)}$
again, we will obtain the following 
\begin{equation}
\begin{split}
H(u) &= 
\frac{1}{2\pi} \oint_{C_{cuts}} dv \frac{U(v)}{ N h(v)(v-u)}
\\
&= 
\frac{1}{2\pi} \oint_{C_{\infty}} dz \frac{U(z)}{ N h(z)(z-u)}
- \frac{iU(u)}{N h(u)}-\sum_{m=1}^r i {\rm Res}_{z=z_k} \frac{U(z)}{Nh(z)(z-u)}.\end{split}
\label{Hucp}
\end{equation}
The last term is the sum of the residues of $U(z)$.
Here we have assumed that $U(z)$ is a meromorphic function of $z$.

Based on $H(u)$ in \eqref{Hucp}, and from
$\Phi^{+}(u) - \Phi^-(u) = 4 \pi \psi(u)$, 
combining with
$\rho(\alpha) = \psi(\alpha) + \rho_{0}(\alpha)$,
we can obtain the eigenvalue density $\rho(\alpha)$.
From next section, 
we will obtain the phase structure of the regular fermion theory,
the critical boson theory and the ${\cal N} = 2$ supersymmetric CS matter 
theory by using the techniques supplied by this subsection.

\section{Higher temperature phases of the regular fermion theory and 
critical boson theory and the duality}
\subsection{Regular fermion theory}
In this subsection we study the level $k$ 
$U(N)$ Chern-Simons theory coupled to massless fundamental 
fermions.
The Lagrangian of the theory is presented in equation (2.1) of 
\cite{Giombi:2011kc}.
From the equation (3.5) of \cite{Jain:2013py}, 
effective potential $V(U)$ is obtained as
\begin{equation}\label{freefer}
 \begin{split}
 V(U)&=-\frac{N^2 \zeta}{6\pi}\left(\frac{\tilde c^3}{\lambda}-\tilde c^3
+3 \int_{-\pi}^{\pi} d \alpha \rho(\alpha) \int_{\tilde c}^{\infty}dy ~ y
(\ln(1+e^{-y-i\alpha})+\ln(1+e^{-y+i\alpha}))\right)
\\
&\equiv V^{r.f}[\rho,N;\tilde{c},\zeta],
 \end{split}
\end{equation}
where ${\tilde c}$ determines the thermal mass of the fermions.%
\footnote{$\Sigma_T = {\tilde c}^2 T^2$ is the thermal mass of the 
fundamental fermions. More precisely, the fermionic self energy is given by 
$\Sigma_{T}(p)=f(\beta p_s)p_s I+i p^{-}g(\beta p_s)\gamma^{-}$, where 
\begin{equation}\label{selffer}\begin{split}
f(y)&=\frac{\lambda}{y}\int_{-\pi}^{\pi} d\alpha~\rho(\alpha)\left(\ln 2\cosh(\frac{\sqrt{y^2+{\tilde c}^2}+i \alpha}{2})+\ln 2\cosh(\frac{\sqrt{y^2+{\tilde c}^2 }-i \alpha}{2})\right)\\
g(y)&=\frac{{\tilde c}^2}{y^2}-f(y)^2.\\
\end{split}
\end{equation} }
The value of ${\tilde c}$ is obtained by extremizing $V(U)$ w.r.t 
${\tilde c}$ at fixed $\rho$, $\zeta$, i.e. ${\tilde c}$ obeys the equation
\begin{equation}\label{tc}
\tilde c=\lambda \int_{-\pi}^{\pi} d\alpha~\rho(\alpha)\left(\ln 2\cosh(\frac{\tilde c+i \alpha}{2})+\ln 2\cosh(\frac{\tilde c-i \alpha}{2})\right).
\end{equation}
The general form of the 
free energy of the regular fermion theory on $S^2 \times S^1$ is 
given as 
\begin{equation}
\begin{split}
F_{r.f}^{N} =& 
V^{r.f}[\rho,N]- N^2 {\cal P}
\int^{\pi}_{-\pi}d \alpha\int^{\pi}_{-\pi} d \beta~
\rho(\alpha)\rho(\beta) \log 
\left|2 \sin \frac{\alpha-\beta}{2}\right|
\\
=&V^{r.f}[\rho,N] + F_2[\rho,N].
\label{F-RF}
\end{split}
\end{equation}
To obtain the free energy in each phase, we
only have to evaluate the eigenvalue density $\rho$ 
and $\tilde{c}$ 
in each phase and just substitute into \eqref{F-RF}.
Note that $\tilde{c}$ can be determined if the eigenvalue density
is determined.
So obtaining the eigenvalue density in each phase
is equivalent to obtaining the 
free energy in each phase.

For later use, we will give the form of $V'(z)$ here.
We obtain $V'(z)$ from \eqref{freefer} as
\begin{equation}
V'(z)
=  -\frac{N \zeta}{2\pi}
\int_{\tilde c}^{\infty}dy ~ y
\left(
\frac{-ie^{-y}}{z+e^{-y}}
+\frac{ie^{-y}}{z^{-1} + e^{-y}}
\right).
\label{Vdash}
\end{equation}

\begin{figure}
  \begin{center}
  \subfigure[]{\includegraphics[scale=.33]{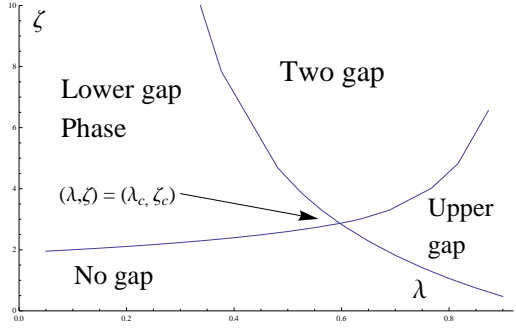}
\label{fig:phase-RF}
}
\caption{Phase diagram of the regular fermion theory.
Here $(\lambda_c,\zeta_c) = (0.596967, 2.86454)$
is the quadruple phase transition point where 
the four phases (no gap, lower gap, upper gap, two gap phases) coexist.}
  \end{center}
\end{figure}

The phase structure of this theory is 
depicted in Fig.~\ref{fig:phase-RF}, 
as obtained in following subsubsections.

\subsubsection{Lower gap phase}
\label{L-RF}
To obtain the 
eigenvalue density in the lower gap phase, 
we will employ the cut region and the cut function described in appendix
\ref{Sec:Low-Cut}.
The function $H(u)$ as well as $\Phi(u) = h(u) H(u)$ in the lower gap phase are obtained by using \eqref{Hucp} with substituting $\rho_{0}(\alpha) = 0$.
$U(z)$ in \eqref{Hucp} is identical to
$V'(z)$ as
\begin{equation}
U(z) = V'(z)
=  -\frac{N \zeta}{2\pi}
\int_{\tilde c}^{\infty}dy ~ y
\left(
\frac{-ie^{-y}}{z+e^{-y}}
+\frac{ie^{-y}}{z^{-1} + e^{-y}}
\right),
\label{VdashRF}
\end{equation}
because $\rho_{0}(\alpha) = 0$. 
By substituting the above into \eqref{Hucp},
we immediately obtain $H(u)$ as well as $\Phi(u)
= h(u) H(u)$ as,
\begin{equation}
 \begin{split}
\Phi(u)=\Phi^{r.f}_{lg}(\tilde{c},b,\zeta;u)
\equiv&
\frac{\zeta}{2\pi}
\int_{\tilde c}^{\infty}dy ~  \frac{yh(u)(u+1)(1+e^{y})}{
\sqrt{(e^{y}+e^{ib})(e^{y}+e^{-ib})}
(u+ e^{y})(u+ e^{-y})}
\\
&+\frac{\zeta}{2\pi}
\left(
\int_{\tilde c}^{\infty}dy ~ y \frac{e^{-y}}{u+e^{-y}}
-
\int_{\tilde c}^{\infty}dy ~ y \frac{e^{-y}}{u^{-1}+e^{-y}}
\right).
\end{split}
\label{Phi-RF-L}
\end{equation}

From $\Phi^{+}(u) - \Phi^-(u) = 4 \pi \rho(u)$
at the unit circle $u = e^{i\alpha}$ with $-b \le \alpha \le b$,
we obtain the eigenvalue density in the lower gap phase as
\begin{equation}
\begin{split}
\rho(\alpha) =& 
\frac{\zeta}{\sqrt{2}\pi^2}
\sqrt{\sin^2 \frac{b}{2} - \sin^2 \frac{\alpha}{2}}
\int^{\infty}_{\tilde{c}} dy\, 
\frac{y \cos \frac{\alpha}{2} \cosh \frac{y}{2}}{
(\cosh y + \cos \alpha)
\sqrt{(\cosh y + \cos b)}}
\\
\equiv& \rho^{r.f}_{lg}(\zeta,\lambda;\tilde{c}, b; \alpha).
\end{split}
\label{eigen-RF-low}
\end{equation}
At $\pi > |\alpha| > b$, $\rho(\alpha) = 0$.
By substituting the eigenvalue density \eqref{eigen-RF-low} into
\eqref{tc} and \eqref{F-RF}, we can obtain $\tilde{c}$ as well as 
the free energy in the lower gap phase of the regular fermion theory.

The condition $\lim_{u \to \infty} \Phi(u) = 1$ requires the 
following condition
\begin{equation}
\tilde{M}^{r.f}_{lg}(\zeta,\tilde{c},b)
\equiv
\frac{\zeta}{2\pi}
\int_{0}^{e^{-\tilde{c}}}dx ~  
\left(
\frac{  \log x}{x}
- \frac{(1+x)}{x}
\frac{\log x}{\sqrt{x^2+ 2 x\cos b  +1}}\right)
= 1.
\label{Cond-low-RF}
\end{equation}
By \eqref{Cond-low-RF} and \eqref{tc}, we can obtain 
$(b,\tilde{c})$ as functions of $(\lambda, \zeta)$ as
$(b,\tilde{c}) = (b(\lambda,\zeta),\tilde{c}(\lambda,\zeta))$.
\eqref{eigen-RF-low}, \eqref{Cond-low-RF} and \eqref{tc}
provide a complete set of the solutions in the lower gap phase of the 
regular fermion theory.

\subsubsection{Upper gap phase}
\label{UP-RF}
Next we will search for a solution with no lower gap and one upper gap.
The domain of cut and the cut function in this upper gap phase 
are defined in appendix \ref{Sec:Up-Cut}.
To obtain the 
eigenvalue density based on \eqref{Hucp},
we will take 
$\rho_{0}(\alpha) = \frac{1}{2 \pi \lambda} \ne 0$ 
at the upper gap region
$-a \le \alpha \le a$.
Then by using \eqref{Hucp} with \eqref{Vdash},
we obtain $H(u)$ as well as $\Phi(u)$ as
\begin{equation}
\begin{split}
\Phi(u) =& \Phi^{r.f}_{ug}(\tilde{c}, a, \zeta;u) + \Phi_{ug}^{\rho_{0}}(\lambda, a; u), 
\qquad {\text {where}}
\\
\Phi^{r.f}_{ug}(\tilde{c}, a, \zeta;u)\equiv&\frac{\zeta}{2\pi}
\int_{\tilde c}^{\infty}dy ~  \frac{yh(u)(1-e^{y})(1-u)}{
\sqrt{(e^{y}+e^{ia})(e^{y}+e^{-ia})}
(e^{y} + u)(e^{-y} + u)}
\\
&+
\frac{\zeta}{2\pi}
\int_{\tilde c}^{\infty}dy ~ y \frac{e^{-y}}{u+e^{-y}}
-\frac{\zeta}{2\pi}
\int_{\tilde c}^{\infty}dy ~ y \frac{e^{-y}}{u^{-1}+e^{-y}},
\\
\Phi_{ug}^{\rho_{0}}(\lambda, a; u)
\equiv&
\frac{i}{\pi \lambda} \int_{L_{ugs}} 
d\omega\, \frac{1}{h(\omega)(\omega - u)}
h(u)
+i{\cal P}\int_{L_{cir}} 
d\omega\, \rho_{0}(\omega)\frac{u+\omega}{\omega(u-\omega)},
\end{split}
\label{Phi-RF-up}
\end{equation}
where $L_{ugs}$ runs 
counterclockwise over upper gap region,
and $L_{cir}$ runs  
counterclockwise over unit circle.

From \eqref{Phi-RF-up}, 
and by taking $\Phi^{+}(u)- \Phi^{-}(u) = 4\pi(\rho(u)- \rho_{0}(u))$
at the cut region $u = e^{i\alpha}$ with 
$\pi \ge |\alpha| \ge a$,
we can obtain the eigenvalue density function as
\begin{equation}
\begin{split}
\rho(\alpha)=&\frac{1}{2\pi \lambda}
-\frac{\zeta}{\sqrt{2}\pi^2}
\sqrt{
\sin^2 \frac{\alpha}{2} 
-\sin^2 \frac{a}{2} 
}
\int_{\tilde c}^{\infty}dy ~  \frac{y |\sin \frac{\alpha}{2}|
\sinh \frac{y}{2}}
{\sqrt{\cosh y + \cos a}(\cos \alpha + \cosh y)}
\\
\equiv& \rho^{r.f}_{ug}(\zeta,\lambda;\tilde{c},a;\alpha).
\end{split}
\label{eigen-RF-up}
\end{equation}
In the region $|\alpha| < |a|$, $\rho(\alpha) = \frac{1}{2\pi\lambda}$. 
To derive \eqref{eigen-RF-up}, we have used a formula
\begin{equation}
\frac{i}{2\pi^2 \lambda} \int_{Lugs} 
d\omega\, \frac{1}{h(\omega)(\omega - u)}
h(u)
= \frac{1}{2\pi \lambda}
\label{Formula up 2}
\end{equation}
where the $u = e^{i\alpha}$ is located at the cut region.
This is proved in appendix \ref{h-one-2}.
By substituting the eigenvalue density \eqref{eigen-RF-up} into
\eqref{tc} and \eqref{F-RF}, we can obtain $\tilde{c}$ as well as
the free energy in the 
upper gap phase of the regular fermion theory.

From the condition, $\lim_{u \to \infty}\Phi(u) 
= 1- \int^{\pi}_{-\pi} d\alpha \rho_{0}(\alpha)$,
we obtain the equation
\begin{equation}
\tilde{M}^{r.f}_{ug}(\zeta,\tilde{c},a)
\equiv \frac{\zeta}{2\pi}
\int_{0}^{e^{-\tilde{c}}}dx ~  
\left(
\frac{  \log x}{x}
- \frac{(1-x)}{x}
\frac{\log x}{\sqrt{x^2+ 2 x\cos a  +1}}\right)
= 1-\frac{1}{\lambda}
.
\label{Cond-up-RF}
\end{equation}
Here we have used the formula
\begin{equation}
\frac{i}{\pi \lambda}\int_{Lugs}d \omega~ \frac{1}{h(\omega)} =  -\frac{1}{\lambda}
\label{Formula up 1}
\end{equation}
which is proved in appendix~\ref{h-1-one}.
By \eqref{Cond-up-RF} and \eqref{tc},
we can obtain $(a, \tilde{c})$ as functions of $(\lambda, \zeta)$ as
$(a,\tilde{c}) = (a(\lambda,\zeta), \tilde{c}(\lambda, \zeta))$.
\eqref{eigen-RF-up}, \eqref{Cond-up-RF} and \eqref{tc} provide
a complete set of solution 
in the upper gap phase of the regular fermion theory.


\subsubsection{Two gap phase}
\label{sec:RF-2}
Now we will search for a solution with one lower gap and one upper gap.
The details of our two cuts and the cut function $h(u)$ are described in 
appendix \ref{Sec:Two-Cut}.

The function $H(u)$ as well as $\Phi(u) = h(u) H(u)$ are 
obtained by using \eqref{Hucp}, 
\begin{equation}
\begin{split}
\Phi(u) =& 
\Phi^{r.f}_{tg}(\zeta, a,b,\tilde{c}; u) + 
\Phi^{\rho_{0}}_{tg}(\lambda, a,b;u) \qquad {\text {where}}
\\
\Phi^{r.f}_{tg}(\zeta, a,b,\tilde{c}; u) \equiv& 
\frac{\zeta}{2\pi}
\int_{\tilde c}^{\infty}dy ~  
\left(\frac{y e^{-y}(u^{-1}-u)}{(u+e^{-y})(u^{-1}+e^{-y})}
- \frac{yh(u)e^{y}}{h(-e^{y})} 
\frac{2u + e^{y}+e^{-y}}{(u+e^{y}) (u+e^{-y})}\right)
\\
\Phi^{\rho_{0}}_{tg}(\lambda, a,b;u)
\equiv&\frac{i}{\pi \lambda} \int_{L_{ugs}} 
d\omega\, \frac{h(u)}{h(\omega)(\omega - u)}
+i{\cal P}\int_{L_{cir}} 
d\omega\, \rho_{0}(\omega)\frac{u+\omega}{\omega(u-\omega)}.
\end{split}
\label{Two-RF-Phi}
\end{equation}

From $\lim_{u \to \infty} \Phi(u) = 1 
- \int^{\pi}_{-\pi} d \alpha~ \rho_{0}(\alpha)$, 
we obtain the following two conditions,
\begin{equation}
\begin{split}
\frac{1}{4\pi\lambda}\Upsilon(a,b)
=& 
\frac{\zeta}{2\pi}{\cal Y}^{r.f}(a,b,\tilde{c}), \qquad {\text {where}}
\\
\Upsilon(a,b)
\equiv&
\int_{-a}^{a} 
d\theta\, \frac{1}{
\sqrt{\sin^2\frac{a}{2}-\sin^2\frac{\theta}{2}}
\sqrt{\sin^2\frac{b}{2}-\sin^2\frac{\theta}{2}}},
\\
{\cal Y}^{r.f}(a,b,\tilde{c})
\equiv& 
\int^{\infty}_{\tilde{c}} dy~\frac{y}{
\sqrt{(\cosh y + \cos a)(\cosh y + \cos b)}
},
\end{split}
\label{RF two u1}
\end{equation}
\begin{equation}
\begin{split}
\frac{1}{4\pi\lambda}\Lambda(a,b) =& 1- 
\frac{\zeta}{4 \pi}{\cal G}^{r.f}(a,b,\tilde{c}),
\qquad
{\text {where}}
\\
\Lambda(a,b) \equiv&
\int_{-a}^{a} 
d\theta\, \frac{\cos \theta}{
\sqrt{\sin^2\frac{a}{2}-\sin^2\frac{\theta}{2}}
\sqrt{\sin^2\frac{b}{2}-\sin^2\frac{\theta}{2}}},
\\
{\cal G}^{r.f}(a,b,\tilde{c})
 \equiv&
\int^{\infty}_{\tilde{c}} dy~
y\left( \frac{e^{y}+e^{-y}}{
\sqrt{(\cosh y + \cos a)(\cosh y + \cos b)}
} - 2\right).
\end{split}
\label{RF two u0}
\end{equation}
By using 
\eqref{RF two u1},
\eqref{RF two u0} and 
\eqref{tc}, $(a,b, \tilde{c})$ are determined as  
functions of $(\lambda, \zeta)$ as
$(a,b, \tilde{c}) = (a(\lambda, \zeta),b(\lambda, \zeta), 
\tilde{c}(\lambda, \zeta))$.

From \eqref{Two-RF-Phi} 
and by taking $\Phi^{+}(u)- \Phi^{-}(u) = 4\pi(\rho(u)- \rho_{0}(u))$ 
at the cuts $|a| < |\alpha| < |b|$, we obtain the eigenvalue density
as
\begin{equation}
\begin{split}
\rho(\alpha) =& \rho^{r.f}_{tg}(\alpha)
= \rho_{1,tg}^{r.f}(\zeta, a,b,\tilde{c};\alpha)
+\rho_{2,tg}(\lambda, a,b;\alpha),
\qquad {\text {where}}
\\
\rho_{1,tg}^{r.f}(\zeta, a,b,\tilde{c};\alpha)
\equiv& 
\frac{\zeta }{\pi^2}
\cF(a,b;\alpha)
\int_{\tilde c}^{\infty}dy  \frac{ye^{-y}}{
\nu_{r.f}(a,b;y)}\left(
\frac{|\sin \alpha|}{\cos \alpha
+  \cosh y }
\right),
\\
\rho_{2,tg}(\lambda, a,b;\alpha)
\equiv
&\frac{\vline\,\sin\alpha\,\vline}{4\pi^2\lambda}
{\cal F}(a,b;\alpha)~I_1(a,b,\alpha),
\\
\cF(a,b,\alpha)
\equiv&
\sqrt{
(\sin^2\frac{\alpha}{2} - \sin^2\frac{a}{2})
(\sin^2\frac{b}{2} - \sin^2\frac{\alpha}{2})
},
\\
\nu_{r.f}(a,b;y)
\equiv& \sqrt{(1+2e^{-y}\cos a + e^{-2y})(1+2e^{-y}\cos b + e^{-2y})},
\\
I_1(a,b;\alpha) \equiv& \int_{-a}^a \frac{d\theta}{(\cos\theta- \cos\alpha)\sqrt{\left(\sin^2\frac{a}{2}-\sin^2\frac{\theta}{2}\right)\left(\sin^2\frac{b}{2}-\sin^2\frac{\theta}{2}\right)}}. 
\end{split}
\label{eigen-RF-two}
\end{equation}
$\rho(\alpha) = 0$ in $b < |\alpha| < \pi$,
and $\rho(\alpha) = \frac{1}{2\pi\lambda}$ in $0 < | \alpha| < a$.
During the calculation of 
\eqref{eigen-RF-two},
we have used \eqref{RF two u1} also.
Here $\rho_{2,tg}$ has the same functional form as the 
eigenvalue density in the GWW type matrix integration (7.6) in \cite{Jain:2013py}.
$\rho_{1,tg}^{r.f}$ can be regarded as 
an additional term depending on the detail of the theory.
By substituing the eigenvalue density \eqref{eigen-RF-two} into
\eqref{tc} and \eqref{F-RF}, we can obtain $\tilde{c}$ as well as 
the free energy in the 
two gap phase. 
$\Upsilon(a,b), \Lambda(a,b)$ and 
$|\sin \alpha|\cF(a,b,\alpha)I_1(a,b,\alpha)$ can be represented 
by the complex line integral 
by taking $\omega = e^{i\theta}$ 
as
\begin{equation}
\begin{split}
\Upsilon(a,b) =& 
-4i \int_{L_{ugs}} d\omega \frac{1}{h(\omega)},
\\
\Lambda(a,b) =& 
-4i \int_{L_{ugs}} d\omega \frac{\omega}{h(\omega)},
\\
|\sin \alpha| \cF(a,b,\alpha) I_1(a,b;\alpha)
=&  
i\int_{L_{ugs}} d\omega~ \frac{h^{+}(u)}{h(\omega)}\left(
\frac{2}{(\omega-u)} + \frac{1}{u}
\right).
\end{split}
\label{Cont-rep}
\end{equation}
These are useful to discuss the level-rank duality later.

The combination of \eqref{RF two u1}, \eqref{RF two u0}, 
\eqref{tc} and \eqref{eigen-RF-two} provides a complete set of 
solutions in the two gap phase.

At the large $\zeta$ limit, as we proved in 
appendix \ref{RF-Z-Pr}, 
the eigenvalue density approaches the 
universal distribution \eqref{evd 0} because 
$\tilde{c}$ remains as a finite positive quantity at the limit. 
In the limit, $a$,$b$ and the eigenvalue 
density behave as
\begin{equation}
\begin{split}
a =& \pi \lambda - 
\frac{\epsilon}{2},
\qquad 
b = \pi \lambda + \frac{\epsilon}{2},
\\
\epsilon =& 8 \sin (\pi \lambda) \exp
\left(
- \frac{\sin (\pi\lambda)}{2}\lambda\zeta
\int^{\infty}_{\tilde{c}} dy~\frac{y}{\cosh y + \cos \pi \lambda}\right) +\ldots,
\\
\rho(\alpha) =&  \frac{1}{\pi^2 \lambda} \cos^{-1} \sqrt{
\frac{\alpha - a}{b - a}
}.
\end{split}
\label{Large z RF}
\end{equation}
At appendix \ref{RF-LZ}, we have demonstrated the above behavior
\eqref{Large z RF}. As we can see in the appendix, 
the eigenvalue density function is dominated by $\rho_{2.tg}$
at the large $\zeta$ limit. The form 
of the eigenvalue density approaches $\cos^{-1}\sqrt{\alpha_1}$ function 
which is the same form as (7.11) in \cite{Jain:2013py} which is the large 
$\zeta$ 
limit of the one of the GWW model.
We can also see that the range of the domain of the cut $\epsilon$ is
smaller than the one in the GWW type matrix integration at the same value of
$(\lambda, \zeta)$.

\subsubsection{Phase transition points}
\paragraph{Phase transition points between no gap and lower gap}
Let us consider the behavior of the lower gap solutions
\eqref{eigen-RF-low} and \eqref{Cond-low-RF}
at the point $b = \pi$ which would correspond to 
the phase transition points from the lower gap to the no gap phase.
If we substitute $b = \pi$ into \eqref{Cond-low-RF},
it becomes
\begin{eqnarray}
1 = -\frac{\zeta}{2\pi}
\int_{0}^{e^{-\tilde{c}}}dx ~  
\frac{2 \log x}{1-x}
= \frac{\zeta}{\pi}
\sum_{n=1}^{\infty}\left(
\frac{1+\tilde{c} n}{n^2}
e^{-n\tilde{c}}
\right).
\label{RF-no-low}
\end{eqnarray}
This is exactly same as the 
condition for the phase transition 
from the no gap to the lower gap discussed in 
\cite{Jain:2013py}, 
which is obtained by substituting $\alpha = \pi$ into the 
third line of (6.33) and by requiring $\rho(\pi) = 0$.
Based on this, if we substitute $b = \pi$ into
\eqref{eigen-RF-low}, 
we can see 
\begin{equation}
\begin{split}
\rho(\alpha)
&=\frac{\zeta}{\sqrt{2}\pi^2}
\sqrt{\sin^2 \frac{b}{2} - \sin^2 \frac{\alpha}{2}}
\int^{\infty}_{\tilde{c}} dy\, 
\frac{y \cos \frac{\alpha}{2} \cosh \frac{y}{2}}{
(\cosh y + \cos \alpha)
\sqrt{(\cosh y + \cos b)}}\biggr|_{b = \pi}
\\
&=\frac{\zeta}{\pi^2}
\int^{\infty}_{\tilde{c}}
dy~ y\left(
\frac{1}{2(1-e^{-y})}
-\frac{1}{4(1+e^{-y+i\alpha})}
-\frac{1}{4(1+e^{-y-i\alpha})}
\right)
\\
&=\frac{1}{2\pi}
- \frac{\zeta}{2\pi^2}
\sum_{n=1}^{\infty} 
(-1)^n\cos n \alpha
\left(\frac{1+\tilde{c} n}{n^2}\right)e^{-n\tilde{c}}.
\end{split}
\label{eigen-RF-low-no}
\end{equation}
In the last step we have used \eqref{RF-no-low}, the Taylor expansion and
\begin{equation}
\int^{\infty}_{\tilde c} dy~ ye^{-ny}
= 
\left(\frac{1+\tilde{c} n}{n^2}\right)e^{-n\tilde{c}}.
\label{eq:Taylor-formula no-low}
\end{equation}
The last line of \eqref{eigen-RF-low-no}
is exactly same as the eigenvalue density function in the no gap phase
described in the third line of Eq.~(6.33) in \cite{Jain:2013py}.
So, at the phase transition points,
the lower gap solutions are smoothly 
connected to the no gap phase solutions.

\paragraph{Phase transition points between no gap and upper gap}

Let us consider the behavior of \eqref{eigen-RF-up} and \eqref{Cond-up-RF}
at $a= 0$ which would correspond to the phase transition points 
from the upper gap to the no gap phase.
If we substitute $a = 0$ into \eqref{Cond-up-RF}, it becomes
\begin{equation}
1- \frac{1}{\lambda}=
\frac{\zeta}{2\pi}
\int_{0}^{e^{-\tilde{c}}}dx ~  
\frac{2\log x}{1+x}
= \frac{\zeta}{\pi}
\sum_{n=1}^{\infty} (-1)^n\frac{1+ \tilde{c}n}{n^2}e^{-n\tilde{c}}.
\label{RF-no-up}
\end{equation}
This is exactly same as the 
condition for the phase transition 
from the no gap to the upper gap phase discussed in 
\cite{Jain:2013py}, 
which is obtained by substituting $\alpha = 0$ into the 
third line of (6.33) and by requiring $\rho(0) = \frac{1}{2\pi\lambda}$.

Based on this, if we substitute $a = 0$ in
\eqref{eigen-RF-up}, 
we can see 
\begin{equation}
\begin{split}
\rho(\alpha)
&=
\frac{1}{2\pi \lambda}
-\frac{\zeta}{\sqrt{2}\pi^2}
\sqrt{
(\sin^2 \frac{\alpha}{2} 
-\sin^2 \frac{a}{2} )
}
\int_{\tilde c}^{\infty}dy ~  \frac{y \sin \frac{\alpha}{2}
\sinh \frac{y}{2}}
{\sqrt{\cosh y + \cos a}(\cos \alpha + \cosh y)}\biggr|_{a=0}
\\
&=
\frac{1}{2\pi\lambda}
+\frac{\zeta}{\pi^2}
\int^{\infty}_{\tilde{c}}
dy~ y\left(
\frac{1}{2(1+e^{-y})}
-\frac{1}{4(1+e^{-y+i\alpha})}
-\frac{1}{4(1+e^{-y-i\alpha})}
\right)
\\
&=\frac{1}{2\pi}
- \frac{\zeta}{2\pi^2}
\sum_{n=1}^{\infty} 
(-1)^n\cos n \alpha
\left(\frac{1+\tilde{c} n}{n^2}\right)e^{-n\tilde{c}}.
\end{split}
\label{eigen-RF-up-no}
\end{equation}
In the last step we have used \eqref{RF-no-up}, the 
Taylor expansion and 
the formula \eqref{eq:Taylor-formula no-low}.
The last line of \eqref{eigen-RF-up-no}
is exactly same as the eigenvalue density function 
in the no gap phase,
described in the third line of Eq.~(6.33) in \cite{Jain:2013py}.
So 
the upper gap solutions are smoothly
connected to the ones in the no gap phase
at the phase transition points.

\begin{figure}
  \begin{center}
  \subfigure[]{\includegraphics[scale=.33]{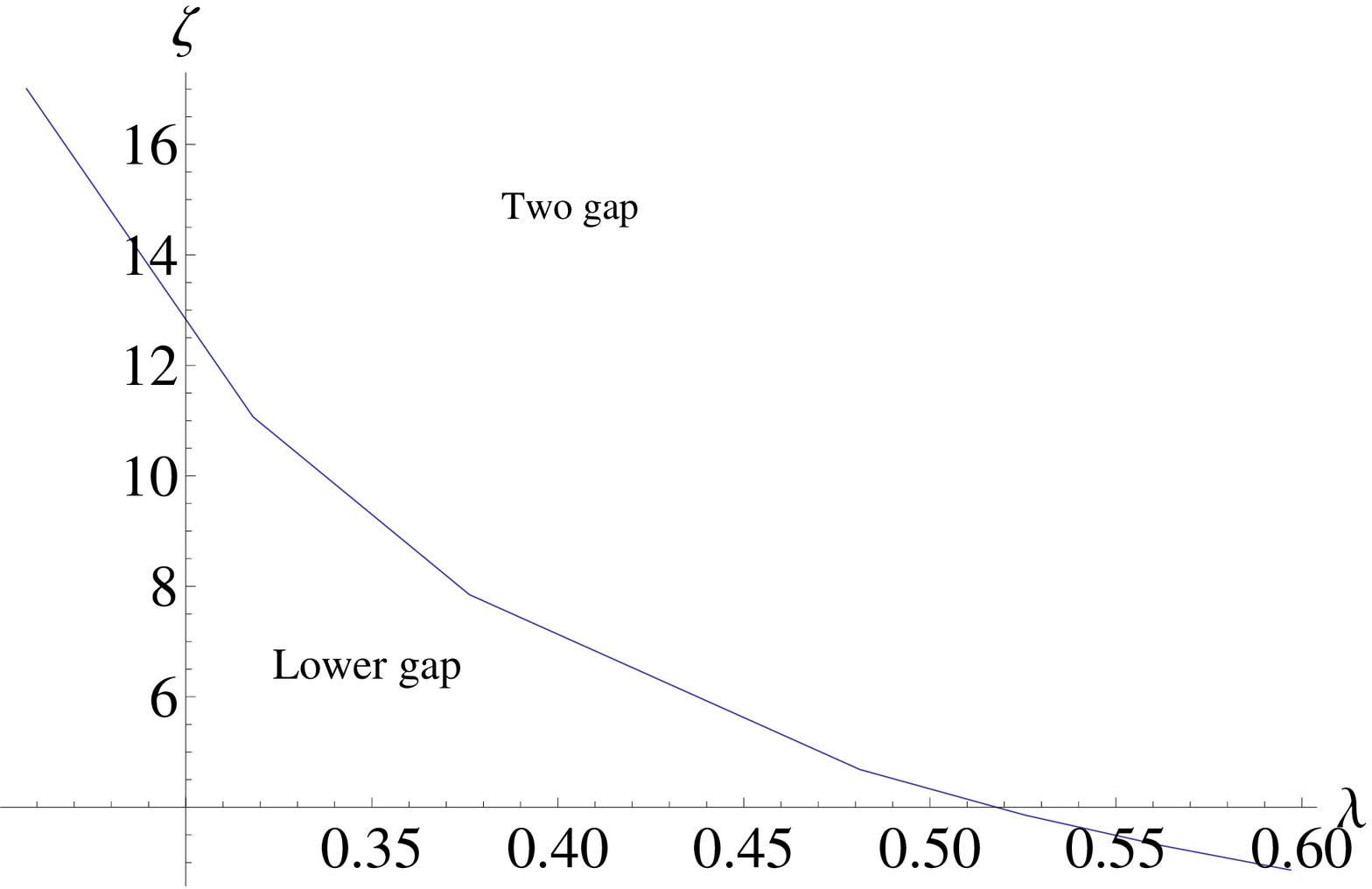}
\label{fig:low-two-RF}}
  \qquad\qquad
  \subfigure[]{\includegraphics[scale=.33]{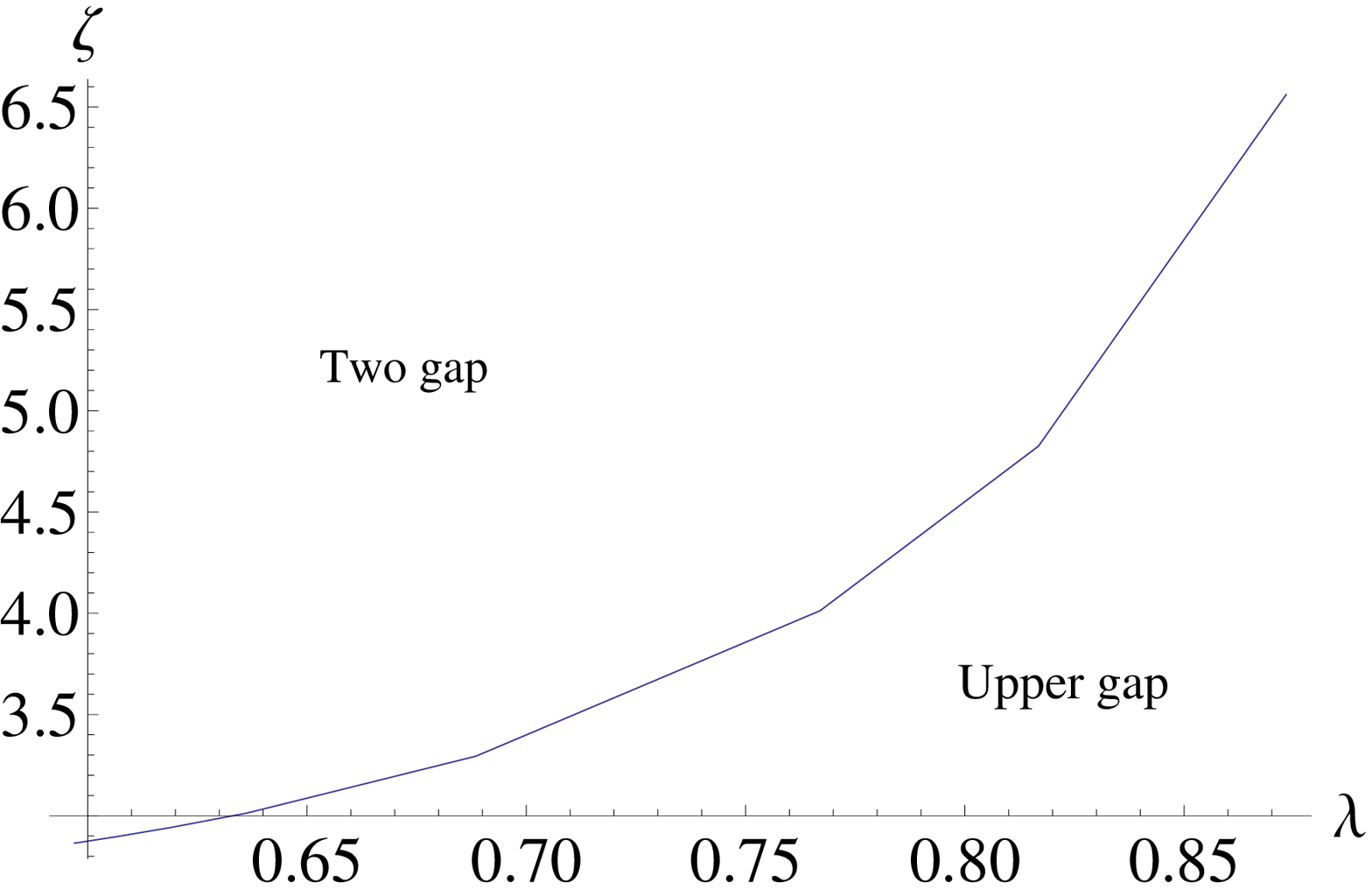}
\label{fig:up-two-RF}}
\caption{These are the plots of phase transition points in the regular 
fermion theory.
Fig.~\ref{fig:low-two-RF} shows the plots of 
the phase transition points from the lower gap to
the two gap phase, and Fig.~\ref{fig:up-two-RF} shows the ones between the 
upper gap and the two gap phase.}
  \end{center}
\end{figure}

\paragraph{Phase transition from the lower gap to the two gap}
In the lower gap phase at fixed $\lambda$, 
the phase transition from the lower gap phase to the two gap phase will occur
when the maximum of the eigenvalue density $\rho(0)$ reaches  
$\rho(0) = \frac{1}{2\pi \lambda}$.
The condition would be represented as 
\begin{equation}
\rho(0)
= 
\frac{\zeta}{\sqrt{2}\pi^2}
\sin \frac{b}{2} 
\int^{\infty}_{\tilde{c}} dy\, 
\frac{y \cosh \frac{y}{2}}{
(\cosh y + 1)
\sqrt{(\cosh y + \cos b)}}
= \frac{1}{2\pi \lambda}.
\label{low-to-2 RF}
\end{equation}
The combination of \eqref{low-to-2 RF}, \eqref{tc}
and \eqref{Cond-low-RF} provides the 
phase transition points from the lower gap to the two gap.
By numerical calculations based on \eqref{low-to-2 RF}, \eqref{tc}
and \eqref{Cond-low-RF}, we obtain the phase transition points 
plotted in Fig.~\ref{fig:low-two-RF}.

Let us see the phase transition points from the 
standpoint of the two gap phase.
If we substitute $a = 0$ 
into \eqref{RF two u1}, 
it becomes the same as \eqref{low-to-2 RF}, 
which is the condition for the phase transition points.
If we set $a=0$ at \eqref{RF two u0},
it becomes the same as \eqref{Cond-low-RF}.
Let us check whether eigenvalue density in the two gap phase 
\eqref{eigen-RF-two} becomes the one in the lower gap phase
\eqref{eigen-RF-low} in $a = 0$ limit.
By using \eqref{low-to-2 RF}, we can replace 
$\rho_{2,tg}(\lambda,0,b;\alpha)$ of \eqref{eigen-RF-two} by 
\begin{equation}
\rho_{2,tg}(\lambda,0,b;\alpha) = \frac{\zeta}{2\pi^2} 
\int^{\infty}_{\tilde{c}}
dy~ \frac{y \cos \frac{\alpha}{2}
\sqrt{\sin^2 \frac{b}{2} - \sin^2 \frac{\alpha}{2}}
}{\sqrt{(1+\cosh y)(\cosh y + \cos b)}}.
\label{RF-l-2s}
\end{equation}
Then by summing up with $\rho^{r.f}_{1,tg}(\zeta,0,b,\tilde{c};\alpha)$, 
\eqref{eigen-RF-two} at $a =0$ 
becomes
\eqref{eigen-RF-low}.
So we can see that solutions of two gap phase are
smoothly connected to solutions of the lower gap phase
at the phase transition point.

\paragraph{Phase transition from the upper gap to the two gap}
In the upper gap phase, 
if the minimum of the eigenvalue density $\rho(\pi)$ reaches to 
$\rho(\pi) = 0$
the phase transition from the upper gap phase to the two gap phase will occur.
The condition would be represented as 
\begin{equation}
\frac{\zeta}{\pi^2}
\cos \frac{a}{2}
\int_{\tilde c}^{\infty}dy ~  \frac{y e^{-y}}{
\sqrt{(e^{-2y}+2 e^{-y}\cos a +1)}
(1- e^{-y})}
=\frac{1}{2\pi\lambda }.
\label{Up-to-2-RF}
\end{equation}
The combination of \eqref{Up-to-2-RF}, 
\eqref{Cond-up-RF} and \eqref{tc} 
provides the 
phase transition points from the upper gap to the two gap.
By numerical calculations based on 
\eqref{Up-to-2-RF}, 
\eqref{Cond-up-RF} and \eqref{tc}, 
we have obtained the phase transition points 
plotted in Fig.~\ref{fig:up-two-RF}.

Let us see the phase transition points from the 
standpoint of the two gap phase.
If we substitute $b = \pi$ 
into \eqref{RF two u1}, 
it becomes \eqref{Up-to-2-RF}.
By using the relationship \eqref{Up-to-2-RF},
\eqref{RF two u0} at  $b = \pi$
is reduced to \eqref{Cond-up-RF}.
Let us check that
the 
eigenvalue density in the two gap phase \eqref{eigen-RF-two} at $b = \pi$ 
is smoothly connected 
to the one in the upper gap phase \eqref{eigen-RF-up}.
To confirm it, 
first we apply the following formula
\begin{equation}
I_{1}(a,\pi,\alpha)
= \frac{\pi}{\cos^2\frac{\alpha}{2}}\left( \frac{1}{\sin\frac{\alpha}{2}\sqrt{\sin^2\frac{\alpha}{2}-\sin^2\frac{a}{2}}}- \frac{1}{\cos\frac{a}{2}} \right)
\label{int-form 0},
\end{equation}
and then by using \eqref{Up-to-2-RF} 
we can rewrite $\rho_{2,tg}$ as
\begin{equation}
\rho_{2,tg}(\lambda,a,\pi;\alpha)  
= \frac{1}{2\pi\lambda} - \frac{\zeta}{\pi^2}
\int^{\infty}_{\tilde{c}} dy~
\frac{ye^{-y} \sin \frac{\alpha}{2} \sqrt{\sin^2 \frac{\alpha}{2} 
- \sin^2 \frac{a}{2}}}{(1-e^{-y})\sqrt{1 + 2 \cos a e^{-y} + e^{-2y}}}.
\end{equation}  
Then by summing up with $\rho^{r.f}_{1,tg}(\zeta,a,\pi,\tilde{c},\alpha)$, 
we can see that \eqref{eigen-RF-two} 
at $b = \pi$
becomes 
\eqref{eigen-RF-up}.
So we can see that solutions of two gap phase are
smoothly connected to solutions of the upper gap phase.

\paragraph{Quadruple phase transition 
point in the regular fermion theory}
There is the quadruple phase transition point
\begin{equation}
\lambda_{c}^{r.f} = 0.596967, \qquad \zeta_{c}^{r.f} = 2.864539029, \qquad
(\tilde{c}_{c} = 0.644715)
\label{Critical value RF}
\end{equation}
at which the no gap, the lower gap, the upper gap and 
the two gap phase coexist.
Let us check whether \eqref{Critical value RF}
is the quadruple phase transition point or not.
Note that \eqref{RF-no-low} is the condition for the 
phase transition from the no gap to the lower gap phase.
Eq.~\eqref{RF-no-up} is the condition for the phase transition 
between the no gap and the upper gap,
\eqref{low-to-2 RF} is the one between the lower gap and the two gap,
and \eqref{Up-to-2-RF} is the one between the upper gap and the two gap. 
So if \eqref{Critical value RF} satisfies these four equations 
simultaneously, it would be the quadruple phase transition point.
If we substitute $b = \pi$
into \eqref{low-to-2 RF} 
it becomes
\begin{equation}
\frac{\zeta}{\pi^2}\int^{\infty}_{\tilde c} dy~ y\frac{e^{-y}}{1-e^{-2y}}
= \frac{1}{2\pi\lambda}.
\label{low-2-cri}
\end{equation}
If we substitute $a = 0$ into 
\eqref{Up-to-2-RF}, 
it also becomes the same equation as 
\eqref{low-2-cri}.
We can also see that if both 
\eqref{RF-no-low} 
and 
\eqref{RF-no-up} 
are simultaneously satisfied, 
\eqref{low-2-cri}
is also automatically satisfied.
So if there is a point satisfying
\eqref{low-2-cri}, the point becomes the quadruple phase 
transition point.
By the study in \cite{Jain:2013py},
we already know that
only the point \eqref{Critical value RF} 
simultaneously satisfies both conditions
\eqref{RF-no-low}
and 
\eqref{RF-no-up}. Hence \eqref{Critical value RF} satisfies 
\eqref{low-2-cri}.
So \eqref{Critical value RF} becomes the quadruple phase transition point
in the regular fermion theory.
Then due to the existence of the quadruple point, 
the phase structure of the regular fermion theory becomes as in 
Fig.~\ref{fig:phase-RF}.
We can also check that it becomes 
$a = 0$ and $b = \pi$ at \eqref{Critical value RF},
and then the eigenvalue density functions 
in each phase 
\eqref{eigen-RF-low},
\eqref{eigen-RF-up}  
\eqref{eigen-RF-two} and (6.33) of \cite{Jain:2013py}
coincide.

\subsection{Critical boson theory}
In this subsection we will study the higher temperature phases, 
the lower gap, the upper gap and the two gap phase of the 
level $k$ $U(N)$ Chern-Simons matter theory 
coupled to massless critical bosons in the fundamental representation.
The low temperature no gap phase is already studied in section 6.2 of
\cite{Jain:2013py}.
This is a Chern-Simons gauged version of the $U(N)$ Wilson Fisher theory.
The Lagrangian of the UV theory is 
that of massless minimally coupled fundamental bosons 
deformed by the interaction,
\begin{equation}
 \delta S=\int d^{3}x A \bar\phi\phi,
\end{equation}
where $A$ is a Lagrange multiplier field.
\footnote{See 
subsection 4.3 of \cite{Aharony:2012ns} for details, and in particular 
eq.(4.35) for the Lagrangian; note \cite{Aharony:2012ns} employs the symbol 
$\sigma$ for our field $A$.}
The effective potential in this theory is given by (3.12) of 
\cite{Jain:2013py} and it is
\footnote{Originally in \cite{Jain:2013py}, the effective potential 
is obtained as (3.9) of \cite{Jain:2013py},
\begin{equation}
\label{vcrit}
 v[\rho]=-\frac{N }{6\pi}\left(\sigma^3 +\frac{2 (\sigma^2-A \beta^2)^{\frac{3}{2}}}{\lambda}\right) + \frac{N}{2\pi}\int_{\sigma}^{\infty} y dy \int_{-\pi}^{\pi}\rho(\alpha)d\alpha \left(\ln(1-e^{-y+i\alpha})+\ln(1-e^{-y-i\alpha})\right). 
\end{equation}
From this, constants $A$ and $\sigma$ are obtained by requirement
that they extremize the 
above \eqref{vcrit} at constant $\rho$.
By this, we yield $A = \sigma^2 T^2$, and the equation 
\eqref{selfboscri}. 
By substituting these, the effective potential is simplified to be
\eqref{vcritb}.}
\begin{equation}
\label{vcritb}
\begin{split}
 V(U)=&-\frac{N^2 \zeta}{6\pi}\sigma^3  
+ \frac{N^2 \zeta}{2\pi}\int_{\sigma}^{\infty} dy\int_{-\pi}^{\pi} d\alpha~y \rho(\alpha) \left(\ln(1-e^{-y+i\alpha})+\ln(1-e^{-y-i\alpha})\right) \\
\equiv& V^{c.b}[\rho, N],
\end{split}
\end{equation} 
where $\sigma$ provides the squared thermal mass as $\sigma^2 T^2$.
The $\sigma$ was determined by
\begin{equation}\label{selfboscri}
  \int_{-\pi}^{\pi} \rho(\alpha)     \left(\ln 2\sinh(\frac{\sigma- i \alpha}{2})+\ln 2\sinh(\frac{\sigma+ i \alpha}{2})\right)=0.
\end{equation}
The general form of the 
free energy of the critical boson theory on $S^2 \times S^1$ is 
given as 
\begin{equation}
\begin{split}
F_{c.b}^{N} =&
V^{c.b}[\rho,N]
- N^2 {\cal P}
\int^{\pi}_{-\pi}d \alpha\int^{\pi}_{-\pi} d \beta~
\rho(\alpha)\rho(\beta) \log 
\left|2 \sin \frac{\alpha-\beta}{2}\right|\\
=& V^{c.b}[\rho,N] + F_2[\rho,N].
\label{F-CB}
\end{split}
\end{equation}
To obtain the free energy in each phase, we
only have to evaluate the eigenvalue density $\rho$ and $\sigma$ 
in each phase, and substitute to \eqref{F-CB}.
Note that $\sigma$ can be determined if the eigenvalue density
is determined.
So obtaining the eigenvalue density in each phase 
is equivalent to obtaining the 
free energy in each phase.

For later use we will give the form of $V'(z)$ here.
We obtain $V'(z)$ from \eqref{vcritb},
\begin{equation}
V'(z) 
=  \frac{N \zeta}{2\pi}
\int_{\sigma}^{\infty}dy ~ y
\left(
\frac{-ie^{-y}}{z^{-1}-e^{-y}}
+\frac{ie^{-y}}{z - e^{-y}}
\right)
\end{equation}
where $z = e^{i\alpha}$. 

\begin{figure}
  \begin{center}
  \subfigure[]{\includegraphics[scale=.33]{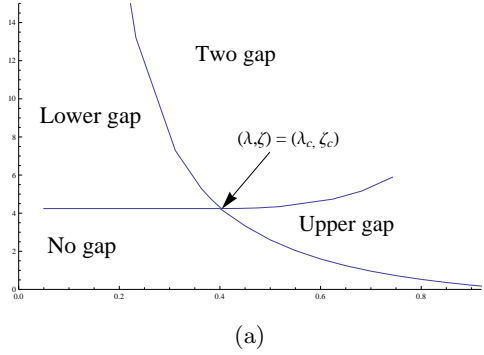}
\label{fig:phase-CB}
}
\caption{Phase diagram of the critical boson theory.
Here $(\lambda_c,\zeta_c) = (0.403033, 4.24292)$
is the quadruple phase transition point where 
the four phases (no gap, lower gap, upper gap, two gap phases) coexist.}
  \end{center}
\end{figure}

The phase structure of this theory is depicted in Fig.~\ref{fig:phase-CB}.
We will elaborate the phase structure in following subsubsections.

\subsubsection{Lower gap phase}
\label{L-CB}
In the lower gap phase of the critical boson theory, 
we use the same procedure as the one in section \ref{L-RF}
to obtain the eigenvalue density $\rho$.
We use the cut region and the cut function 
described in appendix \ref{Sec:Low-Cut}.
$U(z)$ is same as $V'(z)$
\begin{equation}
U(z) = V'(z) = \frac{N \zeta}{2\pi}
\int_{\sigma}^{\infty}dy ~ y
\left(
\frac{-ie^{-y}}{z^{-1}-e^{-y}}
+\frac{ie^{-y}}{z - e^{-y}}
\right),
\end{equation}
because $\rho_{0} = 0$. 
By substituting the above into \eqref{Hucp},
we immediately obtain $H(u)$ as well as $\Phi(u)$, as
\begin{equation}
\begin{split}
\Phi(u)=
\Phi^{c.b}_{lg}(\sigma, b, \zeta; u)
\equiv&\frac{\zeta}{2\pi}
\int_{\sigma}^{\infty}dy ~ y 
\frac{e^{-y}h(u)(1-e^{-y})(1+u^{-1})}{
\sqrt{(e^{-y}-e^{ib})(e^{-y}-e^{-ib})}
(u - e^{-y})(u^{-1} - e^{-y})}
\\
&+\frac{\zeta}{2\pi}
\left(
\int_{\sigma}^{\infty}dy ~ y \frac{e^{-y}}{u-e^{-y}}
-
\int_{\sigma}^{\infty}dy ~ y \frac{e^{-y}}{u^{-1}-e^{-y}}
\right).
\end{split}
\label{Phi-CB-L}
\end{equation}
From $\Phi^{+}(u) - \Phi^-(u) = 4 \pi \rho(u)$
at the cut $u = e^{i\alpha}$ with $-b \le \alpha \le b$,
we obtain the eigenvalue density in the lower gap phase as
\begin{equation}
\begin{split}
\rho(\alpha)
=& \frac{\zeta}{\sqrt{2} \pi^2}
\sqrt{\sin^2\frac{b}{2} - \sin^2 \frac{\alpha}{2}}
\int^{\infty}_{\sigma} dy \frac{y \sinh \frac{y}{2}\cos \frac{\alpha}{2}}
{\sqrt{\cosh y - \cos b}(\cosh y - \cos \alpha)}
\\
\equiv& \rho^{c.b}_{lg}(\zeta, \lambda;\sigma, b; \alpha).
\end{split}
\label{eigen-CB-low}
\end{equation}
At the lower gap phase we get $\rho(\alpha) = 0$ 
in $\pi > |\alpha| > b$.
By substituing the eigenvalue density \eqref{eigen-CB-low} into
\eqref{selfboscri} and \eqref{F-CB}, we can obtain the free energy in the 
lower gap phase in the critical boson theory. 

The condition $\lim_{u \to \infty} \Phi(u) = 1$ requires following
condition
\begin{equation}
\tilde{M}^{c.b}_{lg}(\zeta,\sigma,b)
\equiv 
-\frac{\zeta}{2\pi}
\int_{0}^{e^{-\sigma}}dx ~  
\left(
\frac{  \log x}{x}
- \frac{1-x}{x}
\frac{\log x}{\sqrt{x^2- 2 x\cos b  +1}}
\right)
= 1.
\label{Cond-low-CB}
\end{equation}
By \eqref{Cond-low-CB} and \eqref{selfboscri}, we can obtain 
$(b,\sigma)$ as functions of $(\lambda, \zeta)$ as
$(b,\sigma) = (b(\lambda,\zeta),\sigma(\lambda,\zeta))$.
\eqref{eigen-CB-low}, \eqref{Cond-low-CB} and \eqref{selfboscri}
provide a set of complete solutions in the lower gap phase of the 
critical boson theory.


\subsubsection{Upper gap phase}
\label{UP-CB}
Next we will search for an upper gap solution 
in the critical boson theory. 
The cut region and the cut function $h(u)$ in this case are defined 
in appendix~\ref{Sec:Up-Cut}.
Here the procedure to obtain the 
eigenvalue density function $\rho$ is the same as the one in 
subsection~\ref{UP-RF}.
We will take 
$\rho_{0}(\alpha) = \frac{1}{2 \pi \lambda} \ne 0$ 
at the upper gap region
$-a \le \alpha \le a$.

Functions $H(u)$ and $\Phi(u)$ are obtained based on 
\eqref{Hucp},
\begin{equation}
\begin{split}
\Phi(u) =& \Phi^{c.b}_{ug}(\sigma,a,\zeta;u)
+ \Phi^{\rho_{0}}_{ug}(\lambda,a;u),
\qquad {\text {where}}
\\
\Phi^{c.b}_{ug}(\sigma,a,\zeta;u)\equiv&
\frac{\zeta}{2\pi}
\int_{\sigma}^{\infty}dy ~  \frac{ye^{-y}h(u)
(1+e^{-y})(1-u^{-1})}{
\sqrt{(e^{-y}-e^{ia})(e^{-y}-e^{-ia})}
(u^{-1} - e^{-y})(u - e^{-y})}
\\
&+\frac{\zeta}{2\pi}
\left(
\int_{\sigma}^{\infty}dy ~  \frac{ye^{-y}}{u-e^{-y}}
-
\int_{\sigma}^{\infty}dy ~  \frac{ye^{-y}}{u^{-1}-e^{-y}}
\right).
\end{split}
\label{Phi-CB-U}
\end{equation}
Here we use the same definition of 
$\Phi^{\rho_{0}}_{ug}(\lambda,a;u)$ as \eqref{Phi-RF-up}.
From \eqref{Phi-CB-U}, 
by taking $\Phi^{+}(u)- \Phi^{-}(u) = 4\pi(\rho(u)- \rho_{0}(u))$
at the cut region $u = e^{i\alpha}$ with 
$\pi \ge |\alpha| \ge a$,
we can obtain the eigenvalue density function as
\begin{equation}
\begin{split}
\rho(\alpha)
=&\frac{1}{2\pi \lambda} 
-\frac{\zeta}{\sqrt{2}\pi^2}
\sqrt{\sin^2 \frac{\alpha}{2} - \sin^2 \frac{a}{2}}
\int_{\sigma}^{\infty}dy ~  \frac{y |\sin\frac{\alpha}{2}|\cosh \frac{y}{2}}{
\sqrt{\cosh y -\cos a}(\cosh y -\cos \alpha)}
\\
\equiv&
\rho^{c.b}_{ug}(\zeta,\lambda;\sigma,a;\alpha).
\end{split}
\label{eigen-CB-up}
\end{equation}
In the region $|\alpha| < |a|$, $\rho(\alpha) = \frac{1}{2\pi\lambda}$. 
By substituing the eigenvalue density \eqref{eigen-CB-up} into
\eqref{selfboscri} and \eqref{F-CB}, we can obtain the free energy in the 
upper gap phase. 
To derive \eqref{eigen-CB-up}, we have used the formula
\eqref{Formula up 2}.

From the condition, $\lim_{u \to \infty}\Phi(u) 
= 1- \int^{\pi}_{-\pi} d\alpha \rho_{0}(\alpha)$,
we obtain the condition
\begin{equation}
\tilde{M}^{c.b}_{ug}(\zeta, \sigma,a)
\equiv
-\frac{\zeta}{2\pi}
\int_{0}^{e^{-\sigma}}dx ~  
\left(
\frac{  \log x}{x}
-\frac{1+x  }{x}
\frac{\log x}{\sqrt{x^2- 2 x\cos a  +1}}\right)
=
1- \frac{1}{\lambda}
.
\label{Cond-up-CB}
\end{equation}
Here we have used the formula \eqref{Formula up 1}.
By \eqref{Cond-up-CB} and \eqref{selfboscri},
we can obtain $(a, \sigma)$ as functions of $(\lambda, \zeta)$ as
$(a,\sigma) = (a(\lambda,\zeta), \sigma(\lambda, \zeta))$.
\eqref{eigen-CB-up}, \eqref{Cond-up-CB} and \eqref{selfboscri} provide
a set of complete solutions 
in the upper gap phase in the critical boson theory.


\subsubsection{Two gap phase}
\label{sec:CB-2}
We will search for a two gap solution in the critical boson theory.
Our two cuts and the cut function are defined in 
appendix~\ref{Sec:Two-Cut}.
By the same procedure as the one 
in section~\ref{sec:RF-2}, 
we can obtain $H(u)$ as well as $\Phi(u)$ as
\begin{equation}
\begin{split}
\Phi(u) =&
\Phi^{c.b}_{tg}(\zeta,a,b,\sigma;u)
+ \Phi^{\rho_0}_{tg}(\lambda,a,b;u)
\\
\Phi^{c.b}_{tg}(\zeta,a,b,\sigma;u)
\equiv&
\frac{\zeta}{2\pi}
\int_{\sigma}^{\infty}dy ~  
\left(
\frac{ye^{-y}(u^{-1}-u)}{(u-e^{-y})(u^{-1}-e^{-y})}
+
\frac{ ye^{y}h(u)(e^{y}+e^{-y}-2u)}{
h(e^{y})(e^{y}-u )(e^{-y}-u )}\right).
\end{split}
\label{Two-CB-Phi}
\end{equation}
Here $\Phi^{\rho_0}_{tg}(\lambda,a,b;u)$ is 
already defined in \eqref{Two-RF-Phi}.
From $\lim_{u \to \infty} \Phi(u) = 1
- \int^{\pi}_{-\pi} d \alpha~ \rho_{0}(\alpha)$, 
we obtain following two conditions,
\begin{equation}
\begin{split}
\frac{1 }{4\pi \lambda  } \Upsilon(a,b)
 =& \frac{\zeta}{2\pi} 
{\cal Y}^{c.b}(a,b,\sigma)
\\
{\cal Y}^{c.b}(a,b,\sigma)
\equiv&
\int^{\infty}_{\sigma} dy~\frac{y}{
\sqrt{(\cosh y - \cos a)(\cosh y - \cos b)}
},
\end{split}
\label{CB two u1}
\end{equation}
and 
\begin{equation}
\begin{split}
\frac{1}{4\pi\lambda}\Lambda(a,b) =& 1+ 
\frac{\zeta}{4 \pi}
{\cal G}^{c.b}(a,b,\sigma)
\\
{\cal G}^{c.b}(a,b,\sigma)
\equiv& 
\int^{\infty}_{\sigma} dy~
y\left( \frac{e^{y}+e^{-y}}{
\sqrt{(\cosh y - \cos a)(\cosh y - \cos b)}
} - 2\right).
\end{split}
\label{CB two u0}
\end{equation}
$\Upsilon(a,b)$ and $\Lambda(a,b)$ are already defined in 
\eqref{RF two u1} and \eqref{RF two u0}.
By using 
\eqref{CB two u0},
\eqref{CB two u1} and 
\eqref{selfboscri}, $(a,b, \sigma)$ are determined as 
functions of $(\lambda, \zeta)$ as
$(a,b, \sigma) = (a(\lambda, \zeta),b(\lambda, \zeta), 
\sigma(\lambda, \zeta))$.

From \eqref{Two-CB-Phi} 
and by taking $\Phi^{+}(u)- \Phi^{-}(u) = 4\pi(\rho(u)- \rho_{0}(u))$ 
at the cuts $|a| < |\alpha| < |b|$, we obtain the eigenvalue density function
as
\begin{equation}
\begin{split}
\rho(\alpha) =& \rho^{c.b}_{tg}(\alpha) =
\rho_{1,tg}^{c.b}(\zeta, a,b,\tilde{c};\alpha)
+\rho_{2,tg}(\lambda, a,b;\alpha),
\qquad {\text {where}}
\\
\rho_{1,tg}^{c.b}(\zeta, a,b,\tilde{c};\alpha)
\equiv& -
\frac{\zeta }{\pi^2}
\cF(a,b;\alpha),
\int_{\tilde c}^{\infty}dy  \frac{ye^{-y}}{
\nu_{c.b}(a,b;y)}\left(
\frac{|\sin \alpha|}{\cosh y 
-  \cos \alpha }
\right)
\\
\nu_{c.b}(a,b;y)
\equiv& \sqrt{
(e^{-2y}-2e^{-y}\cos a+1)
(e^{-2y}-2e^{-y}\cos b+1)
}.
\end{split}
\label{eigen-CB-two}
\end{equation}
Here \eqref{eigen-CB-two} shares the same definition of 
$\rho_{2,tg},\cF$ with \eqref{eigen-RF-two}. $\rho_{2,tg}$ 
is the same functional form 
as the eigenvalue density 
in the GWW model (7.6) in \cite{Jain:2013py}.
$\rho_{1,tg}^{c.b}$ can be regarded as an additional term 
depending on the detail of the theory.
By substituing the eigenvalue density \eqref{eigen-CB-two} into
\eqref{selfboscri} and \eqref{F-CB}, we can obtain the free energy in the 
two gap phase of the critical boson theory. 
During the calculation of 
\eqref{eigen-CB-two},
we have used \eqref{CB two u1} also.

The combination of \eqref{CB two u1}, \eqref{CB two u0}, 
\eqref{selfboscri} and \eqref{eigen-CB-two} provides a complete set of 
solutions
in the two gap phase.


At large $\zeta$ limit, as we proved in 
appendix \ref{CB-Z-Pr}, 
the eigenvalue density approaches the 
universal distribution \eqref{evd 0}
because $\sigma$ remains finite positive quantity at the limit.  
In the limit, $a$,$b$ and the eigenvalue 
density behave as
\begin{equation}
\begin{split}
a =& \pi \lambda - \frac{\epsilon}{2},
b = \pi \lambda + \frac{\epsilon}{2},
\\
\epsilon =& 
8 \sin (\pi \lambda) \exp
\left(
- \frac{\sin (\pi\lambda)}{2}\lambda\zeta
\int^{\infty}_{\tilde{c}} dy~\frac{y}{\cosh y - \cos \pi \lambda}\right) \ldots,
\\
\rho(\alpha) =&  \frac{1}{\pi^2 \lambda} \cos^{-1} \sqrt{
\frac{\alpha - a}{b - a}
}.
\end{split}
\label{Large z CB}
\end{equation}
At appendix \ref{CB-LZ}, we have demonstrated the above behavior
\eqref{Large z CB}.
As we can see in the appendix, 
the eigenvalue density is dominated by $\rho_{2.tg}$ 
at the large $\zeta$ limit. The form 
of the eigenvalue density approaches $\cos^{-1}\sqrt{\alpha_1}$ function 
which has 
the same form as Eq.~(7.11) in \cite{Jain:2013py}, which is the large $\zeta$ 
limit of the eigenvalue density of the GWW model.
We can also see that the range of the domain of the cut $\epsilon$ is
smaller than the one in the GWW type matrix integration at the same value of
$(\lambda, \zeta)$.

\subsubsection{Phase transition points}
\paragraph{Phase transition from the lower gap to the no gap}
Let us consider the behavior of lower gap solutions
\eqref{Cond-low-CB} and \eqref{eigen-CB-low}
at $b = \pi$ which would correspond to the 
phase transition points from the lower gap to the no gap.
If we substitute $b = \pi$ into \eqref{Cond-low-CB},
it becomes
\begin{equation}
1 = -\frac{\zeta}{\pi}
\int_{0}^{e^{-\sigma}}dx ~  
\left(
\frac{ \log x}{1+x}
\right)
= \frac{\zeta}{\pi}
\sum_{n=1}^{\infty} (-1)^{n+1}\frac{1+ \sigma n}{n^2}e^{-n\sigma}.
\label{CB-no-low}
\end{equation}
This is exactly same as the 
condition for the phase transition 
from the no gap to the lower gap phase
which is obtained by substituting $\alpha = \pi$ into 
eq.~(6.24) of \cite{Jain:2013py} and by requiring $\rho(\pi) = 0$.
Based on this, if we substitute $b = \pi$ into
\eqref{eigen-CB-low}, 
we obtain
\begin{equation}
\begin{split}
\rho(\alpha)
&=\frac{\zeta}{\sqrt{2} \pi^2}
\sqrt{\sin^2\frac{b}{2} - \sin^2 \frac{\alpha}{2}}
\int^{\infty}_{\sigma} dy \frac{y \sinh \frac{y}{2}\cos \frac{\alpha}{2}}
{\sqrt{\cosh y - \cos b}(\cosh y - \cos \alpha)}\biggr|_{b = \pi}
\\
&=\frac{\zeta}{\pi^2}
\int^{\infty}_{\sigma}
dy~ y\left(
-\frac{1}{2(1+e^{-y})}
+\frac{1}{4(1-e^{-y+i\alpha})}
+\frac{1}{4(1-e^{-y-i\alpha})}
\right)
\\
&=\frac{1}{2\pi}
+ \frac{\zeta}{2\pi^2}
\sum_{n=1}^{\infty} 
\cos n \alpha
\left(\frac{1+\sigma n}{n^2}\right)e^{-n\sigma}.
\end{split}
\label{eigen-CB-low-no}
\end{equation}
In the last step we have used 
\eqref{CB-no-low}, the Taylor expansion
and an integration formula \eqref{eq:Taylor-formula no-low}.
The last line of \eqref{eigen-CB-low-no}
is exactly same as the eigenvalue distribution in the no gap phase,
described at Eq.~(6.24) in \cite{Jain:2013py}.
So, at the phase transition points, 
the lower gap solutions are smoothly connected to the 
the no gap phase solutions. 

\paragraph{Phase transition from the upper gap to the no gap}
Let us consider the behavior of upper gap solutions 
\eqref{eigen-CB-up} and \eqref{Cond-up-CB}
at $a= 0$, which would correspond to the phase transition points 
from the upper gap to the no gap phase.
If we substitute $a = 0$ into \eqref{Cond-up-CB}, it becomes
\begin{equation}
1- \frac{1}{\lambda}
= 
\frac{\zeta}{\pi}
\int_{0}^{e^{-\sigma}}dx ~  
\frac{  \log x}{1-x}
= -\frac{\zeta}{\pi}
\sum_{n=1}^{\infty}\left(
\frac{1+\sigma n}{n^2}
e^{-n\sigma}
\right).
\label{CB-no-up}
\end{equation}
This is exactly same as the 
condition for the phase transition 
from the no gap to the upper gap phase 
discussed in \cite{Jain:2013py}, 
which is obtained by substituting $\alpha = 0$ into 
(6.24) of \cite{Jain:2013py} 
and by requiring $\rho(0) = \frac{1}{2\pi\lambda}$.

Based on this, if we substitute $a = 0$ into
\eqref{eigen-CB-up}, 
we can see 
\begin{equation}
\begin{split}
\rho(\alpha)
&= \frac{1}{2\pi \lambda} 
-\frac{\zeta}{\sqrt{2}\pi^2}
\sqrt{\sin^2 \frac{\alpha}{2} - \sin^2 \frac{a}{2}}
\int_{\sigma}^{\infty}dy ~  \frac{y |\sin\frac{\alpha}{2}|\cosh \frac{y}{2}}{
\sqrt{\cosh y -\cos a}(\cosh y -\cos \alpha)} \biggr|_{a = 0}
\\
&= \frac{1}{2 \pi \lambda}
- \frac{\zeta}{\pi^2}
\int^{\infty}_{\sigma} dy~
y \left(
\frac{1}{2 (1-e^{-y} )}
-\frac{1}{4 (1-e^{-y+i\alpha} )}
-\frac{1}{4 (1-e^{-y-i\alpha} )}
\right)
\\
&=\frac{1}{2\pi}
+ \frac{\zeta}{2\pi^2}
\sum_{n=1}^{\infty} 
\cos n \alpha
\left(\frac{1+\sigma n}{n^2}\right)e^{-n\sigma}.
\end{split}
\label{eigen-CB-up-no}
\end{equation}
In the last step we have used 
\eqref{CB-no-up}, the Taylor expansion
and the formula \eqref{eq:Taylor-formula no-low}.
The last line of \eqref{eigen-CB-up-no}
is exactly same as the eigenvalue distribution in the no gap phase,
described at eq.~(6.24) in \cite{Jain:2013py}.
So, at the phase transition points,
the upper gap solutions are smoothly 
connected to the no gap phase solutions.

\begin{figure}
  \begin{center}
  \subfigure[]{\includegraphics[scale=.33]{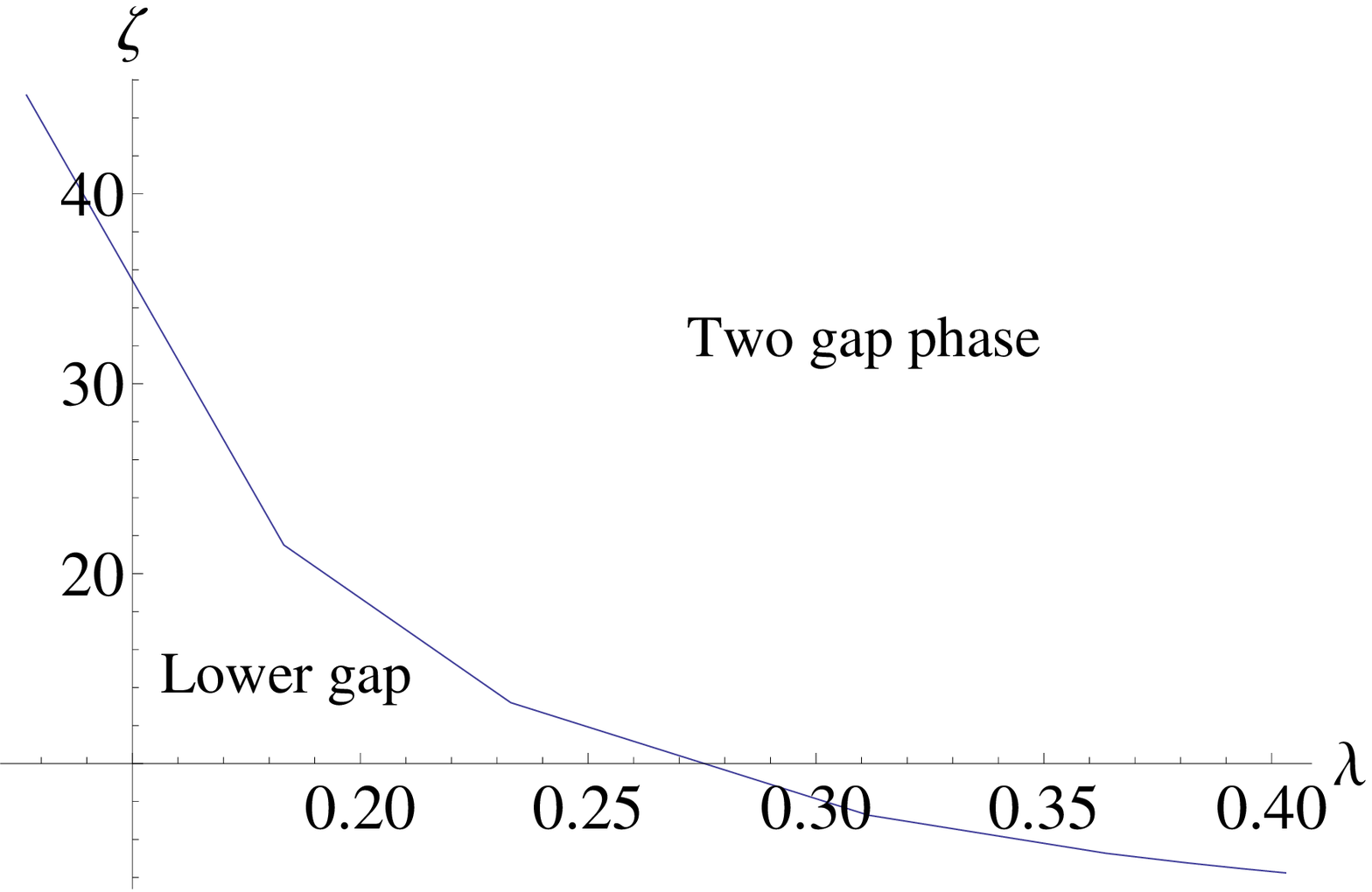}
\label{fig:low-two-CB}}
\qquad \qquad
  \subfigure[]{\includegraphics[scale=.33]{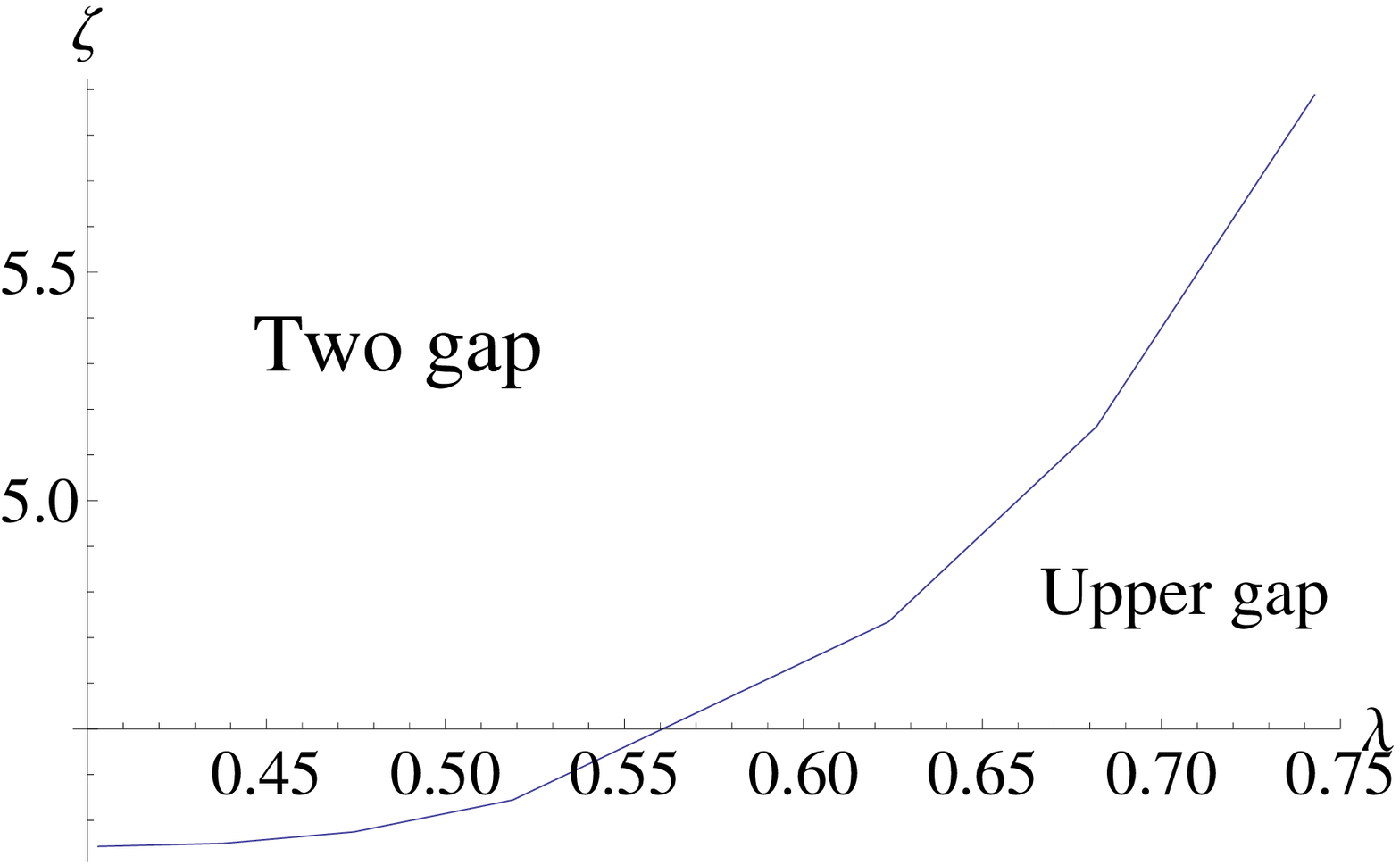}
\label{fig:up-two-CB}}
\caption{These are the plots of phase transition point in the critical 
boson theory.
Fig.~\ref{fig:low-two-CB} shows the plots of 
the phase transition points from the lower gap to
the two gap phase, and Fig.~\ref{fig:up-two-CB} shows the ones between the 
upper gap and the two gap phase.}
  \end{center}
\end{figure}

\paragraph{Phase transition from the lower gap to the two gap}
In the lower gap phase at fixed $\lambda$, 
if the maximum of the eigenvalue density $\rho(0)$ reaches to 
$\rho(0) = \frac{1}{2\pi \lambda}$,
the phase transition from the lower gap phase to the two gap phase will occur.
The conditions would be represented in terms of \eqref{eigen-CB-low}
as 
\begin{equation}
\rho(0) 
= 
\frac{\zeta}{2\pi^2}\sin \frac{b}{2}
\int_{\sigma}^{\infty}dy ~  \frac{y}{
\sqrt{2(\cosh y-\cos b)}
\sinh \frac{y}{2}}
= \frac{1}{2\pi \lambda}.
\label{low-to-2 CB}
\end{equation}
The combination of \eqref{low-to-2 CB}, \eqref{selfboscri}
and \eqref{Cond-low-CB} provides the 
phase transition points from the lower gap to the two gap.
By numerical calculations based on 
\eqref{low-to-2 CB}, \eqref{selfboscri}
and \eqref{Cond-low-CB}, 
we obtain the phase transition points 
plotted in Fig.~\ref{fig:low-two-CB}.

Let us see the phase transition points from the 
standpoint of the two gap phase.
If we substitute $a = 0$ 
into \eqref{CB two u1}, 
it becomes the same as \eqref{low-to-2 CB}, 
which is the condition of the phase transition points.
If we set $a=0$ at \eqref{CB two u0},
it becomes the same as \eqref{Cond-low-CB}.
Let us check if the eigenvalue density in the two gap phase 
\eqref{eigen-CB-two} becomes the one in the lower gap phase
\eqref{eigen-CB-low} in $a = 0$ limit.
By using \eqref{low-to-2 CB},
we can replace 
$\rho_{2,tg}(\lambda,0,b;\alpha)$ of \eqref{eigen-CB-two} 
by
\begin{equation}
\rho_{2,tg}(\lambda,0,b;\alpha) =
\frac{\zeta}{2\pi^2}
\int_{\sigma}^{\infty}dy ~  \frac{y\cos \frac{\alpha}{2}
\sqrt{\sin^2\frac{b}{2}-\sin^2\frac{\alpha}{2}}}{
\sqrt{2(\cosh y-\cos b)}
\sinh \frac{y}{2}}.
\label{CB-l-2}
\end{equation}
Then by summing up with $\rho^{c.b}_{1,tg}(\zeta,0,b,\tilde{c};\alpha)$, 
\eqref{eigen-CB-two} at $a = 0$ becomes
\eqref{eigen-CB-low}.
So, at the phase transition points,
we can see that the solutions of two gap phase 
are smoothly connected to the lower gap phase solutions.

\paragraph{Phase transition from the upper gap to the two gap}
In the upper gap phase, 
if the minimum of the eigenvalue density $\rho(\pi)$ reaches to 
$\rho(\pi) = 0$, the phase transition from the upper 
gap phase to two gap phase will occur.
The condition would be represented as 
\begin{equation}
\frac{1}{2\pi \lambda} = 
\frac{\zeta}{\sqrt{2}\pi^2}
\cos \frac{a}{2}
\int_{\sigma}^{\infty}dy ~  \frac{y \cosh \frac{y}{2}}{
\sqrt{\cosh y -\cos a}(\cosh y +1)}.
\label{CB-UP-TWO}
\end{equation}
The combination of \eqref{CB-UP-TWO}, 
\eqref{Cond-up-CB} and \eqref{selfboscri} 
provides the 
phase transition points from the upper gap to the two gap.
By numerical calculations based on 
\eqref{CB-UP-TWO}, 
\eqref{Cond-up-CB} and \eqref{selfboscri}, 
we obtain the phase transition points 
plotted in Fig.~\ref{fig:up-two-CB}.

Let us see the phase transition points from the 
standpoint of the two gap phase.
If we substitute $b = \pi$ 
into \eqref{CB two u1} 
it becomes \eqref{CB-UP-TWO} which is the phase transition condition.
By using \eqref{CB-UP-TWO},
we can see that \eqref{CB two u0}
at  $b = \pi$ is reduced to \eqref{Cond-up-CB}.
Let us check that
the eigenvalue density in the two gap phase \eqref{eigen-CB-two} at $b = \pi$
is smoothly connected 
to the one in the upper gap phase \eqref{eigen-CB-up}.
To confirm it, 
first we apply the formula \eqref{int-form 0}.
Then by using \eqref{CB-UP-TWO} 
we can rewrite $\rho_{2,tg}$ as
\begin{equation}
\rho_{2,tg}(\lambda,a,\pi;\alpha)  
= \frac{1}{2\pi\lambda}- \frac{\zeta}{\pi^2}
\int^{\infty}_{\sigma} dy~
\frac{ye^{-y} |\sin \frac{\alpha}{2}| \sqrt{\sin^2 \frac{\alpha}{2} 
- \sin^2 \frac{a}{2}}}{(1+e^{-y})\sqrt{1 - 2 \cos a e^{-y} + e^{-2y}}}.
\end{equation}  
Then by summing up with 
$\rho^{c.b}_{1,tg}(\zeta,a,\pi,\tilde{c},\alpha)$, 
we can see that \eqref{eigen-CB-two} 
at $b = \pi$
becomes
\eqref{eigen-CB-up}.
So we can see that the solutions of two gap phase are
smoothly connected to solutions of the upper gap phase.

\paragraph{Quadruple phase transition 
point in the critical boson theory}
There is a quadruple phase transition point
\begin{equation}
\lambda_{c}^{c.b} = 0.403033, \qquad \zeta_{c}^{c.b} = 4.24292, \qquad
(\sigma_{c} = 0.644715),
\label{Critical value CB}
\end{equation}
at which the four phases coexist.
Let us check whether \eqref{Critical value CB}
is the quadruple phase transition point or not.
Note that \eqref{CB-no-low} is the condition for the 
phase transition from the no gap to the lower gap phase.
Eq.~\eqref{CB-no-up} is the the condition for the 
phase transition between the no gap and the upper gap,
\eqref{low-to-2 CB} is the one between the lower gap and the two gap,
and \eqref{CB-UP-TWO} is the one between the upper gap and the two gap. 
So if \eqref{Critical value CB} satisfies these four equations 
simultaneously, it becomes the quadruple phase transition point.
If we substitute $b = \pi$
into \eqref{low-to-2 CB} 
it becomes
\begin{equation}
\frac{\zeta}{\pi^2} \int^{\infty}_{\sigma} dy~ \frac{y e^{-y}}{1 - e^{-2y}}
= \frac{1}{2\pi\lambda},
\label{cb-cri}
\end{equation}
and if we substitute $a = 0$ into 
\eqref{CB-UP-TWO}, 
it also becomes the same equation as 
\eqref{cb-cri}.
We can also see that if both 
\eqref{CB-no-low} 
and 
\eqref{CB-no-up} 
are simultaneously satisfied, 
\eqref{cb-cri}
is also automatically satisfied.
So if there is a point satisfying
\eqref{cb-cri}, the point becomes the quadruple phase 
transition point.
By the study in \cite{Jain:2013py},
we already know that
only the point \eqref{Critical value CB} 
simultaneously satisfies both condition
\eqref{CB-no-low}
and 
\eqref{CB-no-up}. Hence \eqref{Critical value CB} satisfies 
\eqref{cb-cri}.
So \eqref{Critical value CB} becomes the quadruple phase transition point
in the critical boson theory.
Then due to the existence of the quadruple point, 
the phase structure of the critical boson theory becomes as in 
Fig.~\ref{fig:phase-CB}.
We can also check that it becomes 
$a = 0$ and $b = \pi$ at \eqref{Critical value CB},
and then the eigenvalue density functions 
in each phase 
\eqref{eigen-CB-low},
\eqref{eigen-CB-up},  
\eqref{eigen-CB-two} and (6.24) of \cite{Jain:2013py}
coincide. 

\subsection{Duality between the regular fermion theory and the 
critical boson theory}
\label{D-RF-CB}
In this subsection we will verify the level-rank duality
between the level $k$ $U(N)$ 
critical boson theory and the level $k$ $U(k-N)$ regular fermion theory.
We will show that the quantities 
at $(\lambda_{c.b}, \zeta_{c.b})$ in the critical boson theory 
have duality relationships 
to the quantities at $(\lambda_{r.f}, \zeta_{r.f})
= (1-\lambda_{c.b}, \frac{\lambda_{c.b}}{1-\lambda_{c.b}}\zeta_{c.b} )$ 
in the regular fermion theory as
\begin{equation}
\rho^{r.f} (\alpha) = 
\frac{\lambda_{c.b}}{1-\lambda_{c.b}}\left(
\frac{1}{2 \pi \lambda_{c.b}} - \rho^{c.b}(\alpha + \pi) 
\right),
\label{corres-1}
\end{equation}
\begin{equation}
b_{r.f} = \pi - a_{c.b}, \quad a_{r.f} = \pi - b_{c.b},
\quad \tilde{c} = \sigma.
\label{corres-2}
\end{equation}
Here the $\rho^{r.f}$ is the eigenvalue density 
in the regular fermion theory and $\rho^{c.b}$ is the one 
in the critical boson theory.
The relationship between $(\lambda_{c.b},\zeta_{c.b})$
and $(\lambda_{r.f},\zeta_{r.f})$ is
\begin{equation}
\lambda_{r.f} = 1- \lambda_{c.b}, \qquad 
\zeta_{r.f} = \frac{\lambda_{c.b}}{1-\lambda_{c.b}}\zeta_{c.b},\qquad
\left(\frac{N}{k} = \lambda_{c.b}, \quad \frac{k-N}{k} = \lambda_{r.f}
\right).
\label{corres-0}
\end{equation}
If the above duality relationships 
\eqref{corres-1} $\sim$ \eqref{corres-0}
are valid, 
the free energy at 
$(\lambda_{c.b}, \zeta_{c.b})$ in the critical boson  
theory agrees with the free energy at
$(\lambda_{r.f}, \zeta_{r.f})
= \left(1-\lambda_{c.b}, 
\frac{\lambda_{c.b}}{1-\lambda_{c.b}}\zeta_{c.b} \right)$ 
in the regular fermion theory. 
So if we verify the above relationships 
\eqref{corres-1} $\sim$ \eqref{corres-0},
it would be the proof of the level-rank duality
between the regular fermion theory and the critical boson theory.

In this subsection, we will first demonstrate the relationships 
\eqref{corres-1} $\sim$ \eqref{corres-0} 
between the lower gap phase of the regular fermion theory and 
the upper gap phase of the critical boson theory.
Next we will verify the relationships
between 
the upper gap phase of the regular fermion theory 
and the lower gap phase of the critical boson theory.
As the third step, we will demonstrate the duality relationships between 
the two gap phase of both theories.
We complete the proof of the relationships
\eqref{corres-1} $\sim$ \eqref{corres-0} 
between both theories by these three steps.
As the final part of this subsection, 
we will completely 
verify the duality by evaluating the free energy in both theories.

\subsubsection{Between the lower gap phase of regular fermion and 
upper gap phase of critical boson}
\label{RF-L-CB_U}
Let us define sets of quantities ${\cal S}^{c.b}_{ug}, {\cal S}^{r.f}_{lg}$ 
as
\begin{equation}
\begin{split}
{\cal S}^{c.b}_{ug} =& (\lambda_{c.b}, \zeta_{c.b};
a_{c.b}, \sigma, \rho_{ug}^{c.b}(\alpha)), \\
{\cal S}^{r.f}_{lg} = &
(\lambda_{r.f}, \zeta_{r.f};
b_{r.f}, \tilde{c}, \rho_{lg}^{r.f}(\alpha))
\\
=&\left(1-\lambda_{c.b}, \frac{\lambda_{c.b}}{1-\lambda_{c.b}}\zeta_{c.b};
\pi-a_{c.b}, \sigma, 
\frac{\lambda_{c.b}}{1-\lambda_{c.b}}
\left(
\frac{1}{2\pi\lambda_{c.b}} - \rho^{c.b}_{ug}
( \pi+\alpha)\right)\right).
\end{split}
\end{equation}
Note that the eigenvalue density 
\eqref{eigen-CB-up}
and the two equations \eqref{Cond-up-CB}, \eqref{selfboscri}
provide a complete set of solutions in the upper gap phase of the 
critical boson model,
while the eigenvalue density \eqref{eigen-RF-low}
and the two equations \eqref{Cond-low-RF}, \eqref{tc}
provide a complete set of solutions in the  
lower gap phase of the regular fermion theory.
So we can verify the duality relationships by 
checking that ${\cal S}_{lg}^{r.f}$ becomes a solution provided 
by the three equations in the regular fermion theory 
if ${\cal S}_{ug}^{c.b}$ is a solution provided by the three equations 
in the critical boson side.

First we will confirm the relationship \eqref{corres-1}
between the eigenvalue densities
\eqref{eigen-CB-up}
and \eqref{eigen-RF-low} under the correspondence 
\eqref{corres-0},\eqref{corres-2}.
We can show the relationship by directly substituting the 
\eqref{corres-0},\eqref{corres-2} into \eqref{eigen-RF-low} as
\footnote{Steps of the calculation to verify 
\eqref{RF_L-CB_U-e} are basically combinations of 
$\cos (x+\pi) = -\cos x, \, \sin\frac{x+\pi}{2} = \cos \frac{x}{2},$ as follows
\begin{equation}
\begin{split}
&\frac{\lambda_{c.b}}{1-\lambda_{c.b}}
\left(
\frac{1}{2\pi\lambda_{c.b}} - \rho^{c.b}_{ug}
\left(\zeta_{c.b},\lambda_{c.b};
\sigma,a_{c.b}; \alpha+\pi\right)\right)
\\
=& \frac{\lambda_{c.b}}{1-\lambda_{c.b}}
\frac{\zeta_{c.b}}{\sqrt{2}\pi^2}
\sqrt{\sin^2 \frac{\alpha + \pi}{2} - \sin^2 \frac{a_{c.b}}{2}}
\int_{\sigma}^{\infty}dy ~  
\frac{y |\sin\frac{\alpha + \pi}{2}|\cosh \frac{y}{2}}{
\sqrt{\cosh y -\cos a_{c.b}}(\cosh y -\cos (\alpha+\pi))}
\\
=& 
\frac{\lambda_{c.b}}{1-\lambda_{c.b}}
\frac{\zeta_{c.b}}{\sqrt{2}\pi^2}
\sqrt{-\cos^2 \frac{\pi-a_{c.b}}{2}+\cos^2 \frac{\alpha}{2}}
\int_{\sigma}^{\infty}dy ~  
\frac{y \cos\frac{\alpha}{2}\cosh \frac{y}{2}}{
\sqrt{\cosh y +\cos(\pi - a_{c.b})}(\cosh y +\cos \alpha)}
\\
=& 
\frac{\lambda_{c.b}}{1-\lambda_{c.b}}\frac{\zeta_{c.b}}{\sqrt{2}\pi^2}
\sqrt{\sin^2 \frac{\pi-a_{c.b}}{2} - \sin^2 \frac{\alpha}{2}}
\int_{\sigma}^{\infty}dy ~  
\frac{y \cos\frac{\alpha}{2}\cosh \frac{y}{2}}{
\sqrt{\cosh y +\cos (\pi - a_{c.b})}(\cosh y +\cos \alpha)}
\\
=& \rho^{r.f}_{lg}
\left(\frac{\lambda_{c.b}}{1-\lambda_{c.b}}\zeta_{c.b},1-\lambda_{c.b};
\sigma,\pi-a_{c.b}; \alpha\right).
\end{split}
\label{D-1-1}
\end{equation}
Basically, almost all steps of calculations to verify the duality relationships
are just repeating the similar steps to the above.
So from here we will not 
give a blow-by-blow description of similar steps
to avoid to be lengthy.}
\begin{equation}
\begin{split}
&\rho^{r.f}_{lg}
\left(\frac{\lambda_{c.b}}{1-\lambda_{c.b}}\zeta_{c.b},1-\lambda_{c.b};
\sigma,\pi-a_{c.b}; \alpha\right)
\\
=&  \frac{\lambda_{c.b}}{1-\lambda_{c.b}}
\left(
\frac{1}{2\pi\lambda_{c.b}} - \rho^{c.b}_{ug}
\left(\zeta_{c.b},\lambda_{c.b};
\sigma,a_{c.b}; \pi+\alpha\right)\right).
\end{split}
\label{RF_L-CB_U-e}
\end{equation}

We can also see that 
${\cal S}^{r.f}_{lg}$ becomes a 
solution of \eqref{Cond-low-RF} 
if 
${\cal S}^{c.b}_{ug}$ is a solution 
of \eqref{Cond-up-CB}
by 
\begin{equation}
\tilde{M}^{r.f}_{lg}\left(
\frac{\lambda_{c.b}}{1-\lambda_{c.b}}\zeta_{c.b}, \sigma, \pi-a_{c.b}
\right) = -\frac{\lambda_{c.b}}{1-\lambda_{c.b}}
\tilde{M}^{c.b}_{ug}\left(
\zeta_{c.b}, \sigma, a_{c.b}
\right).
\label{RF_L-CB_U-c}
\end{equation}
You can check this by a straightforward calculation similar to
\eqref{D-1-1}.

We will also see that ${\cal S}^{r.f}_{lg}$ becomes a 
solution of \eqref{tc} if ${\cal S}^{c.b}_{ug}$
is a solution of \eqref{selfboscri}.
This is already verified in section 3.6.2 of 
\cite{Jain:2013py} in the general case.
We can check it as follows:
\begin{equation}
\begin{split}
0 &= \int^{\pi}_{-\pi} d \alpha \rho_{ug}^{c.b}(\alpha) \left(
\log \left(
2 \sinh \left( 
\frac{\sigma - i \alpha}{2}
\right)
\right)
+ c.c
\right) 
\\
\Leftrightarrow
0 &=
\int^{0}_{2\pi} d \tilde{\alpha} 
\left( 
\frac{1}{2\pi \lambda_{r.f}}
- \rho_{lg}^{r.f}(\tilde{\alpha})
\right)
\left(
\log \left(
2 \cosh \left( 
\frac{\tilde{c} + i \tilde{\alpha}}{2}
\right)
\right)
+c.c
\right) 
\\
\Leftrightarrow 
\tilde{c} &=
\lambda_{r.f}
\int^{\pi}_{-\pi} d \alpha 
\rho_{r.f}^{lg}(\alpha)
\left(
\log \left(
2 \cosh \left( 
\frac{\tilde{c} + i \alpha }{2}
\right)
\right)
+
c.c
\right) .
\end{split}
\label{tc-self}
\end{equation}
Here we have used 
\begin{equation}
\int^{2\pi}_{0} d\alpha~\left(\log(1 + 2 \cos \alpha e^{-\tilde{c}}
+ e^{-2\tilde{c}}
\right) = 0 \qquad (\tilde{c} > 0)
\end{equation}
confirmed by the Taylor expansion and 
$\int^{2\pi}_{0} d\alpha e^{i\alpha} =0$.

By the above discussion, we can 
conclude that ${\cal S}^{r.f}_{lg}$ becomes a solution 
provided by 
the three equations \eqref{eigen-RF-low}, \eqref{Cond-low-RF} 
and \eqref{tc} if ${\cal S}^{c.b}_{ug}$ is a solution 
provided by the equations \eqref{eigen-CB-up}, \eqref{Cond-up-CB},
and \eqref{selfboscri}.
Thus we have verified the 
duality relationships between the lower gap phase of the 
regular fermion theory and the upper gap phase of the critical boson theory.

\subsubsection{Between upper gap phase of regular fermion and 
lower gap phase of critical boson}
\label{RF-U-CB_L}

Let us define sets of quantities ${\cal S}^{c.b}_{lg}, {\cal S}^{r.f}_{ug}$ 
as
\begin{equation}
\begin{split}
{\cal S}^{c.b}_{lg} =& (\lambda_{c.b}, \zeta_{c.b};
b_{c.b}, \sigma, \rho_{lg}^{c.b}(\alpha)), \\
{\cal S}^{r.f}_{ug} = &
(\lambda_{r.f}, \zeta_{r.f};
a_{r.f}, \tilde{c}, \rho_{ug}^{r.f}(\alpha))
\\
=&\left(1-\lambda_{c.b}, \frac{\lambda_{c.b}}{1-\lambda_{c.b}}\zeta_{c.b};
\pi-b_{c.b}, \sigma, 
\frac{\lambda_{c.b}}{1-\lambda_{c.b}}
\left(
\frac{1}{2\pi\lambda_{c.b}} - \rho^{c.b}_{lg}
( \pi+\alpha)\right)\right).
\end{split}
\end{equation}
By similar reasoning to that in section \ref{RF-L-CB_U},
we can verify the duality relationships by 
checking that ${\cal S}_{ug}^{r.f}$ becomes a solution provided by the 
three equations 
\eqref{eigen-RF-up}
\eqref{Cond-up-RF} and \eqref{tc}
in the regular fermion theory 
if ${\cal S}_{lg}^{c.b}$ is a solution provided by the three equations
\eqref{eigen-CB-low},\eqref{Cond-low-CB} and \eqref{selfboscri}
in the critical boson side.

We will confirm the relationship \eqref{corres-1}
between the eigenvalue densities
\eqref{eigen-CB-low}
and \eqref{eigen-RF-up} under the correspondence 
\eqref{corres-0},\eqref{corres-2}.
By a similar calculation to \eqref{D-1-1},
we can show the relationship 
\eqref{corres-1}
as
\begin{equation}
\begin{split}
&\rho^{r.f}_{ug}
\left(\frac{\lambda_{c.b}}{1-\lambda_{c.b}}\zeta_{c.b},1-\lambda_{c.b};
\sigma,\pi-b_{c.b}; \alpha\right)
\\
=&  \frac{\lambda_{c.b}}{1-\lambda_{c.b}}
\left(
\frac{1}{2\pi\lambda_{c.b}} - \rho^{c.b}_{lg}
\left(\zeta_{c.b},\lambda_{c.b};
\sigma,b_{c.b}; \pi+\alpha\right)\right).
\end{split}
\label{RF_U-CB_L-e}
\end{equation}

By checking 
\begin{equation}
\tilde{M}^{r.f}_{ug}\left(
\frac{\lambda_{c.b}}{1-\lambda_{c.b}}\zeta_{c.b}, \sigma, \pi-b_{c.b}
\right) = -\frac{\lambda_{c.b}}{1-\lambda_{c.b}}
\tilde{M}^{c.b}_{lg}\left(
\zeta_{c.b}, \sigma, b_{c.b}
\right)
\label{RF_U-CB_L-c}
\end{equation}
in a similar way to \eqref{RF_L-CB_U-c},
we can also see that ${\cal S}_{ug}^{r.f}$ becomes a solution of 
\eqref{Cond-up-RF} if ${\cal S}_{lg}^{c.b}$ is a solution of 
\eqref{Cond-low-CB}.

By repeating the analogous calculation to \eqref{tc-self},
we can see that ${\cal S}_{ug}^{r.f}$ becomes a solution of 
\eqref{tc} if ${\cal S}_{lg}^{c.b}$ is a solution of 
\eqref{selfboscri}.

So by the above discussion we can 
conclude that ${\cal S}^{r.f}_{ug}$ becomes a solution 
of the three equations \eqref{eigen-RF-up}, \eqref{Cond-up-RF} 
and \eqref{tc} if ${\cal S}^{c.b}_{lg}$ is a solution 
of the equations \eqref{eigen-CB-low}, \eqref{Cond-low-CB},
and \eqref{selfboscri}.
Thus we have verified the 
duality relationships between the upper gap phase of the 
regular fermion theory and the lower gap phase of the critical boson theory.

\subsubsection{Between two gap phases of regular fermion and 
critical boson theory}
\label{RF-CB_2}
Let us define sets of quantities ${\cal S}^{c.b}_{tg}, {\cal S}^{r.f}_{tg}$ 
as
\begin{equation}
\begin{split}
{\cal S}^{c.b}_{tg} =& (\lambda_{c.b}, \zeta_{c.b};
a_{c.b},b_{c.b}, \sigma, \rho_{tg}^{c.b}(\alpha)), \\
{\cal S}^{r.f}_{tg} = &
(\lambda_{r.f}, \zeta_{r.f};
a_{r.f}, b_{r.f},\tilde{c}, \rho_{tg}^{r.f}(\alpha))
\\
=&\left(1-\lambda_{c.b}, \frac{\lambda_{c.b}}{1-\lambda_{c.b}}\zeta_{c.b};
\pi-b_{c.b}, \pi-a_{c.b}, \sigma, 
\frac{\lambda_{c.b}}{1-\lambda_{c.b}}
\left(
\frac{1}{2\pi\lambda_{c.b}} - \rho^{c.b}_{tg}
( \pi+\alpha)\right)\right).
\end{split}
\end{equation}
By similar reasoning to that in section \ref{RF-L-CB_U},
we can verify the duality relationships by 
checking that ${\cal S}_{tg}^{r.f}$ becomes a solution provided by the 
four equations 
\eqref{eigen-RF-two},
\eqref{RF two u1},\eqref{RF two u0} and \eqref{tc}
in the regular fermion theory 
if ${\cal S}_{tg}^{c.b}$ is a solution provided by the four equations
\eqref{eigen-CB-two},\eqref{CB two u1}, \eqref{CB two u0} 
and \eqref{selfboscri}
in the critical boson side.

\paragraph{Correspondence \eqref{corres-1} between 
the eigenvalue density functions}
First we will confirm the relationship \eqref{corres-1}
between the eigenvalue densities
\eqref{eigen-CB-two}
and \eqref{eigen-RF-two} under the correspondence 
\eqref{corres-0},\eqref{corres-2}.
We can directly check 
\begin{equation}
\rho^{r.f}_{1,tg}\left(
\frac{\lambda_{c.b}}{1-\lambda_{c.b}}\zeta_{c.b},
\pi-b_{c.b}, \pi-a_{c.b}, \sigma; \alpha
\right)
= -\frac{\lambda_{c.b}}{1-\lambda_{c.b}}
\rho^{c.b}_{1,tg}
\left(
\zeta_{c.b},a_{c.b},b_{c.b}, \sigma; \pi + \alpha
\right)
\label{rho-D-1}
\end{equation}
in a similar calculation to \eqref{D-1-1}.
From \eqref{Cont-rep}, we can see 
\footnote{We can see it from
\begin{equation}
\begin{split}
I_{1}(\pi-b,\pi-a;\pi+\alpha)
=& -2\int_{b}^{\pi} \frac{d\theta}{(\cos\theta- \cos\alpha)\sqrt{\left(\sin^2\frac{\theta}{2}-\sin^2\frac{a}{2}\right)\left(\sin^2\frac{\theta}{2}-\sin^2\frac{b}{2}\right)}},
\\
\cF(\pi-b,\pi-a;\pi+\alpha) =& \cF (a,b,\alpha).
\end{split}
\label{a-b filp}
\end{equation}
These provide the complex line integration along the lower gap.
}
\begin{equation}
\rho_{2,tg}\left(1-\lambda_{c.b}, 
a_{r.f}, b_{r.f}; \alpha\right)
= \frac{ih^{+}(u)}{4\pi^2(1-\lambda_{c.b})}
\int_{L_{ugs}^{r.f}} \frac{d\omega}{h(\omega)}\left(
\frac{(\omega+u)}{u(\omega-u)}
\right),
\end{equation}
\begin{equation}
\rho_{2,tg}\left(\lambda_{c.b}, 
\pi-b_{r.f}, \pi-a_{r.f}; \pi+\alpha\right)
= \frac{ih^{+}(u)}{4\pi^2\lambda_{c.b}}
\int_{L_{lgs}^{r.f}} \frac{d\omega}{h(\omega)}\left(
\frac{(\omega+u)}{u(\omega-u)}
\right).
\end{equation}
Here $L_{ugs}^{r.f}$ denotes the complex line integration along the 
upper gap running counterclockwise from $e^{-ia_{r.f}}$ to $e^{ia_{r.f}}$.
On the other hand $L_{lgs}^{r.f}$
denotes the complex line integration along the 
lower gap running counterclockwise from $e^{ib_{r.f}}$ to $e^{-ib_{r.f}}$.
By using the formula
proved in appendix \ref{omega-u},
\begin{equation}
\int_{L^{r.f}_{ugs}} 
\frac{d\omega}{h(\omega)}\frac{\omega+u}{u(\omega - u)}
+\int_{L^{r.f}_{lgs}} 
\frac{d\omega}{h(\omega)}\frac{\omega+u}{u(\omega - u)}
= \frac{-2\pi i}{h^{+}(u)},
\label{Formula two 1}
\end{equation}
we can see that
\begin{equation}
\begin{split}
&\rho_{2,tg}\left(1-\lambda_{c.b}, 
\pi-b_{c.b}, \pi-a_{c.b}; \alpha\right)\\
=& \frac{1}{2\pi(1-\lambda_{c.b})}
-\frac{\lambda_{c.b}}{1-\lambda_{c.b}}\rho_{2,tg}\left(\lambda_{c.b}, 
a_{c.b}, b_{c.b}; \pi+\alpha\right)
\end{split}
\label{rho-D-2}
\end{equation}
where we use $a_{r.f} = \pi-b_{c.b}, b_{r.f} = \pi-a_{c.b}$.
From 
\eqref{rho-D-1}
and \eqref{rho-D-2},
we can confirm the relationship 
\eqref{corres-1}
between the eigenvalue densities
\eqref{eigen-CB-two}
and \eqref{eigen-RF-two},
\begin{equation}
\rho^{r.f}_{tg} (\alpha) = 
\frac{\lambda_{c.b}}{1-\lambda_{c.b}}\left(
\frac{1}{2 \pi \lambda_{c.b}} - \rho^{c.b}_{tg}(\alpha + \pi) 
\right).
\end{equation}


\paragraph{Relationship between \eqref{CB two u1} and 
\eqref{RF two u1}}
We will check that 
${\cal S}^{r.f}_{tg}$ becomes a 
solution of \eqref{RF two u1}
if ${\cal S}^{c.b}_{tg}$ is a solution of 
\eqref{CB two u1}.
We can easily check
\begin{equation}
{\cal Y}^{r.f}(\pi-b_{c.b}, \pi-a_{c.b}, \sigma)
={\cal Y}^{c.b}(a_{c.b}, b_{c.b}, \sigma).
\label{D-3-00}
\end{equation}
From \eqref{Cont-rep}, 
by a similar calculation to \eqref{a-b filp},
we can see 
\begin{equation}
\Upsilon(a_{c.b}, b_{c.b}) = -4i\int_{L^{c.b}_{ugs}} 
\frac{d\omega}{h(\omega)}, \qquad
\Upsilon(\pi-b_{c.b}, \pi-a_{c.b}) = 4i\int_{L^{c.b}_{lgs}} 
\frac{d\omega}{h(\omega)}.
\end{equation}
Here $L_{ugs}^{c.b}$ denotes the complex line integration along the 
upper gap running counterclockwise from $e^{-ia_{c.b}}$ to $e^{ia_{c.b}}$,
and $L_{lgs}^{c.b}$ denotes the one along the 
lower gap running counterclockwise from $e^{ib_{c.b}}$ to $e^{-ib_{c.b}}$.
From the formula proved in appendix \ref{homega-inv}, we can see
\begin{equation}
\int_{L^{c.b}_{ugs}} 
d\omega\, \frac{1}{
h(\omega)} 
 = -\int_{L^{c.b}_{lgs}} 
d\omega\, \frac{1}{h(\omega)}
\Rightarrow \Upsilon(a_{c.b}, b_{c.b}) = 
\Upsilon(\pi-b_{c.b}, \pi-a_{c.b}).
\label{Formula two 2}
\end{equation}
Hence by taking into account $\zeta_{c.b}\lambda_{c.b} 
= (1-\lambda_{c.b})\zeta_{r.f}$, 
we can conclude
\begin{equation}
\begin{split}
&\frac{1 }{4\pi \lambda_{c.b}  } \Upsilon(a_{c.b},b_{c.b})
= \frac{\zeta_{c.b}}{2\pi} 
{\cal Y}^{c.b}(a_{c.b},b_{c.b},\sigma)
\\
\Rightarrow&
\frac{1}{4\pi(1-\lambda_{c.b})}\Upsilon(\pi-b_{c.b},\pi-a_{c.b})
= 
\frac{\lambda_{c.b}}{1-\lambda_{c.b}}
\frac{\zeta_{c.b}}{2\pi}{\cal Y}^{r.f}(\pi-b_{c.b},\pi-a_{c.b},\sigma).
\end{split} 
\label{D-3-2}
\end{equation}
This \eqref{D-3-2}
shows that ${\cal S}^{r.f}_{tg}$ becomes a 
solution of \eqref{RF two u1}
if ${\cal S}^{c.b}_{tg}$ is a solution of 
\eqref{CB two u1}.


\paragraph{ Relationship between \eqref{CB two u0} and 
\eqref{RF two u0}}
We will verify that 
${\cal S}^{r.f}_{tg}$ becomes a 
solution of the (\ref{RF two u0})
if ${\cal S}^{c.b}_{tg}$ is a solution of 
(\ref{CB two u0}).
By a straightforward calculation, we can see
\begin{equation}
{\cal G}^{r.f}(\pi-b_{c.b},\pi-a_{c.b},\sigma) =
{\cal G}^{c.b}(a_{c.b},b_{c.b},\sigma) .
\label{D-3-01}
\end{equation}
From \eqref{Cont-rep}, 
by a similar calculation to \eqref{a-b filp},
we can see 
\begin{equation}
\Lambda(a_{c.b},b_{c.b}) = 
-4i \int_{L_{ugs}^{c.b}} d\omega \frac{\omega}{h(\omega)}, \qquad
\Lambda(\pi-b_{c.b},\pi-a_{c.b}) = 
-4i \int_{L_{lgs}^{c.b}} d\omega \frac{\omega}{h(\omega)}.
\label{D-3-02}
\end{equation}
By using the formula shown in appendix~\ref{homega-omega},
\begin{equation}
1 = \frac{1}{\pi i}\int_{L_{ugs}^{c.b}} d\omega~ 
\frac{\omega}{h(\omega)}
+\frac{1}{\pi i}\int_{L_{lgs}^{c.b}} d\omega~ 
\frac{\omega}{h(\omega)},
\label{Formula two 3}
\end{equation}
we can see
\begin{equation}
\begin{split}
&\frac{1}{4\pi\lambda_{c.b}}\Lambda(a_{c.b},b_{c.b}) = 1+ 
\frac{\zeta_{c.b}}{4 \pi}{\cal G}^{c.b}(a_{c.b},b_{c.b},\sigma)
\\
\Rightarrow&\frac{1}{4\pi(1-\lambda_{c.b})}\Lambda(\pi-b_{c.b},\pi-a_{c.b}) = 1- 
\frac{\lambda_{c.b}}{1-\lambda_{c.b}}\frac{\zeta_{c.b}}{4 \pi}
{\cal G}^{r.f}(\pi-b_{c.b},\pi-a_{c.b},\sigma).
\end{split}
\label{D-3-3}
\end{equation}
This \eqref{D-3-3}
shows that ${\cal S}^{r.f}_{tg}$ becomes a 
solution of the (\ref{RF two u0})
if ${\cal S}^{c.b}_{tg}$ is a solution of 
(\ref{CB two u0}).

\paragraph{Correspondence between \eqref{tc} and \eqref{selfboscri}}
By repeating the analogous calculation to \eqref{tc-self},
we can see that ${\cal S}_{tg}^{r.f}$ becomes a solution of  
\eqref{tc} if ${\cal S}_{tg}^{c.b}$ is a solution of  
\eqref{selfboscri}.

\paragraph{Summary of the proof}
So by the above discussion, we can 
conclude that ${\cal S}^{r.f}_{tg}$ becomes a solution provided by 
the four equations \eqref{eigen-RF-two},\eqref{RF two u1},\eqref{RF two u0} and \eqref{tc} in the regular fermion theory if ${\cal S}^{c.b}_{tg}$ is a solution provided by
the four equations \eqref{eigen-CB-two},\eqref{CB two u1}, \eqref{CB two u0} 
and \eqref{selfboscri} at the critical boson side.
Thus we have verified the 
duality relationships at the two gap phase between 
the regular fermion theory and the critical boson theory.

\subsubsection{Relationships between the phase transition points}
\label{Sec:PTD-RF-CB}
We have seen that the conditions for the phase transitions
\eqref{RF-no-low}, \eqref{RF-no-up}, \eqref{low-to-2 CB} etc,
can be obtained as a certain limit of 
corresponding equations between which 
we have already seen the duality relationships.
So we have already shown the duality relationship between the 
phase transition points, in fact.
For example, \eqref{low-to-2 CB} is obtained by 
$a_{c.b}\to 0$ limit of the equation \eqref{CB two u1} and 
\eqref{Up-to-2-RF} is obtained by $b_{r.f} \to \pi$
limit of 
\eqref{RF two u1}.
We have already shown that \eqref{RF two u1} maps to 
\eqref{CB two u1} under the duality, 
moreover we can see $a_{c.b} = \pi-b_{r.f} = \pi-\pi = 0$.
Hence the phase transition points between the upper gap to 
two gap in the regular fermion maps to the one between the lower gap
and the two gap phase in the critical boson theory.
Now 
we will write the phase transition points between the $A$ phase to $B$ phase
in the regular fermion theory as $(\text{A,B,RF})$, 
and the transition points in the 
critical boson theory as $(\text{A,B,CB})$.
We can see following dual relationships represented by 
these symbols 
\begin{equation}
\begin{split}
&(\text{No,Low,RF} ) \Leftrightarrow(\text{No,Up,CB} ), \quad
(\text{No,Up,RF} ) \Leftrightarrow(\text{No,Low,CB} ), 
\\
&(\text{Low,Two,RF} ) \Leftrightarrow(\text{Up,Two,CB} ), \quad
(\text{Up,Two,RF} ) \Leftrightarrow(\text{Low,Two,CB} ).
\label{PT-CB-RF}
\end{split}
\end{equation}

\subsubsection{Free energy and completing the proof of the duality}
Now we have proved that for any value of 
$(\lambda_{c.b}, \zeta_{c.b})$,
there are duality relationships
\eqref{corres-1}$\sim$\eqref{corres-0}
between the critical boson theory and the 
regular fermion theory.

Under these correspondences
\eqref{corres-1}$\sim$\eqref{corres-0},
we will show the free energy of the level $k$ $U(N)$  
critical boson theory at
$(\lambda_{c.b}, \zeta_{c.b})$ (see \eqref{F-CB})
agrees with the free energy of
the level $k$ $U(k-N)$ regular fermion theory 
at $(\lambda_{r.f}, \zeta_{r.f})
= (1-\lambda_{c.b}, 
\frac{\lambda_{c.b}}{1-\lambda_{c.b}}\zeta_{c.b} )$
(see \eqref{F-RF}),
\begin{equation}
F^{N}_{c.b} = V^{c.b}[\rho^{c.b}, N] + F_2[\rho^{c.b}, N]
= V^{r.f}[\rho^{r.f}, k-N] + F_2[\rho^{r.f}, k-N] = F^{k-N}_{r.f}
\end{equation}
where $\rho^{r.f}$ can be represented by $\rho^{c.b}$ 
through the relationship \eqref{corres-1}.

As already shown in (3.37) of \cite{Jain:2013py},
by a straightforward calculation similar to \eqref{tc-self},
we can show $V^{c.b}[\rho^{c.b}, N] = V^{r.f}[\rho^{r.f}, k-N]$.

Under \eqref{corres-1} ,
$F_{2}[\rho^{r.f},k-N]$ is expanded in terms of $\rho^{c.b}$ as
\begin{align}
F_2[\rho^{r.f},k-N] =&
- N^2 {\cal P}
\int^{\pi}_{-\pi}d \alpha\int^{\pi}_{-\pi} d \beta~
\rho^{c.b}(\pi+\alpha)\rho^{c.b}(\pi+\beta) 
\log \left|2 \sin \frac{\alpha-\beta}{2}\right|
\nonumber \\
&+2 N^2 \frac{1}{2\pi\lambda_{c.b}}
{\cal P}\int^{\pi}_{-\pi}d \alpha\int^{\pi}_{-\pi} d \beta~
\rho^{c.b}(\pi+\beta) \log 
\left|2 \sin \frac{\alpha-\beta}{2}\right|
\nonumber \\
&- N^2 \left(\frac{1}{2\pi\lambda_{c.b}}\right)^2
{\cal P}\int^{\pi}_{-\pi}d \alpha\int^{\pi}_{-\pi} d \beta~
\log \left|2 \sin \frac{\alpha-\beta}{2}\right| .
\label{F-D-1}
\end{align}
By using the periodicity of $\rho^{c.b}(\alpha) = \rho^{c.b}(2\pi + \alpha)$ and by using
\begin{equation}
{\cal P}\int^{\pi}_{-\pi}
d\alpha~ 
\log \left|2 \sin \frac{\alpha-\beta}{2}\right| = 0,
\label{F-D-2}
\end{equation}
we can check 
\begin{align}
F_{2}[\rho^{r.f},k-N] 
=F_{2}[\rho^{c.b},N] .
\end{align}
So we can see
\begin{equation}
F^{N}_{c.b} = F^{k-N}_{r.f},
\end{equation}
under the relationships \eqref{corres-1}$\sim$\eqref{corres-0}.
Now we have completed the proof of the level-rank duality
between the level $k$ $U(N)$ critical boson theory and 
the level $k$ $U(k-N)$ regular fermion theory.

\section{Higher temperature phases of the Supersymmetric Chern-Simons matter 
theory}
Next we will study the higher temperature phases of the
${\cal N} = 2$ supersymmetric level $k$ $U(N)$ Chern-Simons matter theory
with a single fundamental chiral multiplet.
From here we will call the theory ``the SUSY CS matter theory" or 
``the SUSY theory". 
The Lagrangian of the theory we study is presented in 
equations (2.1) and (3.58) of \cite{Jain:2012qi} (see also (6.1) and (6.20) of \cite{Aharony:2012ns}). 
From (3.21) of \cite{Jain:2013py}, the effective potential $V(U)$ in this theory is obtained as
 \beal{\label{vsusy}
  { V(U) } 
=& -\frac{N^2 \zeta }{6 \pi |\lambda|}\left(\tilde c^3 - 6 |\lambda|  \int_{-\pi}^{\pi} d\alpha \rho(\alpha) \text{Re} \int_{\tilde c} ^\infty dy y \log \tanh {y+i\alpha \over 2}\right)\nonumber
\\
\equiv& V^{su}[\rho,N],
}
where the thermal mass (for both the boson as well as the fermion in the 
supermultiplet) is denoted by $\tilde c T$. The constant 
 $\tilde c$ above is determined by the requirement that it extremizes $V(U)$;
i.e. 
\begin{equation} \label{susyct}
\tilde c=2  \biggl | \text{Re} \int_{-\pi}^{\pi} d\alpha \lambda \rho(\alpha) \log \coth {\tilde c + i \alpha \over 2} \biggl |. 
\end{equation}
The effective potential can be interpreted as the sum of the 
regular fermion potential \eqref{freefer} and the 
critical boson potential \eqref{vcritb}.
So the potential can be written as
\begin{equation}
V(U) = V_{r.f}(U) + V_{c.b}(U)
\end{equation}
where
\begin{equation}
 V_{r.f}(U)=-\frac{N^2 \zeta}{6\pi}\left(\frac{\tilde c^3}{\lambda}-\tilde c^3
+3 \int_{-\pi}^{\pi} d \alpha \rho(\alpha) \int_{\tilde c}^{\infty}dy ~ y
(\ln(1+e^{-y-i\alpha})+\ln(1+e^{-y+i\alpha}))\right), 
\label{pot-rf-SUSY}
\end{equation}
\begin{equation}
 V_{c.b}(U)
=-\frac{N^2 \zeta}{6\pi}\tilde{c}^3  
+ \frac{N^2 \zeta}{2\pi}\int_{\tilde{c}}^{\infty} dy\int_{-\pi}^{\pi} d\alpha~y \rho(\alpha) \left(\ln(1-e^{-y+i\alpha})+\ln(1-e^{-y-i\alpha})\right). 
\label{pot-cb-SUSY}
\end{equation}
Hence it is convenient to utilize and combine the results 
in the previous section
to obtain the eigenvalue densities and the free energy in the 
${\cal N} =2$ supersymmetric case.
The general form of the free energy of the 
supersymmetric CS matter theory on $S^2 \times S^1$ is 
given as
\begin{equation}
F_{su}^{N} = V^{su}[\rho,N] +F_2[\rho,N].
\label{F-SUSY}
\end{equation}
As in the other CS matter theories,
to obtain the free energy 
in each phase, we
only have to evaluate the eigenvalue density $\rho$ 
in each phase 
and substitute them 
into \eqref{F-SUSY} and \eqref{susyct}.
So obtaining the eigenvalue densities in each phase 
is equivalent to obtaining the 
free energies in each phase.

For later use we will give the form of $V'(z)$ here.
We obtain $V'(z) = V'_{r.f}(z) + V'_{c.b}(z)$ from 
\eqref{pot-rf-SUSY} and
\eqref{pot-cb-SUSY} as
\begin{equation}
\begin{split}
V'_{r.f}(z) =&  -\frac{N \zeta}{2\pi}
\int_{\tilde c}^{\infty}dy ~ y
\left(
\frac{-ie^{-y}}{z+e^{-y}}
+\frac{ie^{-y}}{z^{-1} + e^{-y}}
\right),
\\
V'_{c.b}(z) =& \frac{N \zeta}{2\pi}
\int_{\tilde{c}}^{\infty}dy ~ y
\left(
\frac{-ie^{-y}}{z^{-1}-e^{-y}}
+\frac{ie^{-y}}{z - e^{-y}}
\right).
\end{split}
\label{pot-der}
\end{equation}

\begin{figure}
  \begin{center}
  \subfigure[]{\includegraphics[scale=.33]{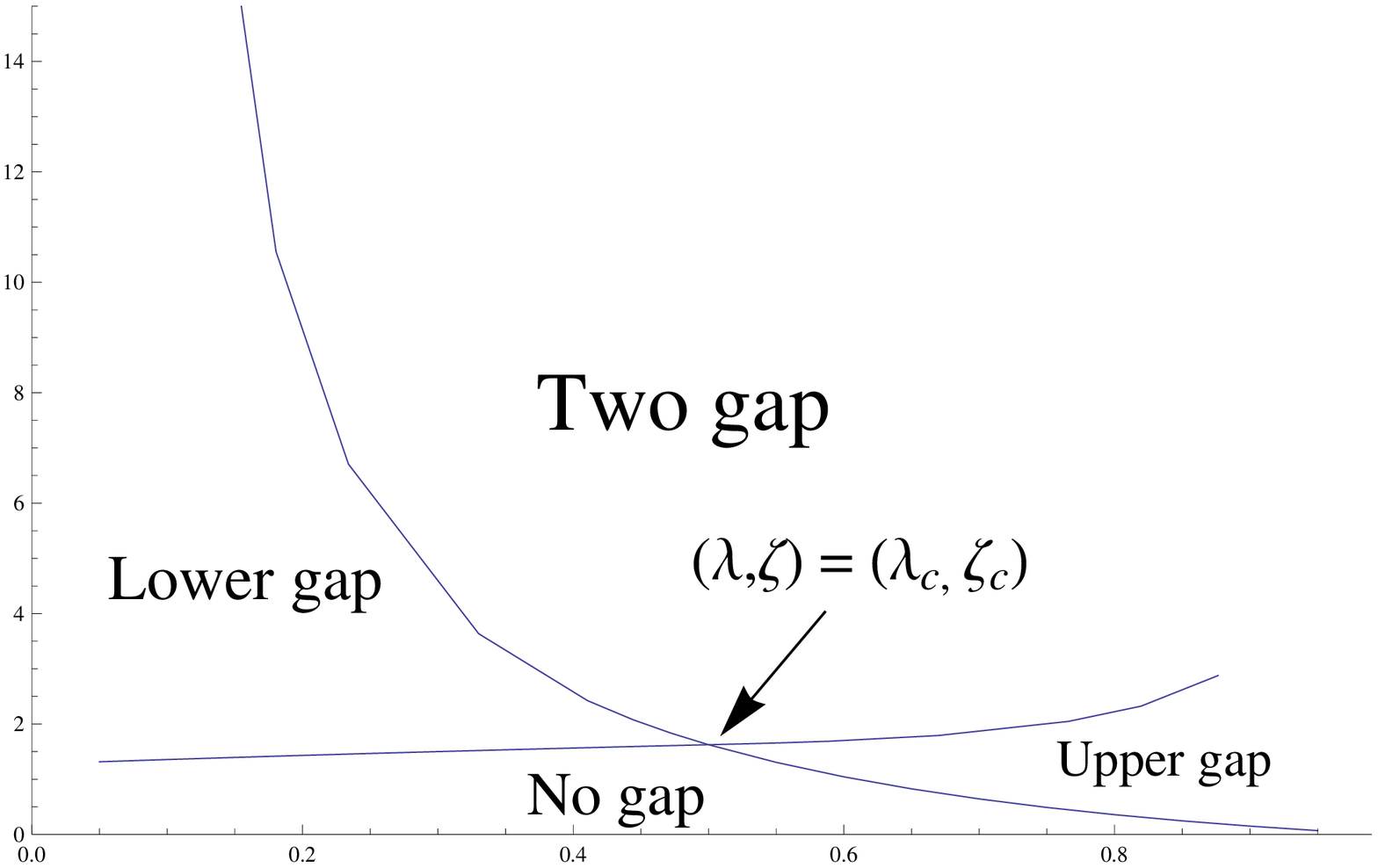}
\label{fig:phase-SUSY}}
\caption{Phase diagram of the SUSY CS matter theory.
Here $(\lambda_c,\zeta_c) = (0.5, 1.62509)$
is the quadruple phase transition point where 
the four phases (no gap, lower gap, upper gap, two gap phases) coexist.}
  \end{center}
\end{figure}

The phase structure of this theory is 
depicted in Fig.~\ref{fig:phase-SUSY}, 
as obtained in following subsubsections.

\subsection{Lower gap phase}
\label{L-SUSY}
In the lower gap phase of the SUSY CS matter theory, 
we employ the same procedure as the one in section \ref{L-RF}
to obtain the eigenvalue density $\rho$.
We use the cut region and the cut function 
defined in appendix \ref{Sec:Low-Cut}.
Because $\rho_{0} = 0$, $U(z)$ becomes the same as $V'(z)$,
and we can immediately obtain 
$H(u)$ as well as $\Phi(u) = h(u) H(u)$
by substituting the above \eqref{pot-der} into \eqref{Hucp}.
Since \eqref{Hucp} is linear in $V'(u) = U(u)$,
the $\Phi(u)$ is obtained as the sum of 
\eqref{Phi-RF-L} and \eqref{Phi-CB-L} with $\sigma = \tilde{c}$.
So $\Phi(u)$ is obtained as 
\begin{equation}
\Phi(u) = 
\Phi^{r.f}_{lg}(\tilde{c},b,\zeta;u) + 
\Phi^{c.b}_{lg}(\tilde{c},b,\zeta;u) \equiv 
\Phi^{su}_{lg}(\tilde{c},b,\zeta;u). 
\end{equation}

From $\Phi^{+}(u) - \Phi^-(u) = 4 \pi \rho(u)$
at the cut $u = e^{i\alpha}$ with $-b \le \alpha \le b$,
we obtain the eigenvalue density in the lower gap phase as
\begin{equation}
\rho(\alpha) =
\rho^{r.f}_{lg}(\zeta,\lambda;\tilde{c},b;\alpha) 
+\rho^{c.b}_{lg}(\zeta,\lambda;\tilde{c},b;\alpha) 
\equiv \rho^{su}_{lg}(\zeta,\lambda;\tilde{c},b;\alpha) ,
\label{eigen-SUSY-low}
\end{equation}
and $\rho(\alpha) = 0$ for $\pi >|\alpha| > b$.
This is the sum of \eqref{eigen-RF-low} and 
\eqref{eigen-CB-low}.
By substituting the eigenvalue density \eqref{eigen-SUSY-low} into
\eqref{susyct} and \eqref{F-SUSY}, we can obtain the free energy in the 
lower gap phase. 

The condition $\lim_{u \to \infty} \Phi(u) = 1$ requires following
equation
\begin{eqnarray}
\tilde{M}^{su}_{lg}(\zeta,\tilde{c},b) \equiv
\tilde{M}^{r.f}_{lg}(\zeta,\tilde{c},b) 
+ \tilde{M}^{c.b}_{lg}(\zeta,\tilde{c},b) =1,
\label{Cond-low-SUSY}
\end{eqnarray}
where $\tilde{M}^{r.f}_{lg}$ and $\tilde{M}^{c.b}_{lg}$
are defined in \eqref{Cond-low-RF} and \eqref{Cond-low-CB}.

By \eqref{Cond-low-SUSY} and \eqref{susyct}, we can obtain 
$(b,\tilde{c})$ as functions of $(\lambda, \zeta)$ as
$(b,\tilde{c}) = (b(\lambda,\zeta),\tilde{c}(\lambda,\zeta))$.
\eqref{eigen-SUSY-low}, \eqref{Cond-low-SUSY} and \eqref{susyct}
provide a set of complete solutions in the lower gap phase of the 
SUSY theory.


\subsection{Upper gap phase}
\label{UP-SUSY}
To look for the upper gap solution in the current 
supersymmetric CS matter theory,
we only have to follow the same procedure employed in 
sections 
\ref{UP-RF} and \ref{UP-CB}.
The upper cut region as well as the cut function are described 
in appendix~\ref{Sec:Up-Cut}.
The function $H(u)$ as well as $\Phi(u)$ are obtained by using 
\eqref{Hucp}, and the function $\Phi(u)$ here is evaluated as 
\begin{equation}
\Phi(u) = 
\Phi_{ug}^{\rho_{0}}(\lambda, a; u) + \Phi^{r.f}_{ug}(\tilde{c}, a, \zeta;u) 
+\Phi^{c.b}_{ug}(\tilde{c}, a, \zeta;u) .
\label{Phi-SUSY-up}
\end{equation}
Here $\Phi^{r.f}_{ug}(\tilde{c}, a, \zeta;u)$ and 
$\Phi_{ug}^{\rho_{0}}(\lambda, a ;u)$ are defined in 
\eqref{Phi-RF-up}, 
$\Phi^{c.b}_{ug}(\tilde{c}, a, \zeta;u)$ is given 
in \eqref{Phi-CB-U}.
From \eqref{Phi-SUSY-up}, 
by taking $\Phi^{+}(u)- \Phi^{-}(u) = 4\pi(\rho(u)- \rho_{0}(u))$
at the cut region $u = e^{i\alpha}$ with 
$\pi \ge |\alpha| \ge a$,
we obtain the eigenvalue density function as
\begin{eqnarray}
\rho(\alpha)&=& \rho^{r.f}_{ug}(\zeta,\lambda;\tilde{c},a;\alpha) + \rho^{c.b}_{ug}(\zeta,\lambda;\tilde{c},a;\alpha) - \frac{1}{2\pi\lambda}
\equiv \rho_{ug}^{su}(\zeta,\lambda,\tilde{c},a;\alpha),
\label{eigen-SUSY-up}
\end{eqnarray}
where $\rho^{r.f}_{ug}$ and $\rho^{c.b}_{ug}$ are 
defined in \eqref{eigen-RF-up} and \eqref{eigen-CB-up} respectively.
In the region $|\alpha| < |a|$, $\rho(\alpha) = \frac{1}{2\pi\lambda}$. 
By substituting the eigenvalue density \eqref{eigen-SUSY-up} into
\eqref{susyct} and \eqref{F-SUSY}, we can obtain the free energy in the 
upper gap phase. 

From the condition, $\lim_{u \to \infty}\Phi(u) 
= 1- \int^{\pi}_{-\pi} d\alpha \rho_{0}(\alpha)$,
we obtain the equation
\begin{eqnarray}
1- \frac{1}{\lambda}
&=& 
\tilde{M}^{r.f}_{ug}(\zeta,\tilde{c},a)
+\tilde{M}^{c.b}_{ug}(\zeta,\tilde{c},a)
\equiv \tilde{M}^{su}_{ug}(\zeta,\tilde{c},a)
\label{Cond-up-SUSY}
\end{eqnarray}
where $\tilde{M}^{r.f}_{ug}(\zeta,\tilde{c},a)$ and
$\tilde{M}^{c.b}_{ug}(\zeta,\tilde{c},a)$ are defined in 
\eqref{Cond-up-RF} and \eqref{Cond-up-CB} respectively.

By \eqref{Cond-up-SUSY} and \eqref{susyct},
we obtain $(a, \tilde{c})$ as functions of $(\lambda, \zeta)$ as
$(a,\tilde{c}) = (a(\lambda,\zeta), \tilde{c}(\lambda, \zeta))$.
\eqref{eigen-SUSY-up}, \eqref{Cond-up-SUSY} and \eqref{susyct} provide
a complete set of solutions in the upper gap phase in the ${\cal N} = 2$ 
supersymmetric CS matter theory.

\subsection{Two gap phase}
\label{sec:SUSY-2}
We will search for a two gap phase solution. 
We only have to follow the same procedure as 
the one in section \ref{sec:RF-2} and \ref{sec:CB-2}.
We will take the two cuts region and the cut function 
defined in appendix \ref{Sec:Two-Cut}.
By the same procedure as section \ref{sec:RF-2}, 
we obtain the functions $H(u)$ and 
$\Phi(u)$ as 
\begin{equation}
\Phi(u) = 
\Phi^{r.f}_{tg}(\zeta, a,b,\tilde{c}; u) + 
\Phi^{c.b}_{tg}(\zeta, a,b,\tilde{c}; u) + 
\Phi^{\rho_{0}}_{tg}(\lambda, a,b;u). 
\label{Two-SUSY-Phi}
\end{equation}
Here $\Phi^{r.f}_{tg}(\zeta, a,b,\tilde{c}; u)$ 
and 
$\Phi^{\rho_{0}}_{tg}(\lambda, a,b;u)$ are
defined in \eqref{Two-RF-Phi}, and 
$\Phi^{c.b}_{tg}(\zeta, a,b,\tilde{c}; u)$ 
is given in \eqref{Two-CB-Phi}.

From this \eqref{Two-SUSY-Phi}, 
by taking  $\Phi^{+}(u)- \Phi^{-}(u) = 4\pi(\rho(u)- \rho_{0}(u))$ 
at the cuts $|a| < |\alpha| < |b|$, we obtain the eigenvalue density function
as 
\begin{equation}
\begin{split}
\rho(\alpha) =& \rho^{su}_{tg}(\alpha) = 
\rho_{2,tg}(\lambda, a,b;\alpha)
+\rho_{1,tg}^{r.f}(\zeta, a,b,\tilde{c};\alpha)
+\rho_{1,tg}^{c.b}(\zeta, a,b,\tilde{c};\alpha)
\\
\equiv&
\rho_{tg}^{su}(\zeta,\lambda; a,b,\tilde{c};\alpha).
\end{split}
\label{eigen-SUSY-two}
\end{equation}
At $b \le |\alpha| \le \pi$, $\rho(\alpha) =0$ and
at $|\alpha| \le a$, $\rho(\alpha) = \frac{1}{2\pi\lambda}$.
$\rho_{2,tg}(\lambda, a,b;\alpha)$
and $\rho_{1,tg}^{r.f}(\zeta, a,b,\tilde{c};\alpha)$ are defined 
in \eqref{eigen-RF-two} and $\rho_{1,tg}^{c.b}(\zeta, a,b,\tilde{c};\alpha)$
is defined in \eqref{eigen-CB-two}.
Similar to \eqref{eigen-RF-two} and \eqref{eigen-CB-two},
\eqref{eigen-SUSY-two} can be interpreted as the sum of 
GWW type eigenvalue density $\rho_{2,tg}$ and 
other additional terms depending on the details of the theory.
By substituting the eigenvalue density \eqref{eigen-SUSY-two} into
\eqref{susyct} and \eqref{F-SUSY}, we can obtain the free energy in the 
two gap phase of the ${\cal N} = 2$ supersymmetric CS matter theory. 

From $\lim_{u \to \infty} \Phi(u) = 1
- \int^{\pi}_{-\pi} d \alpha~ \rho_{0}(\alpha)$, 
we obtain following two conditions,
\begin{equation}
\begin{split}
\frac{1 }{4\pi \lambda  } \Upsilon(a,b)
 =& \frac{\zeta}{2\pi} 
{\cal Y}^{su}(a,b,\tilde{c})
\\
{\cal Y}^{su}(a,b,\tilde{c})
\equiv& {\cal Y}^{r.f}(a,b,\tilde{c}) +
{\cal Y}^{c.b}(a,b,\tilde{c}),
\end{split}
\label{SUSY two u1}
\end{equation}
and 
\begin{equation}
\begin{split}
\frac{1}{4\pi\lambda}\Lambda(a,b) =& 1+ 
\frac{\zeta}{4 \pi}
{\cal G}^{su}(a,b,\tilde{c})
\\
{\cal G}^{su}(a,b,\tilde{c})
\equiv& {\cal G}^{c.b}(a,b,\tilde{c})
-{\cal G}^{r.f}(a,b,\tilde{c}).
\end{split}
\label{SUSY two u0}
\end{equation}
Here $\Upsilon(a,b)$ and ${\cal Y}^{r.f}$ are defined in 
\eqref{RF two u1}, and ${\cal Y}^{c.b}$ is in 
\eqref{CB two u1}.
$\Lambda(a,b)$ and ${\cal G}^{r.f}$ are defined in 
\eqref{RF two u0} and ${\cal G}^{c.b}$ is in 
\eqref{CB two u0}.

By using 
\eqref{SUSY two u0},
\eqref{SUSY two u1} and 
\eqref{susyct}, $(a,b, \tilde{c})$ are determined as 
functions of $(\lambda, \zeta)$ as
$(a,b, \tilde{c}) = (a(\lambda, \zeta),b(\lambda, \zeta), 
\tilde{c}(\lambda, \zeta))$.
The combination of \eqref{SUSY two u0},
\eqref{SUSY two u1}, \eqref{susyct} and \eqref{eigen-SUSY-two}
provide a complete set of solutions
in the two gap phase of the ${\cal N}=2$ SUSY CS matter theory.

At the large $\zeta$ limit, as we prove in 
appendix \ref{SUSY-Z-Pr}, 
the eigenvalue density approaches the 
universal distribution \eqref{evd 0} 
because $\tilde{c}$ remains finite positive quantity at the limit.  
In the limit, $a$,$b$ and the eigenvalue 
density behave as
\begin{equation}
\begin{split}
a =& 
\pi \lambda - \frac{\epsilon}{2},
\qquad 
b = \pi \lambda + \frac{\epsilon}{2},
\\
\epsilon =&
8 
\sin (\pi \lambda)
\exp\left(
- \frac{\sin (\pi \lambda)}{2}
\zeta \lambda
\int^{\infty}_{\tilde{c}}
\frac{d y~ 2y \cosh y}{\cosh^2 y - \cos^2 \pi\lambda}
\right)+\ldots,
\\
\rho(\alpha) =&  \frac{1}{\pi^2 \lambda} \cos^{-1} \sqrt{
\frac{\alpha - a}{b - a}
}.
\end{split}
\label{Large z SUSY}
\end{equation}
In appendix \ref{SUSY-LZ}, we have demonstrated the above behavior
\eqref{Large z SUSY}.
The eigenvalue density function 
\eqref{eigen-SUSY-two} is dominated by $\rho_{2,tg}$, 
and 
approaches $\cos^{-1}\sqrt{\alpha_1}$, 
which is the same functional form as (7.11) in \cite{Jain:2013py}.
The Eq.~(7.11) 
is the large $\zeta$ limit of the eigenvalue density of the GWW model.
We can also see that the range of the domain of the cut $\epsilon$ is
smaller than the one in the regular fermion theory as well as the 
one in the critical boson theory at the same values of
$(\lambda, \zeta)$.


\subsection{Phase transition points}
\paragraph{Phase transition from the lower gap to the no gap}
Let us consider the behavior of lower gap equations
\eqref{Cond-low-SUSY} and \eqref{eigen-SUSY-low}
at $b = \pi$ which would correspond to the
phase transition points from the lower gap to the no gap.
If we substitute $b = \pi$ into \eqref{Cond-low-SUSY},
it becomes
\begin{equation}
1 = -2\frac{\zeta}{\pi}
\int_{0}^{e^{-\tilde{c}}}dx ~  
\left(
\frac{ \log x}{1-x^2}
\right)
= 
\frac{2 \zeta}{\pi}
\sum_{n = 0}^{\infty}
\left( \frac{1 + \tilde{c}(2n +1)}{(2n+1)^2} e^{-(2n+1)\tilde{c}}
\right).
\label{SUSY-no-low}
\end{equation}
This is exactly same as the 
condition for the phase transition 
from the no gap to the lower gap phase 
which is obtained by substituting $\alpha = \pi$ into 
eq.~(6.15) of \cite{Jain:2013py} and by requiring $\rho(\pi) = 0$.
By using the condition
\eqref{SUSY-no-low},
if we substitute 
$b = \pi$ into 
\eqref{eigen-SUSY-low}, we obtain
\begin{align}
\rho_{lg}^{su}(\zeta,\lambda,\tilde{c},\pi;\alpha)
=& \frac{1}{2\pi}
+\frac{\zeta}{\pi^2}\sum_{n=0}^{\infty} 
\cos(2n+1)\alpha\left(
\frac{1 + \tilde{c} (2n+1)}{(2n+1)^2}e^{-(2n+1)\tilde{c}}
\right).
\label{eigen-SUSY-low-no}
\end{align}
This is the same as the eigenvalue density function 
at the no gap phase of the supersymmetric CS matter theory,
which is described by (6.15) of \cite{Jain:2013py}.
Also this can be written as the sum of 
\eqref{eigen-RF-low-no} and \eqref{eigen-CB-low-no}.
So the lower gap solutions are smoothly 
connected to the ones in the no gap phase at the phase transition points. 

\paragraph{Phase transition from the upper gap to the no gap}
Let us consider the behavior of upper gap solutions \eqref{Cond-up-SUSY} and \eqref{eigen-SUSY-up} 
at $a = 0$ which would correspond to the phase transition points 
from the upper gap to the no gap phase.
If we substitute $a = 0$ into \eqref{Cond-up-SUSY},  it becomes
\begin{equation}
1- \frac{1}{\lambda}=
\frac{2\zeta}{\pi}
\int_{0}^{e^{-\tilde{c}}}dx ~  
\left(
\frac{\log x}{1-x^2} 
\right)
= -\frac{2 \zeta}{\pi}\sum_{n=0}^{\infty}
\left( \frac{1 + \tilde{c}(2n+1)}{(2n+1)^2} e^{-(2n+1)\tilde{c}}
\right).
\label{SUSY-no-up}
\end{equation}
This is equivalent to the condition 
for the phase transition 
from the no gap to the upper gap phase 
discussed in \cite{Jain:2013py}, 
which is obtained by substituting $\alpha = 0$ into 
(6.15) of \cite{Jain:2013py} 
and by requiring $\rho(0) = \frac{1}{2\pi\lambda}$.
By using 
\eqref{SUSY-no-up},
the eigenvalue density \eqref{eigen-SUSY-up} at 
$a = 0$  becomes
\begin{align}
\rho_{ug}^{su}(\zeta,\lambda,\tilde{c},0;\alpha)
=& \frac{1}{2\pi}
+\frac{\zeta}{\pi^2}\sum_{n = 0}^{\infty}
\cos(2n+1)\alpha
\left(
\frac{1+\tilde{c}(2n+1)}{(2n+1)^2}
\right)e^{-(2n+1)\tilde{c}}.
\label{eigen-SUSY-up-no}
\end{align}
This is the same as the eigenvalue density function 
at the no gap phase of the SUSY CS matter theory,
which is described by (6.15) of \cite{Jain:2013py}.
So, at the phase transition points, 
the upper gap solutions are smoothly connected to  
the no gap phase solutions.

\begin{figure}
  \begin{center}
  \subfigure[]{\includegraphics[scale=.33]{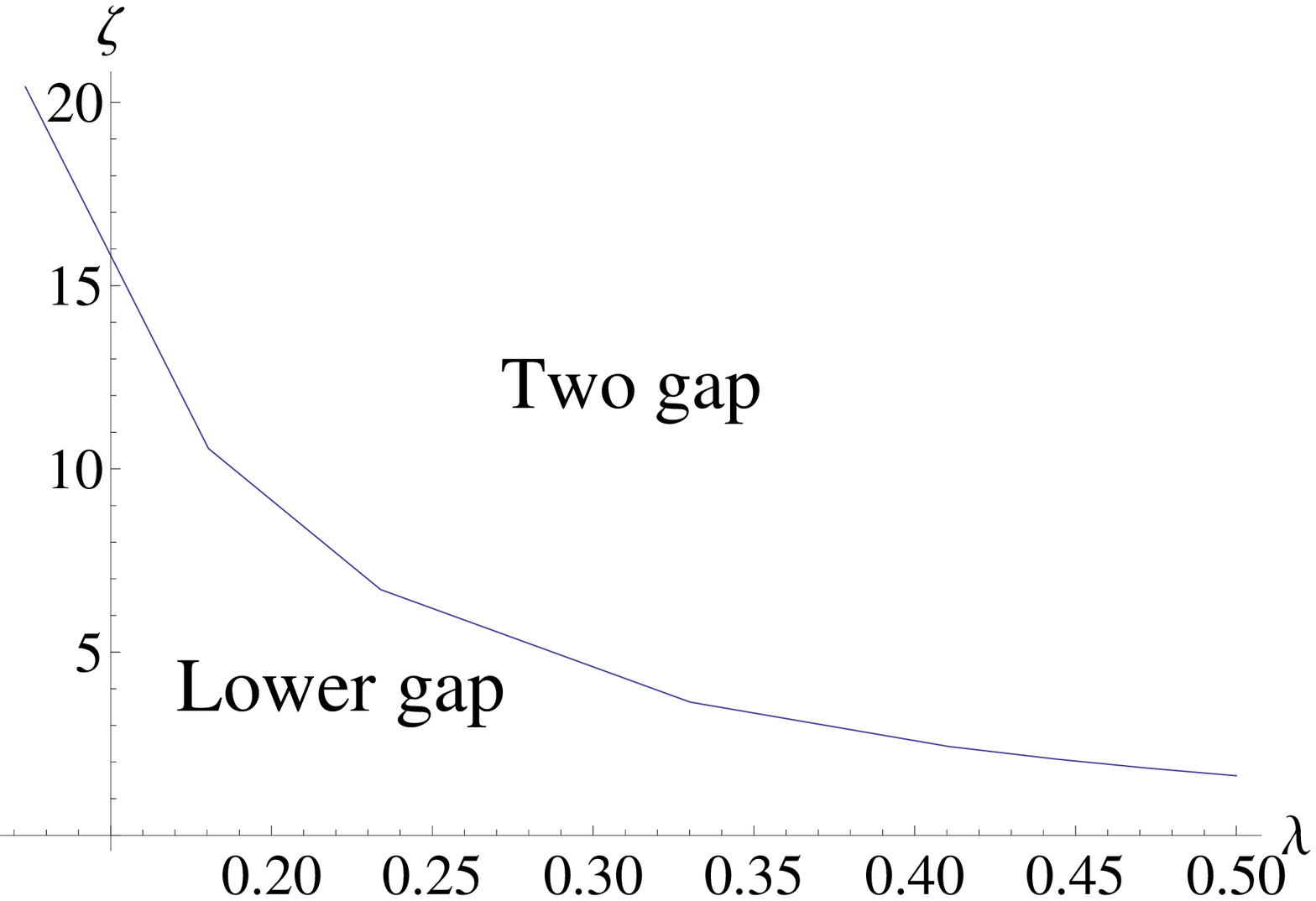}
\label{fig:low-two-SUSY}}
\qquad \qquad
  \subfigure[]{\includegraphics[scale=.3]{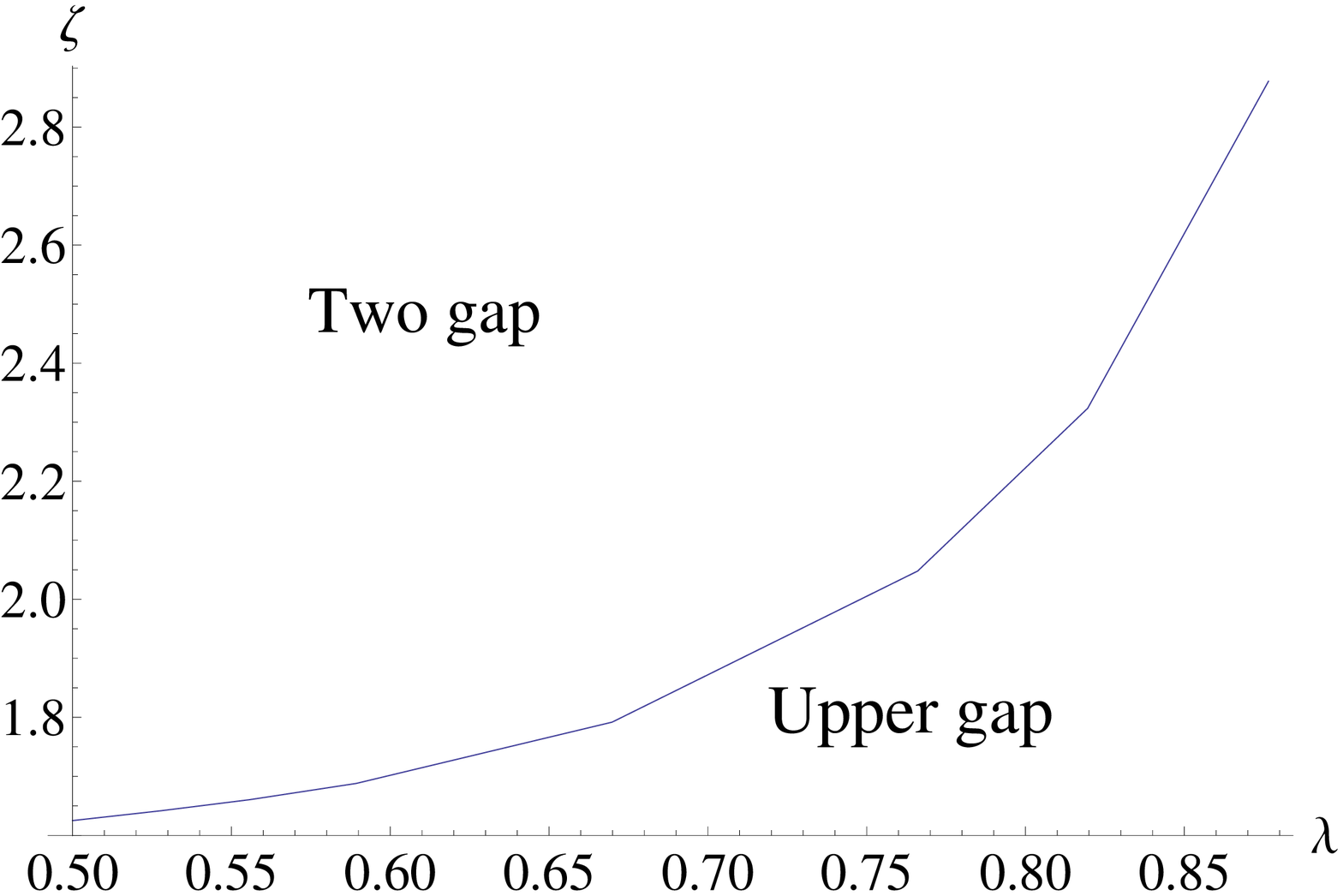}
\label{fig:up-two-SUSY}}
\caption{These are the plots of phase transition points in the SUSY CS matter
theory. Fig.~\ref{fig:low-two-SUSY} shows the plots of the ones 
from the lower gap to the two gap, and Fig.~\ref{fig:up-two-SUSY} is the ones
from the upper gap to the two gap phase.}
  \end{center}
\end{figure}

\paragraph{Phase transition from the lower gap to the two gap}
In the lower gap phase at fixed $\lambda$, 
if the maximum of the eigenvalue density $\rho(0)$ reaches to 
$\rho(0) = \frac{1}{2\pi \lambda}$,
the phase transition from the lower gap phase to the 
two gap phase will occur.
The conditions are represented in terms of \eqref{eigen-SUSY-low}
as 
\begin{equation}
\frac{1}{\lambda}
= \frac{\sqrt{2}\zeta}{\pi}
\int^{\infty}_{\tilde{c}} dy\, 
\left(
\frac{y \sin \frac{b}{2} 
\cosh \frac{y}{2}}{
(\cosh y + 1)
\sqrt{(\cosh y + \cos b)}}
+\frac{y \sin \frac{b}{2} \sinh \frac{y}{2}}{
(\cosh y - 1)
\sqrt{(\cosh y - \cos b)}}\right).
\label{low-to-2 SUSY}
\end{equation}
The combination of \eqref{low-to-2 SUSY}, \eqref{susyct}
and \eqref{Cond-low-SUSY} provide the 
phase transition points from the lower gap to the two gap.
By numerical calculations based on 
\eqref{low-to-2 SUSY}, \eqref{susyct}
and \eqref{Cond-low-SUSY}, 
we obtain the phase transition points 
plotted in Fig.~\ref{fig:low-two-SUSY}.

Let us examine the phase transition points from the standpoint
of the two gap phase.
If we substitute $a = 0$ 
into \eqref{SUSY two u1}, 
it becomes \eqref{low-to-2 SUSY},
and if we set $a = 0$ into 
\eqref{SUSY two u0},
it becomes the same as \eqref{Cond-low-SUSY}.
Let us check whether the eigenvalue density in the two gap phase 
\eqref{eigen-SUSY-two} becomes the one in the lower gap phase
\eqref{eigen-SUSY-low} in $a = 0$ limit.
By using \eqref{low-to-2 SUSY} and by similar calculations 
to \eqref{RF-l-2s} and \eqref{CB-l-2},
we can confirm that \eqref{eigen-SUSY-two} at $a = 0$ becomes
\eqref{eigen-SUSY-low}.
So we can see that the solutions in the two gap phase are smoothly connected
to the ones in the lower gap phase at the phase transition points.

\paragraph{Phase transition from the upper gap to the two gap}
In the upper gap phase at fixed $\lambda$, 
if the minimum of the eigenvalue density $\rho(\pi) = \rho(-\pi)$ 
reaches to 
$\rho(\pi) = 0$,
the phase transition from the upper gap to the two gap phase will occur.
The condition is represented in terms of \eqref{eigen-SUSY-up}
as 
\begin{eqnarray}
\frac{1}{\lambda} &=& 
\frac{\zeta}{\pi}
\int_{\tilde{c}}^{\infty}dy ~  
\left(\frac{\sqrt{2}y\cos \frac{a}{2} \sinh \frac{y}{2}}{
\sqrt{\cosh y +\cos a}(\cosh y -1)}
+
\frac{\sqrt{2}y \cos \frac{a}{2}\cosh \frac{y}{2}}{
\sqrt{\cosh y -\cos a}(\cosh y +1)}\right).
\label{up-to-2 SUSY}
\end{eqnarray}
The combination of \eqref{up-to-2 SUSY}, \eqref{susyct}
and \eqref{Cond-up-SUSY} provide the 
phase transition points from the upper gap to the two gap.
By numerical calculations based on 
\eqref{up-to-2 SUSY}, \eqref{susyct}
and \eqref{Cond-up-SUSY}, 
we obtain the phase transition points 
plotted in Fig.~\ref{fig:up-two-SUSY}.

Let us examine the phase transition points from the standpoint
of the two gap phase.
If we substitute $b = \pi$ 
into \eqref{SUSY two u1} 
and \eqref{SUSY two u0},
they become \eqref{up-to-2 SUSY} and \eqref{Cond-up-SUSY} respectively.
By using these correspondence and by using the same procedure as
the one in section \ref{sec:RF-2} and \ref{sec:CB-2}, 
we can also check that \eqref{eigen-SUSY-two} at $b= \pi$ 
coincides with \eqref{eigen-SUSY-up}.
So, at the phase transition points,
the solutions in the two gap phase are 
smoothly connected to the 
upper gap solutions.

\paragraph{Quadruple phase transition points in the 
SUSY CS matter theory}
There is a quadruple phase transition point
\begin{equation}
\lambda_{c}^{susy} = 0.5, \qquad \zeta_{c}^{susy} = 1.62509, \qquad
(\tilde{c}_{c}^{susy} = 0.542933),
\label{Critical value SUSY}
\end{equation}
at which the four phases coexist.
Let us check whether 
\eqref{Critical value SUSY}
is the quadruple point.
Note that \eqref{low-to-2 SUSY} is the equation for the 
phase transition from the lower gap to the two gap.
Eq.~\eqref{up-to-2 SUSY} is the 
equation for the 
phase transition 
from the upper gap to two gap,
\eqref{SUSY-no-low} is the one from 
the no gap to the lower gap,
and \eqref{SUSY-no-up} is the one from the no gap to the upper gap.
So if the point \eqref{Critical value SUSY}
simultaneously satisfies the four equations, it is indeed the 
quadruple point.
We can see that at $a = 0, b = \pi$, 
\eqref{low-to-2 SUSY} 
as well as \eqref{up-to-2 SUSY} 
become
\begin{equation}
-\frac{2\zeta}{\pi^2} \int^{e^{-\tilde{c}}}_{0} 
dx~ \frac{\log x}{1 - x^2}
= \frac{1}{2\pi\lambda}.
\label{SUSY-cri}
\end{equation}
We can also see that 
at $\lambda = \frac{1}{2}$, 
\eqref{SUSY-no-low} and \eqref{SUSY-no-up} become 
equivalent to \eqref{SUSY-cri}.
By the study at \cite{Jain:2013py}, 
we already know that both \eqref{SUSY-no-low} and \eqref{SUSY-no-up}
are simultaneously satisfied at the point \eqref{Critical value SUSY}. 
So also \eqref{SUSY-cri} is satisfied at 
\eqref{Critical value SUSY}.
Then we see that 
\eqref{Critical value SUSY} is the quadruple phase transition point
in the SUSY CS matter theory.
Due to the existence of the quadruple phase transition point,
the phase structure becomes as in Fig.~\ref{fig:phase-SUSY}.

\subsection{Self-duality of the SUSY CS matter theory}
We will study the self-duality of the supersymmetric CS matter theory
under the Giveon-Kutasov 
type level-rank duality~\cite{Giveon:2008zn,Benini:2011mf}.
The level $k$ $U(N)$ SUSY CS matter theory is dual to the same theory 
with level $k$ $U(k-N)$ gauge group. 
We will show that there are the following relationships
between the quantities in the level $k$ $U(N)$ 
theory at $(\lambda, \zeta)$  and the 
quantities in the level $k$ $U(k-N)$ theory at 
$(\bar{\lambda}, \bar{\zeta}) = (1-\lambda, \frac{\lambda}{1-\lambda}\zeta)$:
\begin{equation}
\bar{\lambda} = 1- \lambda, \qquad 
\bar{\zeta} = \frac{\lambda}{1-\lambda}\zeta,\qquad
\left(\frac{N}{k} = \lambda, \quad \frac{k-N}{k} = \bar{\lambda}
\right),
\label{corres-0-SUSY}
\end{equation}
\begin{equation}
\bar{\rho}^{su} (\alpha) = 
\frac{\lambda}{1-\lambda}\left(
\frac{1}{2 \pi \lambda} - \rho^{su}(\pi+\alpha) 
\right),
\label{corres-1-SUSY}
\end{equation}
\begin{equation}
\bar{b} = \pi - a, \quad \quad \bar{a} = \pi - b, \qquad
\bar{\tilde{c}} = \tilde{c}.
\label{corres-2-SUSY}
\end{equation}
Quantities in the $U(k-N)$ theory are denoted 
with over-lines while the quantities
in the $U(N)$ theory are described without over-lines.
First we will show the duality between the quantities in the lower gap phase
of the $U(N)$ theory and the ones in the upper gap phase of the $U(k-N)$ 
theory.
Next we will demonstrate the duality in the two gap phase between the $U(N)$ 
and the $U(k-N)$ theories.
At the end of this subsection, 
we will show the agreement of the free energy under the 
duality.


\subsubsection{Between the lower gap phase of the $U(N)$ theory and the upper gap phase of the $U(k-N)$ theory}
Let us define sets of quantities
\begin{equation}
\begin{split}
{\cal S}^{su}_{lg} =& (\lambda, \zeta;
b, \tilde{c}, \rho^{su}_{lg}), \\
{\cal S}^{su}_{ug} = &
\left(1-\lambda, \frac{\lambda}{1-\lambda}\zeta;
\pi-b, \tilde{c} ,
\frac{\lambda}{1-\lambda}
\left(
\frac{1}{2\pi\lambda} - \rho^{su}_{lg}
( \pi+\alpha)\right)\right).
\label{SU_L-U-e}
\end{split}
\end{equation}
Please note that the the eigenvalue density 
\eqref{eigen-SUSY-low}
and the two equations \eqref{Cond-low-SUSY}, \eqref{susyct}
provide a complete set of solutions in the lower gap phase,
while the eigenvalue density \eqref{eigen-SUSY-up}
and \eqref{Cond-up-SUSY}, \eqref{susyct}
provide the one in the upper gap phase. 
So we can verify the duality relationships by 
checking that ${\cal S}_{ug}^{su}$ becomes a solution provided 
by the corresponding three equations in the upper gap phase
if ${\cal S}_{lg}^{c.b}$ is a solution 
provided by the other three equations in the lower gap.

First we will confirm the relationship \eqref{corres-1-SUSY}
between the eigenvalue densities
\eqref{eigen-SUSY-low}
and \eqref{eigen-SUSY-up} under the correspondence 
\eqref{corres-0-SUSY},\eqref{corres-2-SUSY}.
Here by utilizing relationship
\eqref{RF_U-CB_L-e} and \eqref{RF_L-CB_U-e},
we can check
\begin{equation}
\rho^{su}_{ug}(\bar{\zeta},\bar{\lambda};\tilde{c},\pi-b;\alpha)
= \frac{\lambda}{1-\lambda}\left(
\frac{1}{2\pi\lambda} - \rho^{su}_{lg}(\zeta,\lambda;\tilde{c},b;\pi+\alpha)
\right).
\label{SU_L-U-c}
\end{equation}
This shows the relationship 
\eqref{corres-1-SUSY}
between the eigenvalue densities
\eqref{eigen-SUSY-low}
and \eqref{eigen-SUSY-up}.

We can also see that 
${\cal S}^{su}_{ug}$ becomes a 
solution of \eqref{Cond-up-SUSY} 
if 
${\cal S}^{su}_{lg}$ is a solution 
of \eqref{Cond-low-SUSY}
by showing
\begin{equation}
\tilde{M}^{su}_{lg}(\zeta,\tilde{c},b)
= -\frac{1-\lambda}{\lambda}\tilde{M}^{su}_{ug}
\left(\frac{\lambda}{1-\lambda}\zeta,\tilde{c},\pi-b\right).
\end{equation}
To show this
we have used the combination of \eqref{RF_L-CB_U-c}
and \eqref{RF_U-CB_L-c}.

We will also see that the ${\cal S}^{su}_{ug}$ becomes a 
solution of the \eqref{susyct} if ${\cal S}^{su}_{lg}$
is a solution of it. 
By using 
\eqref{SU_L-U-c}, we can directly check as
\begin{eqnarray}
\tilde{c}&=&
2  \biggl | \text{Re} \int_{-\pi}^{\pi} d\alpha
\lambda \rho^{su}_{lg}(\alpha) 
\log \coth {\tilde{c} + i \alpha \over 2} \biggl | 
\nonumber \\
\Leftrightarrow
\tilde{c}
&=& -2  \biggl | \text{Re} \int_{-\pi}^{\pi} d\alpha
(1-\lambda) \rho^{su}_{ug}(\pi + \alpha) 
\log \coth {\tilde{c} + i \alpha \over 2} \biggl | 
\nonumber\\
\Leftrightarrow
\tilde{c}
&=& -2  \biggl | \text{Re} \int_{-\pi}^{\pi} d\alpha
(1-\lambda) \rho^{su}_{ug}(\alpha) 
\log \coth {\tilde{c} + i \alpha \over 2} \biggl | .
\label{Lower c check}
\end{eqnarray}
Here we have used 
\begin{equation}
 \int_{-\pi}^{\pi} d\alpha \log \coth {x + i \alpha \over 2} =0, \qquad
\coth ({x + i\pi \over 2}) = \tanh ({x \over 2})  = -\coth ({x \over 2}).
\end{equation}
Eq.~\eqref{Lower c check}
shows that ${\cal S}^{su}_{ug}$ becomes a 
solution of \eqref{susyct} if ${\cal S}^{su}_{lg}$
is a solution of it. 

By the above discussion, we can 
conclude that ${\cal S}^{su}_{ug}$ becomes an upper gap solution 
provided by 
the three equations \eqref{eigen-SUSY-up}, \eqref{Cond-up-SUSY} 
and \eqref{susyct} in the $U(k-N)$ theory 
if ${\cal S}^{su}_{lg}$ is a lower gap solution 
provided by the equations \eqref{eigen-SUSY-low}, \eqref{Cond-low-SUSY},
and \eqref{susyct} in the $U(N)$ theory.
Thus we have verified the 
self-duality relationships between the lower gap phase of the 
level $k$ $U(N)$ SUSY CS matter theory 
and the upper gap phase of the level $k$ $U(k-N)$ SUSY CS matter theory.

\subsubsection{Self-duality in the two gap phase}
Let us define sets of quantities
\begin{equation}
\begin{split}
{\cal S}^{su}_{tg} =& (\lambda, \zeta;
b, \tilde{c}, \rho_{tg}^{su}(\alpha)), \\
\bar{\cal S}^{su}_{tg} =&
(\bar{\lambda}, \bar{\zeta};
\bar{a},\bar{b}, \bar{\tilde{c}}, \bar{\rho}_{tg}^{su}(\alpha))
\\
=& \left(1-\lambda, \frac{\lambda}{1-\lambda}\zeta;
\pi-b, \pi-a, \tilde{c}, 
\frac{\lambda}{1-\lambda}
\left(
\frac{1}{2\pi\lambda} - \rho^{su}_{tg}
( \pi+\alpha)\right)\right).
\label{SU_2-e}
\end{split}
\end{equation}
Here $\bar{\rho}_{tg}^{su}(\alpha)$ is originally defined as
the eigenvalue density function of the dual variable
\begin{equation}
\bar{\rho}_{tg}^{su}(\zeta,\lambda; a,b, \tilde{c};\alpha) \equiv 
\rho_{tg}^{su}(\bar{\zeta},\bar{\lambda}; \bar{a},\bar{b},\bar{\tilde{c}};
\alpha+\pi).
\end{equation}

Please note that the the eigenvalue density 
\eqref{eigen-SUSY-two}
and the three equations
\eqref{SUSY two u1}, 
\eqref{SUSY two u0} and 
\eqref{susyct}
provide a complete set of solutions in the two gap phase.
So we can verify the duality relationships by 
checking the following: 
If ${\cal S}_{tg}^{su}$ is a two gap phase solution of the $U(N)$ theory
provided by the four equations,
also $\bar{\cal S}_{tg}^{su}$ becomes a two gap solution of 
the $U(k-N)$ theory provided 
by the same equations.

\paragraph{Correspondence \eqref{corres-1-SUSY} between 
the eigenvalue density functions}
First we will confirm the relationship \eqref{corres-1-SUSY}
between the eigenvalue densities $\rho^{su}_{tg}$ and 
$\bar{\rho}^{su}_{tg}$.
By using \eqref{rho-D-1} and \eqref{rho-D-2},
we can see
\begin{equation}
\begin{split}
\bar{\rho}_{tg}^{su}(\zeta,\lambda; a,b, \tilde{c};\alpha) \equiv
&\rho^{su}_{tg}
(\frac{\lambda}{1-\lambda}\zeta, 1-\lambda;
\pi-b, \pi-a, \tilde{c}; \alpha)
\\
=&\frac{1}{2\pi(1-\lambda)}
- \frac{\lambda}{1-\lambda}\rho^{su}_{tg}
(\zeta, \lambda;
a, b, \tilde{c}; \pi+\alpha).
\end{split}
\end{equation}
So there is a relationship \eqref{corres-1-SUSY} between
$\rho^{su}_{tg}$ and $\bar{\rho}^{su}_{tg}$.

\paragraph{About \eqref{SUSY two u1}.}
Let us check $\bar{\cal S}^{su}_{tg}$ becomes a solution of 
\eqref{SUSY two u1} if ${\cal S}_{tg}^{su}$ is a solution of it.
By using $\lambda\zeta = (1-\lambda) \bar{\zeta}$ and 
\eqref{Formula two 2}, \eqref{D-3-00}
we can see
\begin{equation}
\begin{split}
&\frac{1 }{4\pi \lambda  } \Upsilon(a,b)
= \frac{\zeta}{2\pi} 
{\cal Y}^{su}(a,b,\tilde{c})
\\
\Rightarrow&
\frac{1}{4\pi(1-\lambda)}\Upsilon(\pi-b,\pi-a)
= 
\frac{\lambda}{1-\lambda}
\frac{\zeta}{2\pi}{\cal Y}^{su}(\pi-b,\pi-a,\tilde{c}).
\end{split} 
\label{D-3-SUSY-2}
\end{equation}
This shows that 
$\bar{\cal S}^{su}_{tg}$ becomes a solution of 
\eqref{SUSY two u1} if ${\cal S}_{tg}^{su}$ is a solution of it.

\paragraph{About \eqref{SUSY two u0}.}
Let us check $\bar{\cal S}^{su}_{tg}$ becomes a solution of 
\eqref{SUSY two u0} if ${\cal S}_{tg}^{su}$ is a solution of it.
By using 
\eqref{D-3-01}, \eqref{D-3-02} and \eqref{Formula two 3},
we can see 
\begin{equation}
\begin{split}
&\frac{1}{4\pi\lambda}\Lambda(a,b) = 1+ 
\frac{\zeta}{4 \pi}{\cal G}^{su}(a,b,\tilde{c})
\\
\Rightarrow&\frac{1}{4\pi(1-\lambda)}\Lambda(\pi-b,\pi-a) = 1+ 
\frac{\lambda}{1-\lambda}\frac{\zeta}{4 \pi}
{\cal G}^{su}(\pi-b,\pi-a,\tilde{c}).
\end{split}
\label{D-3-3-SUSY}
\end{equation}
Please note that
\begin{equation}
{\cal G}^{su}(\pi-b,\pi-a,\tilde{c})
= -{\cal G}^{su}(a,b,\tilde{c}).
\end{equation}
This \eqref{D-3-3-SUSY}
shows that $\bar{\cal S}^{su}_{tg}$ becomes a 
solution of the (\ref{SUSY two u0})
if ${\cal S}^{su}_{tg}$ is a solution of it.

\paragraph{About \eqref{susyct}.}
By repeating the analogous calculation to \eqref{Lower c check},
we can see that $\bar{\cal S}_{tg}^{su}$ becomes a solution of the 
\eqref{susyct} if ${\cal S}_{tg}^{su}$ is a solution of it.

\paragraph{Summary}
By the above discussion we can 
conclude that $\bar{\cal S}^{su}_{tg}$ becomes a solution provided by 
the four equations \eqref{eigen-SUSY-two},\eqref{SUSY two u1},\eqref{SUSY two u0} and \eqref{susyct} in 
the level $k$ $U(k-N)$ SUSY CS matter theory 
if ${\cal S}^{su}_{tg}$ is a solution 
of the same equations in the 
the level $k$ $U(N)$ SUSY CS matter theory.
Thus we have verified the 
self-duality relationships at the two gap phase 
between the level $k$ $U(N)$ SUSY CS matter theory and the 
same theory with level $k$ $U(k-N)$ gauge group.

\subsubsection{Relationships between the phase transition points}
\label{Sec:PTD-SUSY}
We have seen that the conditions for the phase transitions
\eqref{SUSY-no-low}, \eqref{SUSY-no-up}, 
\eqref{low-to-2 SUSY} and 
\eqref{up-to-2 SUSY} 
can be obtained as a certain limit of 
corresponding equations between which 
we have already verified the duality relationships.
So, 
by the same logic as in section \ref{Sec:PTD-RF-CB},
we have already shown the duality relationships between the 
phase transition points.
By using the analogous symbols to the ones used in \eqref{PT-CB-RF}, 
we can see the following duality relationships 
\begin{equation}
(\text{No,Low,SUSY} ) \Leftrightarrow(\text{No,Up,SUSY} ), \quad
(\text{Low,Two,SUSY} ) \Leftrightarrow(\text{Up,Two,SUSY} ) .
\end{equation}

\subsubsection{Free energy and completing the proof of the duality}
Now we have proved that 
there are self-duality relationships
\eqref{corres-0-SUSY}$\sim$\eqref{corres-2-SUSY} 
in the SUSY CS matter theory.
Under these 
correspondences
\eqref{corres-0-SUSY}$\sim$\eqref{corres-2-SUSY},
we will show that 
the free energy of the level $k$ $U(N)$ SUSY CS matter theory 
at $(\lambda, \zeta)$ agrees with the free energy of
the same theory with level $k$ $U(k-N)$ gauge group 
at $(1-\lambda, 
\frac{\lambda}{1-\lambda}\zeta )$,
namely,
\begin{equation}
F^{k-N}_{su} = V^{su}[\bar{\rho}^{su}, k-N] + F_2[\bar{\rho}^{su},k- N]
= V^{su}[\rho^{su}, N] + F_2[\rho^{su}, N] = F^{N}_{su}.
\end{equation}
Here we have used \eqref{F-SUSY}.
As already shown in (3.30) of \cite{Jain:2013py},
by a straightforward calculation similar to \eqref{Lower c check},
we can show $V^{su}[\bar{\rho}^{su}, k-N] = V^{su}[\rho^{su}, N]$.
We can also check that 
$F_2[\bar{\rho},k- N] = F_2[\rho,N]$ by almost the same calculation as
\eqref{F-D-1} and \eqref{F-D-2}.

So we can see
\begin{equation}
F^{N}_{su} = F^{k-N}_{su},
\end{equation}
under the relationships \eqref{corres-0-SUSY}$\sim$\eqref{corres-2-SUSY}.
Now we have completed the proof of the level-rank self-duality
between the level $k$ $U(N)$ SUSY CS matter theory and 
the level $k$ $U(k-N)$ SUSY CS matter theory.

\section{Analytic proof of the duality in the two gap phase of the GWW type 
matrix integral}
In \cite{Jain:2013py}, the phase structure of the 
toy large $N$ GWW type matrix integral has been already investigated.
Although some numerical
evidence have been provided, 
the analytic proof 
of the level-rank duality in the two gap 
phase of the GWW model
has not been completed.
Then we will complete the analytic proof of the duality 
by using the three formulae \eqref{Formula two 1},
\eqref{Formula two 2} and \eqref{Formula two 3}.

This proof would be also useful for analysis of general 
CS matter theories on $S^2 \times S^1$.
As we can see at
\eqref{Large z RF}, \eqref{Large z CB}, \eqref{Large z SUSY}, 
and appendix \ref{Z-gen},
the eigenvalue densities converge to the one in the GWW model
at large $\zeta$ limit.
Moreover, every eigenvalue density of CS matter theories 
at the two gap phase includes the term $\rho_{2,tg}$ which is 
the same functional form as the eigenvalue density in the 
GWW model.
So the duality relationships 
shown in this section play one of the fundamental roles in
establishing 
the level-rank duality in CS matter theories on $S^2 \times S^1$. 

Let us focus on 
the proof of the level-rank duality in the two gap phase 
of the GWW model.
As described in section C.4 of \cite{Jain:2013py},
we can regard $(\lambda,\zeta)$ as functions
of the edge of the cuts $(a,b)$.
To complete the proof of the self-duality of the 
GWW type model in terms of this, 
we should show the following:
\begin{equation}
\lambda(a,b)
\zeta(a,b) = \lambda(\pi-a,\pi-b)\zeta(\pi-a,\pi-b),
\label{eq:lz}
\end{equation}
\begin{equation}
\lambda(\pi- b , \pi -a ) = 1 - \lambda(a,b),
\label{eq:l}
\end{equation}
\begin{equation}
\rho_{tg}\left(
(1-\lambda), \frac{\zeta \lambda}{1-\lambda}, \alpha
\right)
= \frac{\lambda}{1-\lambda}
\left(
\frac{1}{2\pi\lambda}
- \rho_{tg}(\lambda, \zeta, \alpha+\pi)
\right).
\label{eq:r}
\end{equation}
If we can verify the above three equations,
it becomes the proof of the 
level-rank duality in the two gap phase of the GWW type matrix integral.

\subsection{Proof of \eqref{eq:lz}}
From (C.17) of \cite{Jain:2013py} and by identifying $\omega = e^{i\theta}$,
we can see 
\begin{equation}
\zeta(a,b) \lambda(a,b) 
= 
\frac{1}{i \pi}\int_{L_{ugs}} \frac{d \omega}{h(\omega)}.
\end{equation}
From this, by a direct substitution, 
we can also see
\begin{equation}
\zeta(\pi-b,\pi-a) \lambda(\pi-b,\pi-a) 
= 
-\frac{1}{i \pi}\int_{L_{lgs}} \frac{d \omega}{h(\omega)}.
\end{equation}
So by using
\eqref{Formula two 2}, we can immediately verify
\eqref{eq:lz}.
\subsection{Proof of \eqref{eq:l}}
From (C.17) and by the identification $\omega = e^{i\theta}$,
we can see that $\lambda(a,b)$ can be written as
\begin{equation}
\lambda(a,b) = 
\frac{1}{i\pi } \int_{L_{ugs}} \frac{d \omega}
{h(\omega)} \left(\omega - \frac{1}{2}(\cos a + \cos b)
\right).
\end{equation}
By the same way, we can also see
\begin{equation}
\lambda(\pi-b,\pi-a) = 
\frac{1}{i\pi } \int_{L_{lgs}} \frac{d \omega}
{h(\omega)} \left(\omega - \frac{1}{2}(\cos a + \cos b)
\right).
\end{equation}
So by using both \eqref{Formula two 2} and 
\eqref{Formula two 3},
we can show that
\begin{equation}
\lambda(\pi-b,\pi-a)+ \lambda(a,b) =1.
\end{equation}
This is the proof of (\ref{eq:l}).

\subsection{Proof of \eqref{eq:r}}
The eigenvalue density (7.6) of 
\cite{Jain:2013py} can be described by the complex line integral as
\begin{equation}
\rho(\lambda(a,b), \zeta(a,b),\alpha) 
= \frac{ih^{+}(u)}{4 \pi^2\lambda}\int_{L_{ugs}}d\omega~
\frac{1}{h(\omega)}\left(
\frac{2}{\omega - u} + \frac{1}{u}
\right),
\end{equation}
here we identify $\omega = e^{i\theta}$.
In the same way, by using (\ref{eq:l}) and (\ref{eq:lz}),
we can also check that
\begin{equation}
\rho\left(\lambda(\pi-b,\pi-a), \zeta(\pi-b,\pi-a),\pi+\alpha\right)
= \frac{ih^{+}(u)}{4 \pi^2(1-\lambda)}\int_{L_{lgs}}
\frac{d\omega}{h(\omega)}\left(
\frac{2}{\omega - u} + \frac{1}{u}
\right).
\end{equation}
So by using \eqref{Formula two 1},(\ref{eq:l}) and (\ref{eq:lz}), we can check
\begin{equation}
\rho(\lambda, \zeta, \alpha)
+\frac{1-\lambda}{\lambda}
\rho
\left(
1-\lambda, \frac{\lambda}{1-\lambda}\zeta,\pi+\alpha
\right) 
=
\frac{1}{2\pi\lambda}.
\end{equation}
This is the proof of (\ref{eq:r}).

Then we have verified (\ref{eq:lz}),(\ref{eq:l}) and (\ref{eq:r}).
Hence we have completed the analytic proof of the level-rank duality
in the two gap phase of the GWW type model.


\section{Summary and discussions}
In this paper, we have elaborated the higher temperature phases of the 
Chern-Simons matter theories with fundamental representation on 
$S^2 \times S^1$. 
Here we have studied 
the lower gap, the upper gap and the two gap phases of the 
regular fermion theory, the critical boson theory, and the 
${\cal N} = 2$ supersymmetric Chern-Simons matter theory.
We have obtained the eigenvalue density function and the free energy 
in each phase, and we have also obtained the phase transition points and 
the phase diagrams.
Based on these obtained quantities, we have completely analytically
verified the level-rank duality 
(Giveon-Kutasov type duality)
between the 
regular fermion theory and the critical boson theory,
and the self-duality of the SUSY CS matter theory.
We have also supplied the analytic proof of the duality
in the two gap phase of the GWW type matrix integration 
which is suggested in \cite{Jain:2013py}.

We can see that the proof of the duality in the two gap phase of the 
GWW model as well as the three formulae 
\eqref{Formula two 1},\eqref{Formula two 2} and \eqref{Formula two 3}
proved in appendix \ref{Ap:2-gap dual}
are based on the cut structure of the two gap phase, 
which would be shared by all CS matter theories on $S^2 \times S^1$. 
Please note that $\rho_{2,tg}$ in the eigenvalue density functions
at the two gap phase has the same functional form as the eigenvalue density 
of the GWW model, and it is shared by all the eigenvalue densities in the 
two gap phase of the all CS matter theories dealt with in this paper.
The three formulae play the role of establishing the level-rank duality 
with respect to the $\rho_{2,tg}$.
Hence the duality structure of the GWW model 
and the three formulae would play 
the most fundamental role in the level-rank 
duality of general CS matter theories on $S^2 \times S^1$.

\begin{figure}
  \begin{center}
  \subfigure[]{\includegraphics[scale=.28]{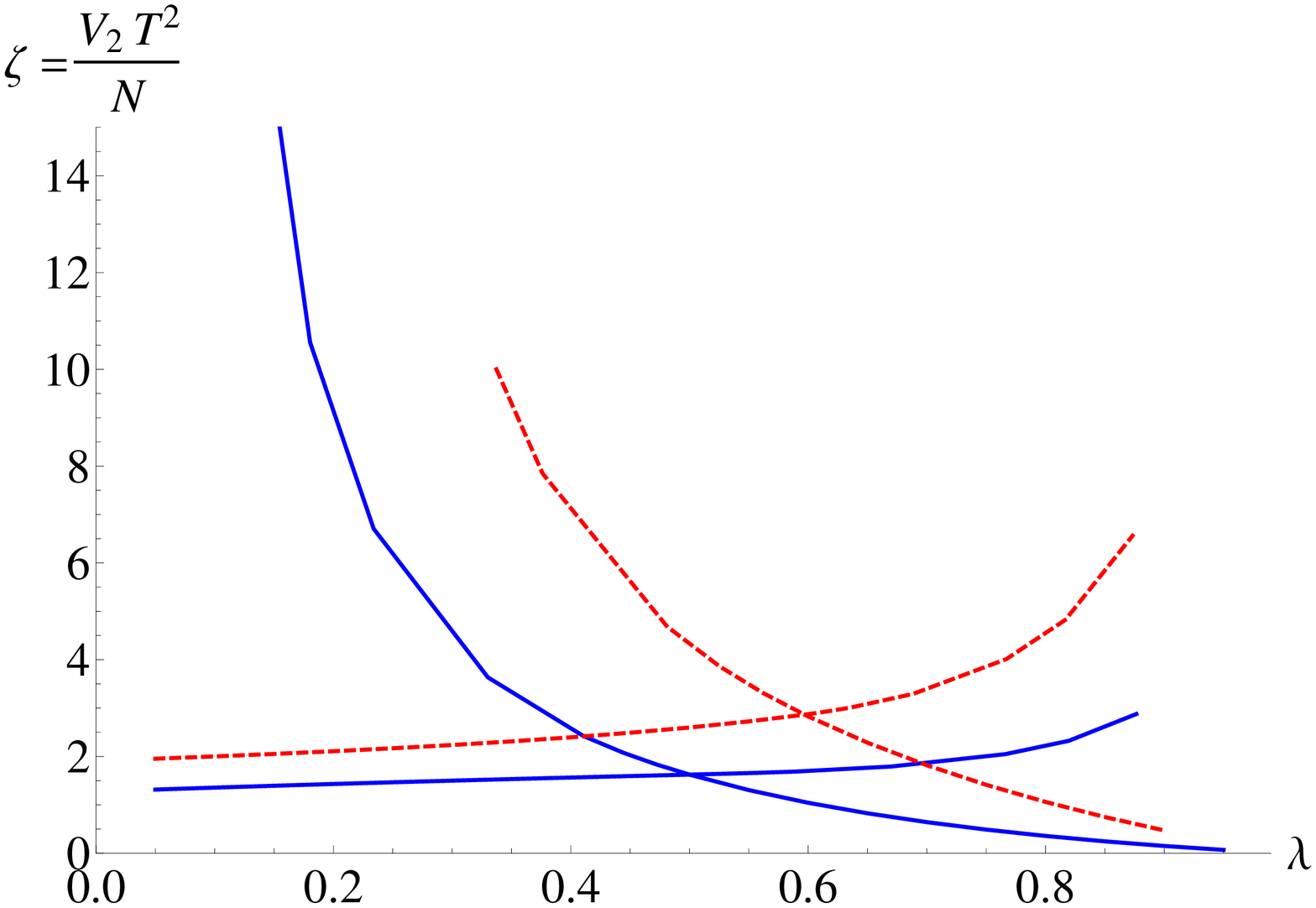}
\label{fig:SUSYvsRF}}
\qquad\qquad
  \subfigure[]{\includegraphics[scale=.28]{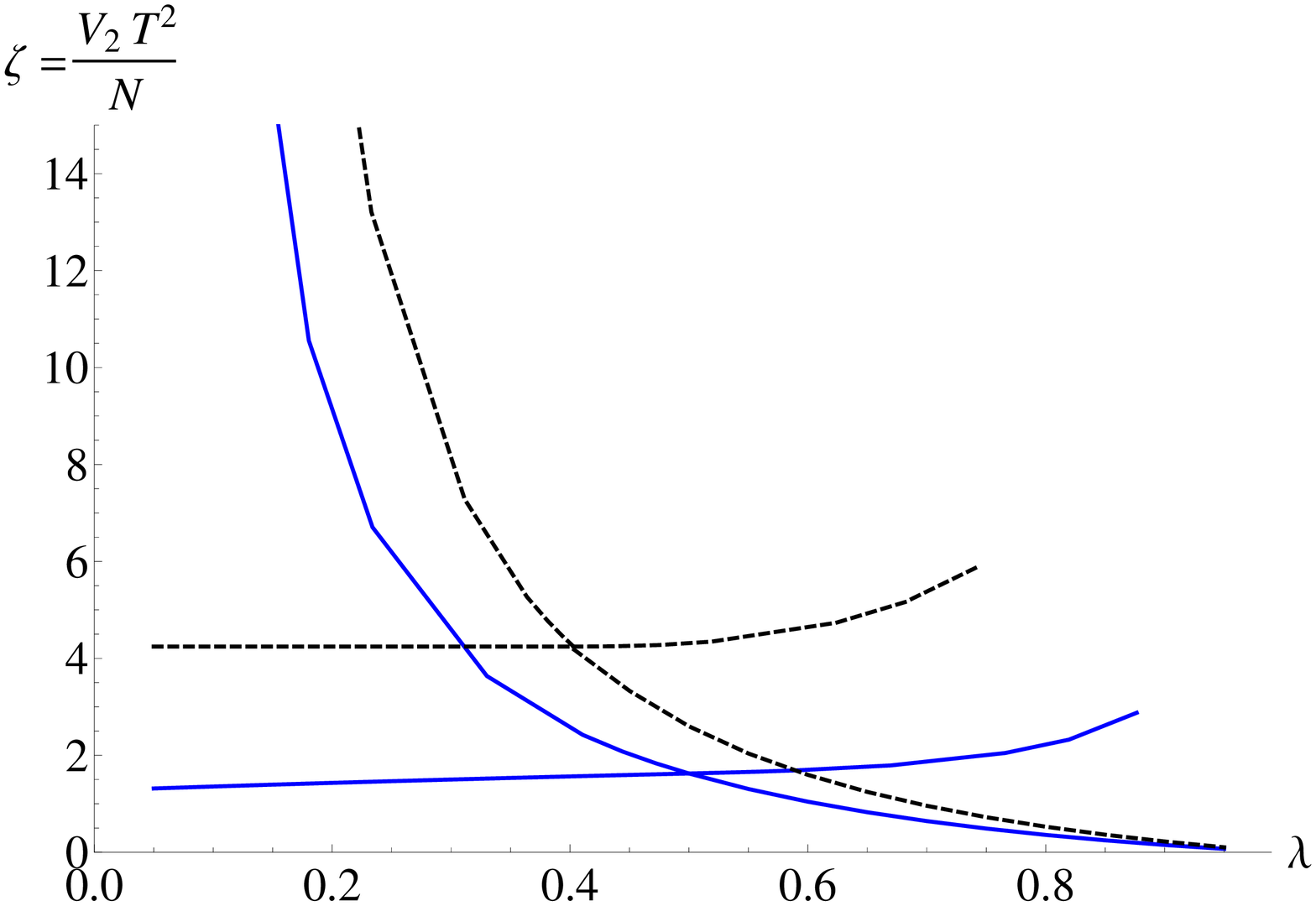}
\label{fig:SUSYvsCB}}
\caption{Fig.\ref{fig:SUSYvsRF} is 
the phase transition points of the SUSY theory v.s the ones
in the regular fermion theory.
Fig.\ref{fig:SUSYvsCB} is 
the phase transition points of the SUSY theory v.s the ones
in the critical boson theory.
Here bold blue line in the both figure is the plots in the SUSY theory.
The red dashed line in Fig.~\ref{fig:SUSYvsRF} is the points in the regular 
fermion theory and the black dashed line in 
in Fig.~\ref{fig:SUSYvsCB} is the points in the critical 
boson theory. From these we can see the phase transition points
in the SUSY theory locate in lower temperature than the other
two theories.
}
  \end{center}
\end{figure}

It is also interesting to compare the three CS matter theories in this paper.
As we have seen at 
\eqref{Large z RF}, \eqref{Large z CB}, \eqref{Large z SUSY}, 
and appendix \ref{Z-gen},
in the supersymmetric theory, 
the range of the domain of the cut $\epsilon$ is
smaller than the one in the regular fermion theory as well as the 
one in the critical boson theory at the same value of
$(\lambda, \zeta)$.
We can also see that the phase transitions at the 
SUSY theory occur at lower temperatures than the ones in the 
other two CS matter theories. 
(See Fig.~\ref{fig:SUSYvsRF},\ref{fig:SUSYvsCB}.)
From this, we can guess that there should be stronger attractive forces between
the eigenvalues in the SUSY theory than the one in the other two theories.
This may be because the SUSY theory has more degree of freedom of matter 
fields than 
the other two theories. Then the effective potential in the 
SUSY theory obtained after 
integrating out the matter fields, which causes 
the attractive force between the eigenvalues, are written by the sum
of the ones of the regular fermion and the critical boson theory.
Hence the attractive force would be enlarged by the summation.

At the very high temperature limit, the eigenvalue densities of 
these three theories and the one of the GWW model converge to
$\cos^{-1}\sqrt{\alpha}$, 
and they 
finally 
become the same universal distribution \eqref{evd 0}
if we further increase the temperature.
We have given the analysis in 
\eqref{Large z RF}, \eqref{Large z CB}, \eqref{Large z SUSY} and 
appendix \ref{Ap:LZ-sec}.
So at the very high temperature limit, the results coincide with the 
ones in \cite{Aharony:2012ns}. We can regard their results 
as the very high temperature limit of our results.

It is easy and straightforward to study the critical fermion and the regular
boson theory on $S^2 \times S^1$ 
in the same way as this paper.

One of the interesting application of this paper would be to analyze the 
phase structure of the dual gravity description governed by parity violating
Vasiliev's higher spin 
equations~\cite{Vasiliev:1990en,Vasiliev:2003ev,Giombi:2012ms}
at finite values of the 
't~Hooft coupling $\lambda=\frac{N}{k}$ \cite{Giombi:2011kc,Chang:2012kt}. 
The thermal system studied in this paper is dual to the Euclidean 
Vasiliev system in global $AdS$ space 
compactified on a circle of circumference $2 \pi R=\frac{1}{T}$.
The field theory analysis performed in this paper implies that this Vasiliev 
system admits a classical limit in large $N$ limit with $V_{2}T^2=\zeta N$ 
with $\zeta$ held fixed. {\it The saddle points obtained in this paper should
map to classical solutions of this bulk description; however the equations 
of motion governing this classical description are  not necessarily
Vasiliev's equations.} This is because quantum corrections to Vasiliev's 
equations proportional $\frac{V_{2}T^2}{N}$ are not suppressed compared to 
classical effects in the combined high temperature and large $N$ limit 
studied in this paper; such terms could modify Vasiliev's classical 
equations. 
It would be fascinating to determine the effective
3 dimensional bulk equations dual to the large $N$ limit of this paper, 
and to study the classical solutions dual to the saddle points 
described in this paper.

\acknowledgments
The author would like to thank 
S.~Bhattacharyya, S.~Jain, S.~Minwalla, 
T.~Sharma, S.~Wadia, S.~Yokoyama
for their collaboration in the early stage of this work and 
helpful discussions. 
He especially thanks Shiraz Minwalla 
for reading manuscript and giving helpful comments.
He is also really grateful for his TIFR friends' kind help
(D.~Bardhan, N.~Kundu, A.~Lytle, M.~Mandlik, T.~Sharma and N.~Sircar) 
to correct English grammar.
T.T would also like to thank T.~Morita for the discussion on the 
basic part of the matrix model and K.~Ohta and H.~Suzuki for 
helpful discussions on the partition function of two dimensional 
gauge theories in the old days.
He would also like to acknowledge our debt to the people of India for 
their generous and steady support to research in the basic sciences.

\appendix
\section{Cut function $h(u)$}
\label{Ap:CUT}
\begin{figure}
  \begin{center}
  \subfigure[]{\includegraphics[scale=.33]{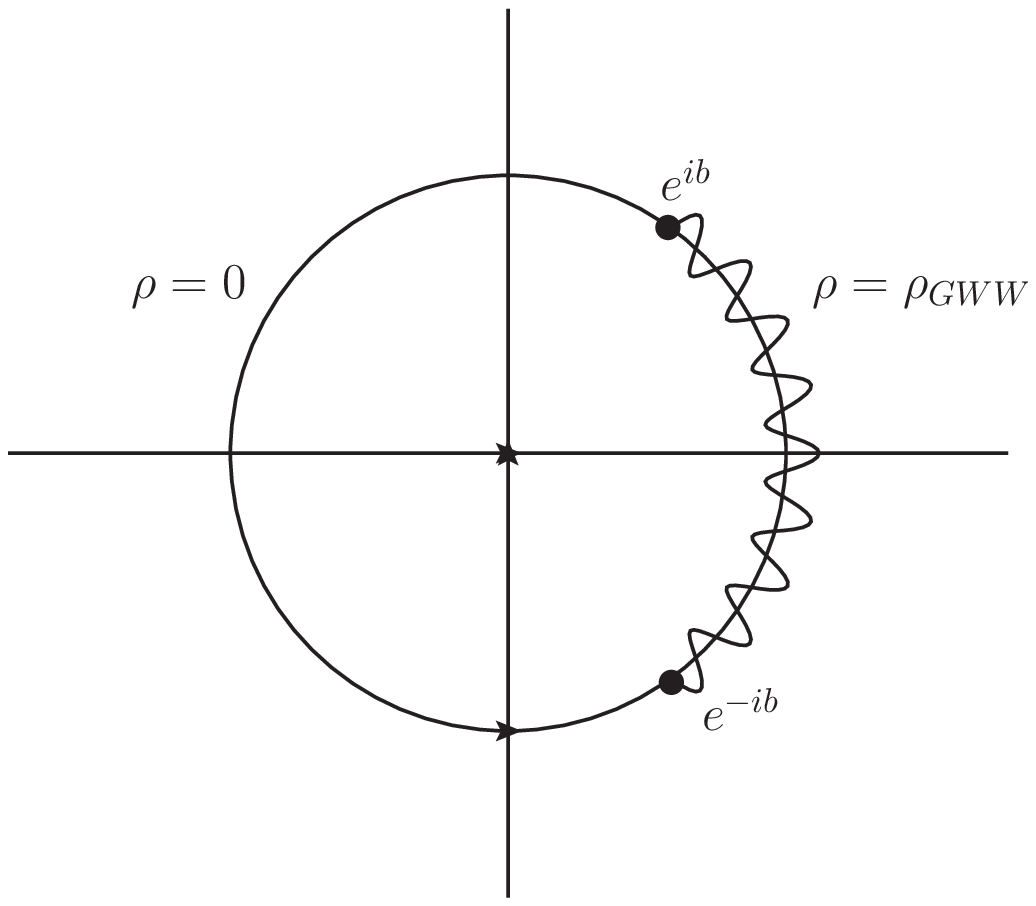}
\label{1-cutGWW}
  }
  \qquad\qquad
  \subfigure[]{\includegraphics[scale=.33]{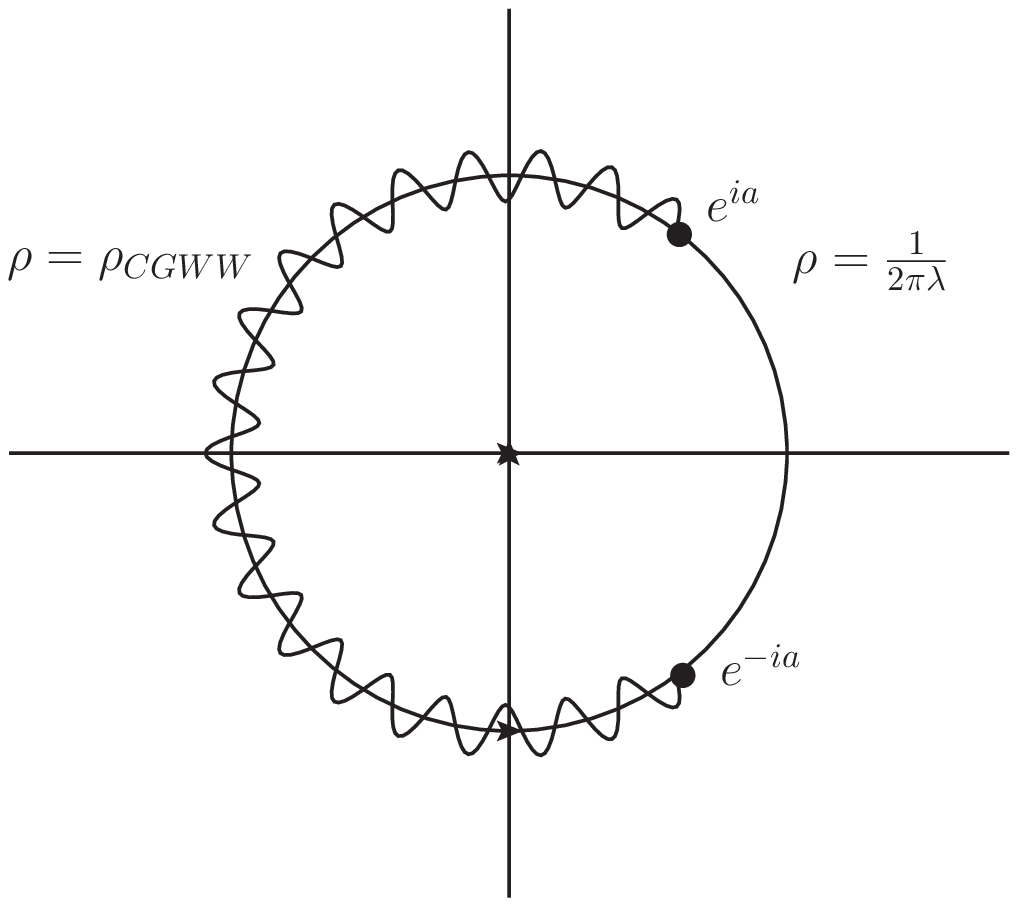}
\label{capped-GWW}
  }
  \qquad\qquad
  \subfigure[]{\includegraphics[scale=.33]{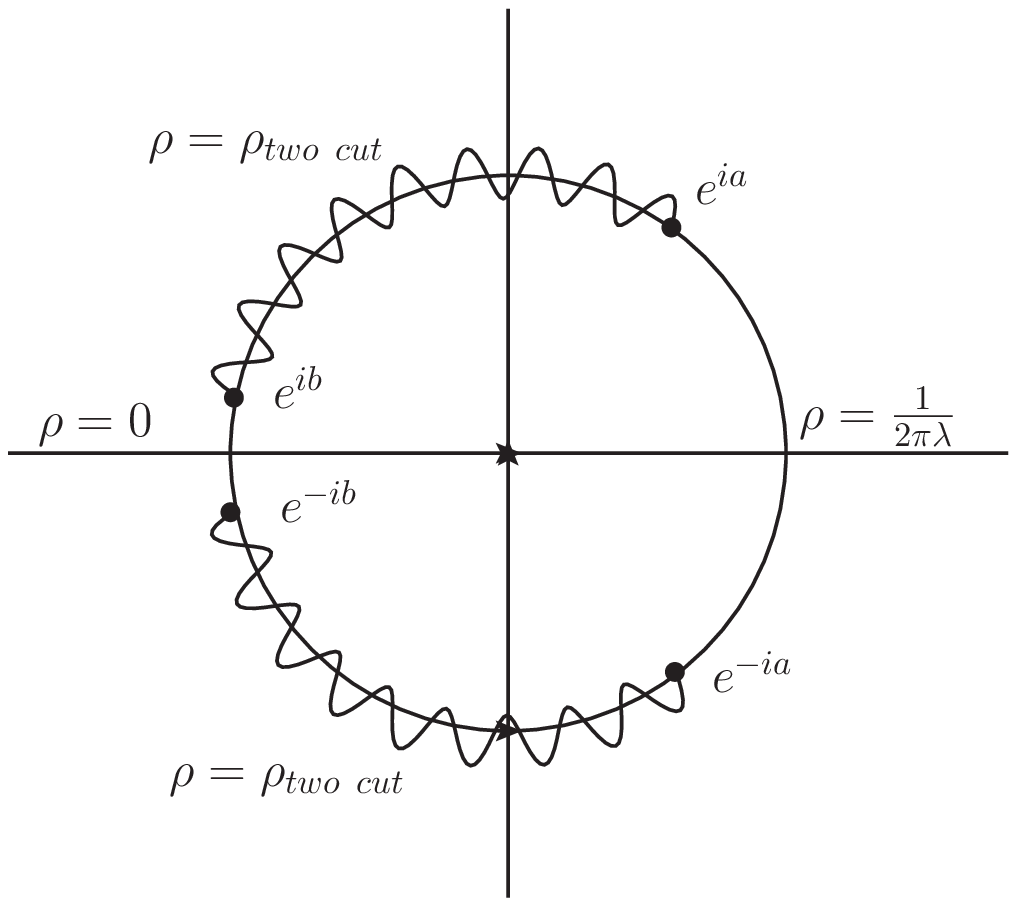}
\label{2cut.eps}
  }
\caption{Fig.\ref{1-cutGWW}, \ref{capped-GWW}, \ref{2cut.eps} show 
the cut region in the lower gap phase, 
in upper gap phase, in the two gap phase, respectively. }
  \end{center}
\end{figure}

We will list the cut region and the cut function
in this appendix.
The cut region in each phase 
is depicted in Fig.\ref{1-cutGWW}, \ref{capped-GWW} and \ref{2cut.eps}.

\subsection{Lower gap case}
\label{Sec:Low-Cut}
In the lower gap phase, there is a single cut from 
$A_{1} = e^{-ib}$, to $B_{1} = e^{ib}$ counterclockwise.
As depicted in figure \ref{1-cutGWW}, 
the center of the cut is located at the point 1.

Now let us define the cut function $h(u)$ carefully.
Let $u=|u|e^{i \theta}$ with $\theta \in (-\pi, \pi]$, then
we will define the $h(u)$ as
\begin{equation}
h(u) = \left((u-e^{ib })
(u - e^{-ib})
\right)^{\frac{1}{2}}.
\label{h-L 1}
\end{equation}
The single cut of this function is taken to lie in an arc on the 
unit circle extending from  $A_1$ to 
$B_1$. The function $h$ is discontinuous on this arc with   
\begin{equation}
h^{\pm}(z) = \pm 2 e^{\frac{i\theta}{2}}
\sqrt{
\sin^2 \frac{b}{2} - 
\sin^2 \frac{\theta}{2}  }
\label{Behaviour in the cut}
\end{equation}
where $z = e^{i\theta}$.
Away from this cut (i.e. on the gap) on the unit circle
\begin{eqnarray}
&h(z) = 2i e^{\frac{i\theta}{2}}
\sqrt{\sin^2 \frac{\theta}{2} - 
\sin^2 \frac{b}{2}  } \qquad &\theta \ge 0, \nonumber \\
&h(z) = -2i e^{\frac{i\theta}{2}}
\sqrt{\sin^2 \frac{\theta}{2} - 
\sin^2 \frac{b}{2}  } \qquad &\theta \le 0. 
\end{eqnarray}
Recall that $\sin^2 \frac{\theta}{2} \geq \sin^2 \frac{b}{2}$
everywhere on the gap. On the real axis $h(x)$ is everywhere real and  
is positive for $x>1$, but is negative for $x<1$.  In particular $h(0)=-1$.
In more detail along the real axis
\begin{eqnarray}
&h(x) = 
\sqrt{\left(
x - \cos b
\right)^2 + 
\sin^2 b  } > 0 \qquad &x > 1, \nonumber \\
&h(x) = -\sqrt{\left(
x - \cos b
\right)^2 + 
\sin^2 b  } < 0 \qquad &x < 1.
\end{eqnarray}

\subsection{Upper gap case}
\label{Sec:Up-Cut}
Let us consider the case with one upper gap without lower gap.
In this case there is a single cut 
extending from $A_{1} = e^{ia}$ to $B_1 = e^{-ia}$ counterclockwise along
the unit circle. The cut centers on the point -1 in the complex plane, 
see Fig.~\ref{capped-GWW}.
We can also regard this region as the complement of the cut in the 
lower gap phase. This interpretation would have important role to 
show the duality later.

In this case, the function $h(z) = \sqrt{(e^{ia}-z)(e^{-ia}-z)}$ has a branch cut running counterclockwise from $e^{ia}$ to $e^{-ia}$. 
The function $h(u)$ obeys followings:
\begin{equation}\begin{split} 
-h^-(e^{i\alpha})= 
h^+(e^{i\alpha})=& + 2i e^{i\frac{\alpha}{2}}\sqrt{\sin^2\frac{\alpha}{2} - \sin^2\frac{a}{2}} {\rm~~~if~~~} \pi>\alpha>a, \\
                   & -2i e^{i\frac{\alpha}{2}}\sqrt{\sin^2\frac{\alpha}{2} - \sin^2\frac{a}{2}} {\rm~~~if~~~} -\pi<\alpha<-a ,\\     
                   h(e^{i \alpha}) =& 2e^{i\frac{\alpha}{2}}\sqrt{\sin^2\frac{a}{2}-\sin^2\frac{\alpha}{2}} {\rm ~~~if~~~} a>\alpha>-a ,\\
h(x) =& +\sqrt{(x-\cos a)^2 +\sin^2a} {\rm ~~~for~~~} x>1, \\
      & -\sqrt{(x-\cos a)^2 +\sin^2a} {\rm ~~~for~~~} x<-1 ,\\
h(z\rightarrow \infty) \rightarrow & z ,\\
h(0)=&1.
\end{split}
\label{cut-func Up}
\end{equation}

\subsection{Two gap case}
\label{Sec:Two-Cut}
In the two gap phase, 
our two cuts extend along unit circle 
from $A_1 = e^{ia}$ to $B_1 = e^{ib}$, 
and from $A_2 = e^{-ib}$ to $B_2 = e^{-ia}$ counterclockwise 
respectively, 
see the Fig.\ref{2cut.eps}.
The cut function $h(u)$ in this two gap case will be
\begin{equation}
h(u) = \left((u-e^{ia})(u-e^{-ia})(u-e^{ib})(u-e^{-ib})
\right)^{\frac{1}{2}},
\label{Cut func 2}
\end{equation}
where this function has branch cuts along the arcs enumerated above.
We will summarize some of properties of the analytic function $h(u)$.
Along the unit circle outside the cuts, $h$ is analytic as

\begin{align}
h(e^{i\theta}) =&  4 e^{i\theta} \sqrt{ ( \sin^2{a \over 2}-\sin^2{\theta \over 2}) 
(\sin^2{b \over 2} - \sin^2{\theta \over 2}) }, \, \quad -a<\theta<a, 
\label{two cut func 1}
\\
h(e^{i\theta}) = & - 4 e^{i\theta} \sqrt{ (\sin^2{\theta \over 2} - \sin^2{a \over 2}) 
(\sin^2{\theta \over 2} - \sin^2{b \over 2}) }, \, \quad -\pi<\theta<-b, \, b<\theta<\pi .
\end{align}

Along the cuts, $h$ becomes discontinuous with 
\beal{
h^{\pm}(e^{i\theta}) =& \pm 4 i e^{i\theta} \sqrt{ (\sin^2{\theta \over 2} - \sin^2{a \over 2}) 
(\sin^2{b \over 2} - \sin^2{\theta \over 2}) }, \, \quad a<\theta<b, \\
h^{\pm}(e^{i\theta}) = & \mp 4 i e^{i\theta} \sqrt{ (\sin^2{\theta \over 2} - \sin^2{a \over 2}) 
(\sin^2{b \over 2} - \sin^2{\theta \over 2}) }, \, \quad -b<\theta<-a, 
}
where $h = h^{+}$ on the outside of the circle and 
$h = h^{-}$ on the inside. Moreover $h$ is real and positive along the 
real axis; explicitly for real $x$.
 \beal{
h(x) =&  \sqrt{ (1-2x \cos a + x^2)  (1-2x \cos b + x^2) }, \, \quad -\infty<x<\infty  .
}
Note in particular that $h(0)=1$. At large $u$ we have  
\beal{
h(u) \sim +u^2 - u (\cos a + \cos b), \quad u \sim \infty.
\label{two cut func 2}
}

\section{Important formula}
\label{Ap:Form}
\subsection{Important formula in the upper gap phase}
In this subsection, we will derive several formulae
in the upper gap phase with cut function 
$h(z) = \sqrt{(e^{ia}-z)(e^{-ia}-z)}$. 
This has a branch cut 
running counterclockwise from $e^{ia}$ to $e^{-ia}$, and
has the property \eqref{cut-func Up}.

\subsubsection{Proof of \eqref{Formula up 1}}
\label{h-1-one}
Here we will prove the formula
\eqref{Formula up 1}.
Since $\frac{1}{h(\omega)}$ is everywhere holomorphic except
the cuts, from the Cauchy's theorem
\begin{equation}
 0 = \oint_{|\omega| = \infty} \frac{d \omega}{h(\omega)}
- \oint_{Ccuts} \frac{d \omega}{h(\omega)}.
\end{equation}
Note that from $h(\omega) \to \omega$ as $|\omega| \to \infty$, 
\begin{equation}
\oint_{|\omega| = \infty} \frac{d \omega}{h(\omega)}
= 2\pi i,
\end{equation}
then
\begin{equation}
\begin{split}
2\pi i = 
\oint_{Ccuts} \frac{d \omega}{h(\omega)}
=& \oint_{|\omega| = 1+|\epsilon|} \frac{d \omega}{h(\omega)}
-\oint_{|\omega| = 1-|\epsilon|} \frac{d \omega}{h(\omega)}
\\
=& \oint_{|\omega| = 1+|\epsilon|} \frac{d \omega}{h(\omega)}
+\oint_{|\omega| = 1-|\epsilon|} \frac{d \omega}{h(\omega)}
\\
=& 2\int_{Lugs} \frac{d \omega}{h(\omega)}.
\end{split}
\end{equation}
From first to second line we have used the fact 
that $\frac{1}{h(\omega)}$ is holomorphic inside the circle
$|\omega| = 1 - |\epsilon|$.
From the second to third, 
we have used $\frac{1}{h^{+}(\omega)} =
-\frac{1}{h^{-}(\omega)}$ in the cut region.
We finally obtain
\begin{equation}
\pi i = 
\int_{Lugs} \frac{d \omega}{h(\omega)}.
\end{equation}
\subsubsection{Proof of \eqref{Formula up 2}}
\label{h-one-2}
Here we will prove the formula
\eqref{Formula up 2}.

Since $\frac{1}{h(\omega)(\omega -u)}$ is everywhere holomorphic except
the cut, from the Cauchy's theorem
\begin{equation}
 0 = \oint_{|\omega| = \infty} \frac{d \omega}{h(\omega)(\omega -u)}
- \oint_{C_{cuts}} \frac{d \omega}{h(\omega)(\omega -u)}.
\end{equation}
From $\frac{1}{h(\omega)(\omega - u)} \to \omega^{-2}$, 
at $|\omega| \to \infty$,
\begin{equation}
\oint_{|\omega| = \infty} \frac{d \omega}{h(\omega)(\omega -u)}
= 0,
\end{equation}
so
\begin{equation}
\begin{split}
0 = 
\oint_{C_{cuts}} \frac{d \omega}{h(\omega)(\omega -u)}
=& \oint_{|\omega| = 1+|\epsilon|} \frac{d \omega}{h(\omega)(\omega -u)}
-\oint_{|\omega| = 1-|\epsilon|} \frac{d \omega}{h(\omega)(\omega -u)}
\\
=& \oint_{|\omega| = 1+|\epsilon|} \frac{d \omega}{h(\omega)(\omega -u)}
+\oint_{|\omega| = 1-|\epsilon|} \frac{d \omega}{h(\omega)(\omega -u)}
\\
=& 
2\int_{L_{ugs}} \frac{d \omega}{h(\omega)(\omega -u)}
+ \int_{L_{arcs}}\frac{d \omega}{h^{+}(\omega)}\left(
\frac{1}{(\omega - u)^{+}} -
\frac{1}{(\omega - u)^{-}} \right).
\end{split}
\end{equation}
Here $L_{arcs}$ denotes the complex line integration counterclockwise 
along the cuts.
Note that on the cut region,  
\footnote{We should note from the standpoint 
of the counterclockwise complex integration,
the sign of the delta function is flipped due to the opposite direction of the 
integration measure direction.}
\begin{equation}
\frac{1}{(\omega - u)^{+}} -
\frac{1}{(\omega - u)^{-}} = 2 \pi i \delta(\omega - u),
\end{equation}
so it becomes,
\begin{equation}
\int_{L_{arcs}}\frac{d \omega}{h^{+}(\omega)}\left(
\frac{1}{(\omega - u)^{+}} -
\frac{1}{(\omega - u)^{-}} \right)
= \frac{2 \pi i }{h^{+}(u)}.
\end{equation}
We finally obtain
\begin{equation}
\int_{L_{ugs}} \frac{d \omega}{h(\omega)(\omega - u)}
= -\frac{\pi i }{h^{+}(u)}.
\end{equation}
\subsection{Important formulae for the proof of level-rank 
duality in the two gap phase}
\label{Ap:2-gap dual}
In this subsection, we will derive several formulae
in the two gap phase with cut function $h(u)$ s.t.
\begin{equation}
h(u) = \left((u-e^{ia})(u-e^{-ia})(u-e^{ib})(u-e^{-ib})
\right)^{\frac{1}{2}},
\end{equation}
where this function has branch cuts 
extending along unit circle from $A_1 = e^{ia}$ to $B_1 = e^{ib}$
and from $A_2 = e^{-ib}$ to $B_2 = e^{-ia}$ running counterclockwise.
This function satisfies \eqref{two cut func 1} $\sim$ \eqref{two cut func 2}.
The formulae derived here are important 
for the proof of the duality in the two gap phase.

\subsubsection{Proof of \eqref{Formula two 1}}
\label{omega-u}
We will show \eqref{Formula two 1},
\begin{equation}
\frac{i h^{+}(u)}{4\pi^2 \lambda}\left[
\int_{L_{ugs}} d \omega~ F(\omega) 
+\int_{L_{lgs}} d \omega~ F(\omega) 
\right] = \frac{1}{2\pi \lambda}
\label{Purpose}
\end{equation}
where we define $F(\omega)$ as
\begin{equation}
F(\omega) = \frac{1}{h(\omega)}\left(
\frac{2}{\omega - u} + \frac{1}{u}
\right).
\end{equation}
Throughout this subsection
$L_{lgs}$ is the arc running counterclockwise along the lower gap 
on the unit circle, 
i.e, arc $e^{i\theta}$ s.t. $b < |\theta| < \pi$. 
$L_{ugs}$ is the arc running counterclockwise along the upper gap
on the unit circle, i.e, arc $e^{i\theta}$ s.t. $-a < \theta < a$.

Since $F(\omega)$ is holomorphic everywhere
except the cut region. Applying the Cauchy's theorem we get
\begin{equation}
0 = \oint_{|\omega| = \infty} d \omega~ F(\omega) - 
\oint_{C_{cuts}} d \omega~F(\omega) .
\end{equation}
Since $F(\omega) \to \frac{1}{\omega^2}$ as $|\omega| \to \infty$,
we have
\begin{equation}
0 = 
\oint_{|\omega| = \infty}d \omega~ F(\omega)
= \oint_{C_{cuts}} d \omega~F(\omega) .
\end{equation}
From this, we can rewrite 
\begin{align}
0 = \oint_{C_{cuts}} d \omega~F(\omega) 
=& 
\oint_{|\omega| = 1+ |\epsilon|} d \omega~F(\omega) 
-
\oint_{|\omega| = 1- |\epsilon|} d \omega~F(\omega) 
\nonumber \\
=& 
\oint_{|\omega| = 1+ |\epsilon|} d \omega~F(\omega) 
+
\oint_{|\omega| = 1- |\epsilon|} d \omega~F(\omega) 
\nonumber \\
=& \oint_{|\omega| = 1} d \omega~ (F^{+}(\omega) + F^{-}(\omega)).
\end{align}
From the first line to second line we have used the fact that
$F(\omega)$ is  holomorphic inside 
the circle $|\omega| = 1- |\epsilon|$,
\begin{equation}
\oint_{|\omega| = 1- |\epsilon|} d \omega~F(\omega) 
= 0 = -\oint_{|\omega| = 1- |\epsilon|} d \omega~F(\omega).
\end{equation}
Note that at the cuts, 
\begin{equation}
F^{+}(\omega) + F^{-}(\omega) = \frac{1}{h^{+}(\omega)}\left(
\frac{2}{(\omega - u)^{+}} -
\frac{2}{(\omega - u)^{-}} \right)
= 
\frac{1}{h^{+}(\omega)} \left( 4\pi i \delta(\omega - u)\right),
\end{equation}
and 
$F^{+}(\omega)= F^{-}(\omega) = F(\omega)$
in the upper gap as well as in the lower gap region.
Then 
\begin{equation}
\begin{split}
0 =& 
\oint_{|\omega| = 1} 
d \omega~ (F^{+}(\omega) + F^{-}(\omega))
\\
=&\int_{L_{arcs}} d \omega~ 
\frac{1}{h^{+}(\omega)} \left( 4\pi i \delta(\omega - u)\right)
+ 2\int_{L_{ugs}} d \omega~ F(\omega)
+ 2\int_{L_{lgs}} d \omega~ F^{+}(\omega)
\\
=&
\frac{4 \pi i}{h^{+}(u)} 
+ 2 \left[
\int_{L_{ugs}} d \omega~ F(\omega) 
+ \int_{L_{lgs}} d \omega~ F(\omega) \right].
\end{split}
\label{Semi-final-two-1}
\end{equation}
So by multiplying $\frac{ih^{+}(u)}{8\pi^2 \lambda}$
to the last line of 
\eqref{Semi-final-two-1}, we 
obtain \eqref{Purpose}.

\subsubsection{Proof of \eqref{Formula two 2}}
\label{homega-inv}
Let us prove the formula 
\eqref{Formula two 2},
\begin{equation}
\int_{L_{ugs}} d \omega~ \frac{1}{h(\omega)} 
= - \int_{L_{lgs}} d \omega~ \frac{1}{h(\omega)} .
\end{equation}

Because $\frac{1}{h(\omega)}$ is holomorphic everywhere
except the cuts region, from the Cauchy's theorem, we get
\begin{equation}
0 = \oint_{|\omega| = \infty} d \omega~ \frac{1}{h(\omega)} - 
\oint_{C_{cuts}} d \omega~\frac{1}{h(\omega)} .
\end{equation}
Using $\frac{1}{h(\omega)} \to \frac{1}{\omega^2}$ as $|\omega| \to \infty$,
we also get
\begin{equation}
0 = \oint_{|\omega| = \infty}d \omega~ \frac{1}{h(\omega)}
= 
\oint_{C_{cuts}} d \omega~\frac{1}{h(\omega)} .
\end{equation}
Hence, we can rewrite
\begin{align}
0=\oint_{C_{cuts}} d \omega~\frac{1}{h(\omega)} 
=& 
\oint_{|\omega| = 1+ |\epsilon|} d \omega~\frac{1}{h(\omega)} 
-
\oint_{|\omega| = 1- |\epsilon|} d \omega~\frac{1}{h(\omega)}  
\nonumber \\
=&
\oint_{|\omega| = 1+ |\epsilon|} d \omega~\frac{1}{h(\omega)} 
+
\oint_{|\omega| = 1- |\epsilon|} d \omega~\frac{1}{h(\omega)}  
\nonumber \\
=& 
\oint_{|\omega| = 1} d \omega~ \left(\frac{1}{h^{+}(\omega)} + 
\frac{1}{h^{-}(\omega)}\right)
\nonumber \\
=& 2
\left[\int_{L_{ugs}} d \omega~ \frac{1}{h(\omega)} + 
\int_{L_{lgs}} d \omega~ \frac{1}{h(\omega)} \right]. 
\label{Semi-final-two-2}
\end{align}
From first line to second line, we have used the fact that
$\frac{1}{h(\omega)}$ is holomorphic 
inside the circle $|\omega| = 1- |\epsilon|$,
\begin{equation}
\oint_{|\omega| = 1- |\epsilon|} d \omega~\frac{1}{h(\omega)} 
= 0 = -\oint_{|\omega| = 1- |\epsilon|} d \omega~\frac{1}{h(\omega)} .
\end{equation}
In the last step from the third to the forth line of 
\eqref{Semi-final-two-2}, 
we have used that
$\frac{1}{h^{+}(\omega)} = -\frac{1}{h^{-}(\omega)}$ 
in the cuts region 
while
$\frac{1}{h^{+}(\omega)} = \frac{1}{h^{-}(\omega)}
= \frac{1}{h(\omega)}$ in the gaps region.
Hence
we can see from the last line of 
\eqref{Semi-final-two-2},
\begin{equation}
\int_{L_{ugs}} d \omega~ \frac{1}{h(\omega)} 
= - \int_{L_{lgs}} d \omega~ \frac{1}{h(\omega)} .
\end{equation}
This completes the proof of \eqref{Formula two 2}.

\subsubsection{Proof of \eqref{Formula two 3}}
\label{homega-omega}
Since $\frac{\omega}{h(\omega)}$ is holomorphic everywhere 
except the cuts, from the Cauchy's theorem, we get 
\begin{equation}
0 = 
\frac{1}{2\pi i}\oint_{|\omega| = \infty} d\omega~ \frac{\omega}{h(\omega)}
-\frac{1}{2\pi i}\oint_{C_{cuts}} d\omega~ \frac{\omega}{h(\omega)}.
\end{equation}
Using $\frac{1}{h(\omega)} \to \frac{1}{\omega^2}$ 
as $|\omega| \to \infty$, we also get
\begin{equation}
\frac{1}{2\pi i}\oint_{C_{cuts}} d\omega~ \frac{\omega}{h(\omega)}= 
\frac{1}{2\pi i}\oint_{|\omega| = \infty} d\omega~ \frac{\omega}{h(\omega)}
= \frac{1}{2\pi i}\int^{2\pi}_{0} d \theta \frac{i\omega^2}{\omega^2}
= 1.
\end{equation}
Hence
\begin{equation}
\begin{split}
1 = 
\frac{1}{2\pi i}\oint_{C_{cuts}} d\omega~ \frac{\omega}{h(\omega)}
=& 
\frac{1}{2\pi i}\oint_{|\omega| = 1+|\epsilon|} d\omega~ 
\frac{\omega}{h(\omega)}
-\frac{1}{2\pi i}\oint_{|\omega| = 1-|\epsilon|} d\omega~ 
\frac{\omega}{h(\omega)}
\\
=&\frac{1}{2\pi i}\oint_{|\omega| = 1+|\epsilon|} d\omega~ 
\frac{\omega}{h(\omega)}
+\frac{1}{2\pi i}\oint_{|\omega| = 1-|\epsilon|} d\omega~ 
\frac{\omega}{h(\omega)}
\\
=&
\frac{1}{\pi i}\int_{L_{ugs}} d\omega~ 
\frac{\omega}{h(\omega)}
+\frac{1}{\pi i}\int_{L_{lgs}} d\omega~ 
\frac{\omega}{h(\omega)}.
\end{split}
\label{semi-final-two-3}
\end{equation}
From first line to second line
we have used the fact that $\frac{\omega}{h(\omega)}$ is holomorphic
inside the circle $|\omega| = 1- |\epsilon|$.
From the second line to third line we have used the fact that
$\left(
\frac{\omega}{h(\omega)}
\right)^{+} = 
-\left(
\frac{\omega}{h(\omega)}
\right)^{-}$
on the cuts region 
while 
$\left(
\frac{\omega}{h(\omega)}
\right)^{+} = 
\left(
\frac{\omega}{h(\omega)}
\right)^{-} = 
\left(
\frac{\omega}{h(\omega)}
\right)
$ in the gaps region.
From the last line of 
\eqref{semi-final-two-3},
we can see
\begin{equation}
1 = \frac{1}{\pi i}\int_{L_{ugs}} d\omega~ 
\frac{\omega}{h(\omega)}
+\frac{1}{\pi i}\int_{L_{lgs}} d\omega~ 
\frac{\omega}{h(\omega)}.
\label{Formula homega-omega}
\end{equation}
This completes the proof of the formula \eqref{Formula two 3}

\section{Behavior of eigenvalue distribution at large $\zeta$}
\label{Ap:LZ-sec}
We expect that as $\zeta \to \infty$, with fixed $\lambda$,
the eigenvalue densities would behave as 
\begin{equation}
\begin{split}
\rho(\alpha) &= \frac{1}{2 \pi \lambda}~~~(|\alpha| < \pi \lambda)
\\
&= 0 ~~~~~~~(|\alpha| > \pi \lambda).
\end{split}
\label{evd}
\end{equation}
First we will verify that at $\zeta = \infty$, with 
fixed finite $\lambda$, 
the eigenvalue density 
in each CS matter theory
must be this function.
After that we will elaborate more on the behavior of the eigenvalue densities
in this limit more.

\subsection{Proof of the universal distribution of eigenvalue density 
at $\zeta = \infty$}
To verify it, it is useful to note that 
$\Upsilon(a,b) = \infty \Rightarrow a = b$,
at $\zeta = \infty$ with finite $\lambda$. 
Here $\Upsilon(a,b)$ 
is defined in \eqref{RF two u1}.
Note that since 
\begin{equation}
\frac{1}{\sqrt{\sin^2\frac{b}{2}-\sin^2 \frac{\alpha}{2}} }
\le \frac{1}{\cos \frac{\alpha}{2}\sqrt{\sin^2\frac{b}{2}-\sin^2 \frac{a}{2}}} ,
\end{equation}
for every $|\alpha| \le a \le b$, 
it would be
\begin{equation}
\label{detab-2}
\Upsilon(a,b) \le \frac{1}{\sqrt{\sin^2\frac{b}{2}-\sin^2 \frac{a}{2}}} 
\int_{-a}^a d \alpha  
\frac{1}
{\cos \frac{\alpha}{2}\sqrt{\sin^2\frac{a}{2}-\sin^2 \frac{\alpha}{2}}}
= 
\frac{1}{2 \cos \frac{a}{2}\sqrt{\sin^2\frac{b}{2}-\sin^2 \frac{a}{2}}}.
\end{equation}
From \eqref{detab-2}
we can see that 
$\Upsilon = \infty
\Rightarrow 
a = b.$
Also in the case of $a = \pi$, 
because of the constraint $a \le b \le \pi$,
$a = b$ is automatically satisfied.
So we conclude that 
\begin{equation}
\Upsilon(a,b) = \infty
\Rightarrow 
a = b.
\label{Proof-Upsilon}
\end{equation}
In following subsubsections, 
we will verify the statement by using \eqref{Proof-Upsilon}.

\subsubsection{Regular fermion case}
\label{RF-Z-Pr}
If the $\tilde{c}$ remains finite 
in the limit $\zeta \to \infty$ with fixed $\lambda$,
from the third line of 
\eqref{RF two u1}, it follows that 
${\cal Y}^{r.f}(a,b,\tilde{c})$ 
remains non-zero finite.
Then from the first line of \eqref{RF two u1},
$\Upsilon(a,b)$ becomes infinite.
Hence 
it follows $a = b$ 
in the limit
$\zeta \to \infty$ with fixed $\lambda$
if the 
$\tilde{c}$ remains finite.

To complete the proof we have to show 
that $\tilde{c}$ remains finite in this limit.
$\tilde{c}$ is given by
\begin{equation}
\tilde{c}
= f_{\rho;\lambda}(\tilde{c})
\equiv 
\frac{\lambda}{(1-\lambda)}\int^{\pi}_{-\pi}
d \alpha \rho(\alpha)
\log ( 1 + 2 \cos \alpha e^{-\tilde{c}} + e^{-2\tilde{c}}).
\label{dec-c-RF}
\end{equation}
The function $f_{\rho;\lambda}(\tilde{c})$ 
is positive by the property of $\rho$ as a probability function and
by the property that $\rho(\alpha') \le \rho(\tilde{\alpha})$ if 
$\pi \ge |\alpha'| \ge |\tilde{\alpha}|$.
At positive $\tilde{c}$ we can see the following 
inequality
\begin{equation}
\begin{split}
0 
&\le 
f_{\rho;\lambda}(\tilde{c})
\le
\frac{2\lambda}{(1-\lambda)} \log 2.
\end{split}
\label{Ineq}
\end{equation}
Solution of \eqref{dec-c-RF}
is represented by the 
the intersection point of the line $y = x$ and
$y = f_{\rho,\lambda}(x)$
in the $x-y$ plane, which is denoted as 
$(x,y) = (\tilde{c}, \tilde{c})$. 
From the inequality \eqref{Ineq}, we 
immediately see that 
\begin{equation}
\tilde{c} \le \frac{2\lambda}{(1-\lambda)} \log 2.
\end{equation}
$\frac{2\lambda}{(1-\lambda)} \log 2$ is finite
quantity for every $0 < \lambda < 1$. 
\footnote{We can not apply this analysis for the case $\lambda = 1$.
But in case of $\lambda =1$, the eigenvalue density is always
$\rho(\alpha) = \frac{1}{2\pi}$, which already
obeys the universal density.
So we do not have to mind the $\lambda =1$ case.}
From this inequality, 
we can see that the $\tilde{c}$ remains positive finite 
even in the limit $\zeta \to \infty$ with fixed $\lambda$. 
Hence we have shown that the eigenvalue density becomes 
universal distribution \eqref{evd} in the limit 
$\zeta \to \infty$.

\subsubsection{Critical boson case}
\label{CB-Z-Pr}
We can show in a way similar to the regular fermion case.
If the positive number $\sigma$ remains finite
in the limit $\zeta \to \infty$ with fixed $\lambda$,
from the second line of \eqref{CB two u1}, 
${\cal Y}^{c.b}(a,b,\sigma)$ 
remains non-zero finite.
Then from the first line of \eqref{CB two u1}, 
$\Upsilon(a,b)$ becomes 
infinite. 
Therefore 
it follows $a = b$ in the limit $\zeta \to \infty$ with fixed $\lambda$, 
if $\sigma$ remains finite.

In order to complete the proof, we have to show 
that the $\sigma$ remains positive 
finite in the corresponding limit.
$\sigma$ is given by the following equation
\begin{equation}
x
=
g_{\rho}(x)
\equiv
-\int^{\pi}_{-\pi} d \alpha~ \rho(\alpha) 
\log \left(1 - 2 \cos \alpha e^{-x} + e^{-2x}\right).
\label{dec-c-CB}
\end{equation}
Solution of \eqref{dec-c-CB}
is represented by the 
intersection point of the line $y = x$ and
$y = g_{\rho}(x)$
in the $x-y$ plane, which is denoted as 
$(x,y) = (\sigma, \sigma)$.
Note that at $x > 0$, 
\begin{equation}
0 \le g_{\rho}(x)
\le  
-2\log \left(1 -  e^{-x}\right) .
\label{Const-CB-zeta}
\end{equation}
Suppose that 
$x = \hat{x} > 0$ is the solution of the 
equation 
\begin{equation}
x = -2\log \left(1 -  e^{-x}\right),
\label{eq:upper CB zeta}
\end{equation}
from the inequality \eqref{Ineq}, we 
can immediately see that 
\begin{equation}
\sigma \le \hat{x} = \log\left(\frac{3 + \sqrt{5}}{2}\right).
\end{equation}
So 
$\sigma$ remains finite 
even in the limit. 
Then we have shown that the eigenvalue density becomes 
universal distribution \eqref{evd} in the limit $\zeta \to \infty$
with fixed $\lambda$.

\subsubsection{SUSY CS matter theory case}
\label{SUSY-Z-Pr}
We can show in a way similar to the previous cases.
As in the previous cases, 
if the positive number $\tilde{c}$ remains finite
in the limit 
$\zeta \to \infty$ with fixed $\lambda$, 
from the second line of \eqref{SUSY two u1}, 
${\cal Y}^{su}(a,b,\sigma)$ remains non-zero finite. 
Then from the first line of \eqref{SUSY two u1},
$\Upsilon(a,b)$ becomes 
infinite. 
Hence it follows $a = b$ in the limit 
$\zeta \to \infty$ with fixed $\lambda$, 
if the $\tilde{c}$ remains finite. 

In order to complete the proof, we have to show 
that the $\tilde{c}$ remains finite at the corresponding limit.
The $\tilde{c}$ is given by the equation
\begin{equation}
\tilde{c} =
\lambda
 \biggl | \int_{-\pi}^{\pi} d\alpha
\rho(\alpha) 
\log 
\frac{1 + 2 \cos \alpha e^{-\tilde{c}} + e^{-2\tilde{c}} }
{1 - 2 \cos \alpha e^{-\tilde{c}} + e^{-2\tilde{c}} }
 \biggr |. 
\label{dec-c}
\end{equation}
For any $\tilde{c} > 0$, $0\le \lambda\le 1$, 
we see the following inequality,
\begin{eqnarray}
\lambda
 \biggl | \int_{-\pi}^{\pi} d\alpha
\rho(\alpha) 
\log 
\frac{1 + 2 \cos \alpha e^{-\tilde{c}} + e^{-2\tilde{c}} }
{1 - 2 \cos \alpha e^{-\tilde{c}} + e^{-2\tilde{c}} }
 \biggr | 
\le
2 \biggl | 
\log 
\frac{2}{1 - e^{-\tilde{c}}}
 \biggr |.
\label{Const-zeta-ineSUSY}
\end{eqnarray}
Note that the last term is independent of $\zeta$.
Suppose that $x = \tilde{c}_{o} > 0$ is 
the solution of the equation
\begin{equation}
x = 2  \biggl| 
\log 
\frac{2 }{1 - e^{-x}}
 \biggr | ,
\end{equation}
due to the inequality \eqref{Const-zeta-ineSUSY},
we see 
\begin{equation}
0<\tilde{c} \le \tilde{c}_o = \log(3 + 2\sqrt{2}).
\end{equation}
Hence we conclude 
the positive $\tilde{c}$ remains finite 
even in the limit $\zeta \to \infty$,
then we have shown that the eigenvalue density becomes 
the universal distribution \eqref{evd} in the limit 
$\zeta \to \infty$ with fixed $\lambda$.

\subsection{Behavior of eigenvalue density in large $\zeta$ limit,
with $1 \gg \frac{1}{\zeta} >0$}
\label{Z-gen}
We will elaborate how the eigenvalue density deviates from the
universal distribution \eqref{evd} if we gradually decrease the temperature 
from $\zeta = \infty$.


\subsubsection{Regular fermion theory}
\label{RF-LZ}
To consider the behavior in the regular fermion theory, 
we should evaluate behavior of the combination of
\eqref{RF two u1}, \eqref{RF two u0} and \eqref{tc}
at large $\zeta$ where $b \sim a + \epsilon$ with $\epsilon \ll 1$.
(As in section C.2 in \cite{Jain:2013py}.)
In this limit, 
these three equations can be 
expanded as in $\epsilon$ to give
\begin{equation}
\begin{split}
&\frac{1}{2 \pi \lambda}\left(
2 \frac{1}{\sin(a)} \log\left(\frac{8\sin(a)}{\epsilon}\right)
-\epsilon \frac{\cos(a)}{\sin^{2}(a)}\left(-1+\log\left(\frac{8\sin(a)}{\epsilon}\right)\right)\right)+O[\epsilon^2]
\\
&=\frac{\zeta}{2 \pi}
\int^{\infty}_{\tilde{c}} dy~
\left(
\frac{y}{\cosh y + \cos a}
+\epsilon\frac{ y \sin a}{2(\cosh y + \cos a)^2}
\right),
\end{split}
\end{equation}
and 
\begin{equation}
\begin{split}
&\frac{1}{2\pi\lambda}
\left(2~a+2~\cot(a) \log\left(\frac{8\sin(a)}{\epsilon}\right)-\epsilon\frac{1}{\sin^{2}(a)}\left(-1+\log\left(\frac{8\sin(a)}{\epsilon}\right)\right)\right)+O[\epsilon^2]
\\
&= 1+\frac{\zeta}{2 \pi}
\int^{\infty}_{\tilde{c}} dy~\frac{y \cos a}{
\cosh y + \cos a}
-\epsilon \frac{\zeta}{4 \pi}
\int^{\infty}_{\tilde{c}} dy~
\frac{  y\cosh y \sin a}{
(\cosh y + \cos a)^2
} ,
\end{split}
\end{equation}
and 
\begin{equation}
\tilde{c} = \frac{1}{2 \pi (1- \lambda)}
\int^{\pi \lambda}_{-\pi \lambda} d \alpha \log(1+2 \cos \alpha e^{-\tilde{c}} 
+e^{-2\tilde{c}} ),
\end{equation}
respectively.
From these we can obtain $\epsilon$ and $a, b$ as 
\begin{align}
\epsilon &= 8 \sin (\pi \lambda) \exp
\left(
- \frac{\sin (\pi\lambda)}{2}\lambda\zeta
\int^{\infty}_{\tilde{c}} dy~\frac{y}{\cosh y + \cos \pi \lambda}\right) + \ldots,
\label{epsilon-RF} \\
a &= \pi \lambda - \frac{1}{2}\epsilon, \qquad 
b =  \pi \lambda + \frac{1}{2}\epsilon.
\end{align}
Since 
\begin{equation}
Y_{rf} = 
\int^{\infty}_{\tilde{c}} dy~\frac{y}{\cosh y + \cos \pi \lambda} \sim 
{\cal O}(\epsilon^0),
\end{equation}
$\zeta$ is estimated as 
$\zeta \sim - \log \epsilon$ 
from \eqref{epsilon-RF}.

Let us consider the asymptotic behavior of the eigenvalue density
\eqref{eigen-RF-two} in $\epsilon \to 0$ limit.
Here $\rho_{2,tg}$ of \eqref{eigen-RF-two} 
is the same functional form as the one in the 
GWW type model (7.6) of \cite{Jain:2013py},
while $\rho^{r.f}_{1,tg}$ of \eqref{eigen-RF-two} 
is regarded as an additional
term appearing in the regular fermion theory.
We can see that $\rho^{r.f}_{1,tg}$
falls off as $\rho^{r.f}_{1,tg} \sim -\epsilon \log \epsilon \to 0$.
On the other hand $\rho_{2,tg}$
remains finite.
So in the limit, $\rho$ becomes
\begin{equation}
\rho(\alpha) = \rho_{2.tg}(\lambda, \pi\lambda - \frac{1}{2}\epsilon,
\pi\lambda+\frac{1}{2}\epsilon ;\alpha) 
= \frac{1}{\pi^2\lambda}\cos^{-1}\sqrt{\alpha_1}
\end{equation}
where $\alpha = a + \alpha_1 \epsilon$ where $ 0 < \alpha_1 <1$.
This behavior is very similar 
to the one in the GWW type model, 
(7.11) of \cite{Jain:2013py}.

\begin{figure}
  \begin{center}
  \subfigure[]{\includegraphics[scale=.33]{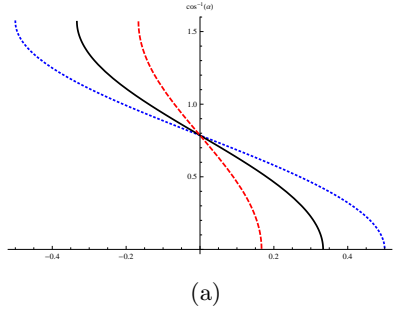}
\label{fig:ep-eigen-RF}}
\caption{Plots of $\cos^{-1}(\sqrt{\alpha_1})$ curves,
which are the eigenvalue density functions on the cut domain 
$\pi\lambda -\frac{\epsilon}{2}\le \alpha \le 
\pi\lambda +\frac{\epsilon}{2}$
at $\zeta \to \infty$. (Horizontal axis is $\alpha$ and vertical axis is
the function 
$\cos^{-1} \sqrt{\frac{\alpha-\pi\lambda}{\epsilon} + \frac{1}{2}}$
We did not plot precise $\epsilon$ here. ) 
By using these schematic graphs, 
we are trying to compare the behaviors of the eigenvalue densities; 
the one of the GWW, the one of the regular fermion (the critical boson) and the one in the SUSY CS matter theory.
Dotted blue 
line is the one in the GWW model and the bold line is 
the regular fermion's one,
and the dashed red line is the one in the SUSY CS matter theory.
The regular fermion theory has a sharper slope than the one in the GWW model,
moreover the SUSY theory has 
a sharper slope than the one in the regular fermion theory. 
At fixed $\lambda$, the center of the cut domain is always fixed at 
$\lambda$.
}
  \end{center}
\end{figure}

Although the functional form of $\rho$ in the large 
$\zeta$ limit is almost same as the 
one in the GWW type model, the range of $\epsilon$ 
in the regular fermion theory 
would be different from $\epsilon$ in the GWW type model. 
To evaluate the $\epsilon$, 
we should calculate the factor
$Y_{rf}$
in \eqref{epsilon-RF},
and compare with  
$\epsilon = 8 \sin (\pi \lambda) 
\exp\left(- \frac{\sin (\pi\lambda)}{2}\lambda\zeta\right)$ 
in the GWW case.
By numerical calculations, we can confirm that
\begin{equation}
Y_{rf} >1
\end{equation}
for every value of $\lambda$.
We plot the 
function $Y_{rf}$ in Fig.~\ref{fig:wide-rang-e SUSY}.
Hence we see that $\epsilon$ at $(\lambda, \zeta)$ 
of the regular fermion theory
\eqref{epsilon-RF} 
is smaller than $\epsilon$ at the same $(\lambda, \zeta)$ 
of GWW model.
This means that the shape of the eigenvalue density function of 
the regular fermion theory has a sharper cliff than the one in 
GWW model.
(Schematic graph of the eigenvalue density is depicted at 
Fig.~\ref{fig:ep-eigen-RF}.)

\subsubsection{Critical boson theory}
\label{CB-LZ}
Also in the critical boson theory,
we consider 
how the eigenvalue density deviates from universal configuration
\eqref{evd} as we decrease the temperature.

To consider the behavior, 
we should investigate the combination of the 
conditions
\eqref{CB two u1}, \eqref{CB two u0}, and \eqref{selfboscri} 
in the large $\zeta$ where $b = a + \epsilon$ with $\epsilon \ll 1$.
In this limit, the asymptotic behavior of 
these three equations are 
\begin{equation}
\begin{split}
&\frac{1}{2 \pi \lambda}\left(
2 \frac{1}{\sin(a)} \log\left(\frac{8\sin(a)}{\epsilon}\right)
-\epsilon \frac{\cos(a)}{\sin^{2}(a)}\left(-1+\log\left(\frac{8\sin(a)}{\epsilon}\right)\right)\right)+O[\epsilon^2].
\\
=&\frac{\zeta}{2 \pi}
\int^{\infty}_{\sigma} dy~
\left(
\frac{y}{\cosh y - \cos a}
-\epsilon\frac{ y \sin a}{2(\cosh y - \cos a)^2}
\right),
\end{split}
\end{equation}
and 
\begin{equation}
\begin{split}
&\frac{1}{2\pi\lambda}
\left(2~a+2~\cot(a) \log\left(\frac{8\sin(a)}{\epsilon}\right)-\epsilon\frac{1}{\sin^{2}(a)}\left(-1+\log\left(\frac{8\sin(a)}{\epsilon}\right)\right)\right)+O[\epsilon^2] \\
=&1+\frac{\zeta}{2 \pi}
\int^{\infty}_{\sigma} dy~\frac{y \cos a}{
\cosh y - \cos a}
-\epsilon \frac{\zeta}{4 \pi}
\int^{\infty}_{\sigma} dy~
\frac{  y\cosh y \sin a}{
(\cosh y - \cos a)^2
} ,
\end{split}
\end{equation}
and 
\begin{equation}
\sigma
= 
-\frac{1}{2 \pi \lambda}
\int^{\pi \lambda}_{-\pi \lambda} 
d \alpha~ 
\log \left(1 - 2 \cos \alpha e^{-\sigma} + e^{-2\sigma}\right),
\label{CB-3-s-decide-2}
\end{equation}
respectively. From these, $\epsilon, a$ and $b$ are evaluated as
\begin{align}
\epsilon =& 8 \sin (\pi \lambda)
\exp\left(
-\frac{1}{2} \sin (\pi \lambda)
\zeta \lambda
\int^{\infty}_{\sigma}
\frac{d y~ y}{\cosh y - \cos \pi\lambda}
\right),  \label{epsilon-CB}\\
a =& \pi \lambda - \frac{\epsilon}{2}, \qquad
b = \pi \lambda + \frac{\epsilon}{2}.
\end{align}
Since
\begin{equation}
Y_{cb} = \int^{\infty}_{\sigma} dy~ \frac{y}{\cosh y - \cos \pi\lambda}
\sim {\cal O}(\epsilon^{0}),
\end{equation}
$\zeta$ is evaluated as 
$\zeta \sim - \log \epsilon$
from \eqref{epsilon-CB}.

Let us consider the asymptotic behavior of the eigenvalue density
\eqref{eigen-CB-two} in $\epsilon \to 0$ limit.
Here $\rho_{2,tg}$ of \eqref{eigen-CB-two} 
is the same functional form as the one in the 
GWW type model (7.6) of \cite{Jain:2013py},
while $\rho^{c.b}_{1,tg}$ of \eqref{eigen-CB-two} 
is regarded as an additional
term appearing in the critical boson theory.
We can see that $\rho^{c.b}_{1,tg}$
falls off as $\rho^{c.b}_{1,tg} \sim -\epsilon \log \epsilon \to 0$.
On the other hand $\rho_{2,tg}$
remains finite.
So in this limit, $\rho$ becomes
\begin{equation}
\rho(\alpha) = \rho_{2.tg}(\lambda, \pi\lambda - \frac{1}{2}\epsilon,
\pi\lambda+\frac{1}{2}\epsilon ;\alpha) 
= \frac{1}{\pi^2\lambda}\cos^{-1}\sqrt{\alpha_1}.
\end{equation}
This behavior is also very similar
to the one in 
the regular fermion case as well as the one in the GWW type 
(7.11) of \cite{Jain:2013py}.

The range of $\epsilon$ 
in the critical boson theory 
would also be different from $\epsilon$ in the GWW type model. 
To evaluate the $\epsilon$, 
we should calculate the factor
$Y_{cb}$
in \eqref{epsilon-CB},
and compare with  
$\epsilon$ 
in the GWW case.
By numerical calculations, we confirm that
\begin{equation}
Y_{cb} >1
\end{equation}
for every value of $\lambda$.
(See the plots of $Y_{cb}$ in Fig.~\ref{fig:wide-rang-e SUSY}).
This means that the shape of the eigenvalue density function of 
the critical boson theory has a sharper cliff than the one in 
GWW model.
(Schematic graph of the eigenvalue density is depicted at 
Fig.~\ref{fig:ep-eigen-RF}.)
We can see that $\epsilon$ in 
\eqref{epsilon-CB} as well as $Y_{cb}$
are dual to $\epsilon$ in \eqref{epsilon-RF} and $Y_{rf}$
under the level-rank duality.
We can also see the duality between $Y_{cb}$ and $Y_{rf}$ from the graph
Fig.~\ref{fig:wide-rang-e SUSY}.

\subsubsection{Supersymmetric CS matter theory}
\label{SUSY-LZ}
Similar to the other cases, 
we consider the behavior of the combination 
of \eqref{SUSY two u1}, \eqref{SUSY two u0} and 
\eqref{susyct} at large $\zeta$.
In this limit these behave as
\begin{equation}
\begin{split}
&\frac{1}{2 \pi \lambda}\left(
2 \frac{1}{\sin(a)} \log\left(\frac{8\sin(a)}{\epsilon}\right)
-\epsilon \frac{\cos(a)}{\sin^{2}(a)}\left(-1+\log\left(\frac{8\sin(a)}{\epsilon}\right)\right)\right)+O[\epsilon^2]
\\
=&\frac{\zeta}{2 \pi}
\int^{\infty}_{\tilde{c}} dy~
\left(
\frac{2y\cosh y}{\cosh^2 y - \cos^2 a}
-\epsilon\frac{ 2y \sin a\cos a \cosh y}{(\cosh^2 y - \cos^2 a)^2}
\right),
\end{split}
\end{equation}
and 
\begin{equation}
 \begin{split}
&\frac{1}{2\pi\lambda}
\left(2~a+2~\cot(a) \log\left(\frac{8\sin(a)}{\epsilon}\right)-\epsilon\frac{1}{\sin^{2}(a)}\left(-1+\log\left(\frac{8\sin(a)}{\epsilon}\right)\right)\right)+O[\epsilon^2] \\
=& 1+\frac{\zeta}{2 \pi}
\int^{\infty}_{\tilde{c}} dy~\frac{2y \cos a \cosh y}{
\cosh^2 y - \cos^2 a}
- \frac{\epsilon\zeta}{4 \pi}
\int^{\infty}_{\tilde{c}} dy~
\frac{ 2 y\cosh y \sin a(\cosh^2 y + \cos^2 a)}{
(\cosh^2 y - \cos^2 a)^2
} ,
\end{split}
\end{equation}
and 
\begin{equation}
\tilde{c} 
=
\frac{1}{2\pi}
\biggl | \int_{-\pi \lambda}^{\pi \lambda} d\alpha
\log 
\frac{1 + 2 \cos \alpha e^{-\tilde{c}} + e^{-2\tilde{c}} }
{1 - 2 \cos \alpha e^{-\tilde{c}} + e^{-2\tilde{c}} }
 \biggr |. 
\label{dec-c 1}
\end{equation}
From these, $\epsilon, a$ and $b$ are evaluated as
\begin{eqnarray}
\epsilon &=& 8 \sin (\pi \lambda)
\exp\left(
- \frac{\sin (\pi \lambda)}{2}
\zeta \lambda
\int^{\infty}_{\tilde{c}}
\frac{d y~ 2y \cosh y}{\cosh^2 y - \cos^2 \pi\lambda}
\right) +\ldots,
\label{SUSY-epsilon}
\\
a &=& \pi \lambda - \frac{\epsilon}{2}, \qquad
b = \pi \lambda + \frac{\epsilon}{2}.
\end{eqnarray}
Since
\begin{equation}
Y_{su} = 
\int^{\infty}_{\tilde{c}}
\frac{d y~ 2y \cosh y}{\cosh^2 y - \cos^2 \pi\lambda}
\sim {\cal O}(\epsilon^{0}),
\end{equation}
we can see 
$\zeta \sim - \log \epsilon.$

Let us consider the asymptotic behavior of the eigenvalue density
\eqref{eigen-SUSY-two} in the limit.
%
Here $\rho_{2,tg}$ of \eqref{eigen-SUSY-two} 
is the same functional form as the one in the 
GWW type model (7.6) of \cite{Jain:2013py},
while $\rho^{susy}_{1,tg}$ of \eqref{eigen-SUSY-two} 
is regarded as an additional
term appearing in the SUSY CS matter theory.
We can see that $\rho^{susy}_{1,tg}$
falls off as $\rho^{susy}_{1,tg} \sim -\epsilon \log \epsilon \to 0$.
On the other hand $\rho_{2,tg}$
remains finite.
So in the limit, $\rho$ becomes
\begin{equation}
\rho(\alpha) = \rho_{2.tg}(\lambda, \pi\lambda - \frac{1}{2}\epsilon,
\pi\lambda+\frac{1}{2}\epsilon ;\alpha) 
= \frac{1}{\pi^2\lambda}\cos^{-1}\sqrt{\alpha_1}.
\end{equation}
This behavior is very similar to other 
CS matter theories. 

\begin{figure}
  \begin{center}
  \subfigure[]{\includegraphics[scale=.38]{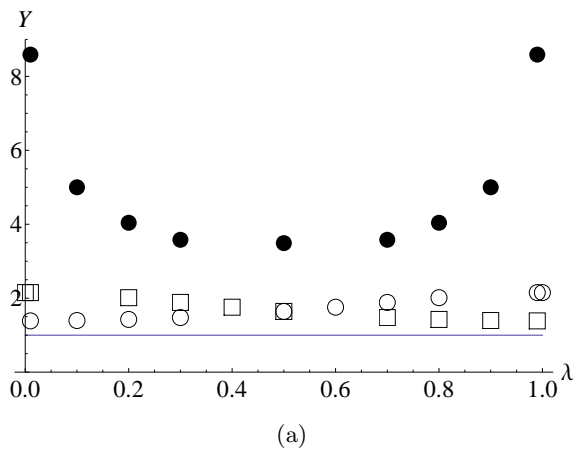}
\label{fig:wide-rang-e SUSY}}
\caption{The points with black filled circle are
plots of $Y=Y_{su}$ in the SUSY CS matter theory. 
The circle and square plots without being filled are 
$Y=Y_{rf} = \int^{\infty}_{\tilde c} dy 
\frac{y}{\cosh y + \cos \pi \lambda}$ in the regular fermion theory 
and $Y=Y_{cb} = \int^{\infty}_{\sigma} dy 
\frac{y}{\cosh y - \cos \pi \lambda}$ at the critical boson theory respectively. Dotted horizontal line indicates $Y = 1$.
We can see that for any $\lambda$,
$Y_{su} > Y_{rf} >1 $ and $Y_{su} > Y_{cb} >1$.}
  \end{center}
\end{figure}

The range of $\epsilon$ 
in the SUSY CS matter theory  
would be the smallest among the 
CS matter theories which we have already studied in this paper.
We can estimate $\epsilon$ 
by calculating the factor $Y_{su}$
in \eqref{SUSY-epsilon}.
(The plots of $Y_{su}$ is in Fig.~\ref{fig:wide-rang-e SUSY}).
From this, we can see $Y_{su} > Y_{cb},(Y_{rf}) >1$.
This means that $\epsilon$ in the SUSY CS matter theory
is the smallest and shape of the eigenvalue density function of 
the SUSY CS matter theory has the sharpest cliff among the CS matter theories.
(Schematic graph of the eigenvalue density is depicted at 
Fig.~\ref{fig:ep-eigen-RF}.)
We can see that $\epsilon$ at 
\eqref{SUSY-epsilon} as well as $Y_{su}$
are self-dual under the level-rank duality.
We can also see the self-duality of $Y_{su}$ from the graph
Fig.~\ref{fig:wide-rang-e SUSY}.

\bibliographystyle{JHEP}
\bibliography{ccs}
\end{document}